\documentclass[10pt]{article}
\usepackage{fullpage}
\usepackage{amsmath}
\usepackage{amssymb}
\usepackage{amsfonts}
\usepackage{comment}
\usepackage[dvips]{epsfig}
\usepackage{color}
\usepackage{cite}

\DeclareMathOperator{\arccot}{arccot}

\def\##1{\underline{#1}}
\def\=#1{\underline{\underline{#1}}}

\def\+
#1{\underline{\bf #1}}
\def\*#1{\underline{\underline{\bf #1}}}

\def\r#1{(\ref{#1})}
\def\l#1{\label{#1}}
\def\c#1{\cite{#1}}

\def\le{\left(}
\def\ri{\right)}
\def\les{\left[}
\def\ris{\right]}
\def\lec{\left\{}
\def\ric{\right\}}

\def\.{\mbox{ \tiny{$^\bullet$} }}

\def\eps{\varepsilon}

\def\epso{\eps_{\scriptscriptstyle 0}}
\def\lambdao{\lambda_{\scriptscriptstyle 0}}
\def\muo{\mu_{\scriptscriptstyle 0}}
\def\etao{\eta_{\scriptscriptstyle 0}}

\def\ko{k_{\scriptscriptstyle 0}}

\def\curl{\nabla\times}
\def\ux{\hat{\#u}_x}
\def\uy{\hat{\#u}_y}
\def\uz{\hat{\#u}_z}

\def\curl{\nabla\times}

\def\calA{{\cal A}}
\def\calB{{\cal B}}

\def\PAmat{\les\=P_\calA\ris}
\def\PBmat{\les\=P_\calB\ris}

\begin{document}

\begin{center}

\LARGE{ {\bf  Surface-plasmon-polariton wave propagation supported by anisotropic materials: multiple modes and  mixed exponential and linear localization characteristics
}}
\end{center}
\begin{center}
\vspace{10mm} \large

Chenzhang Zhou\\
 {\em NanoMM~---~Nanoengineered Metamaterials Group\\ Department of Engineering Science and Mechanics\\
Pennsylvania State University, University Park, PA 16802--6812, USA}
\\
 \vspace{3mm}
 Tom G. Mackay\footnote{E--mail: T.Mackay@ed.ac.uk.}\\
{\em School of Mathematics and
   Maxwell Institute for Mathematical Sciences\\
University of Edinburgh, Edinburgh EH9 3FD, UK}\\
and\\
 {\em NanoMM~---~Nanoengineered Metamaterials Group\\ Department of Engineering Science and Mechanics\\
Pennsylvania State University, University Park, PA 16802--6812,
USA}\\
 \vspace{3mm}
 Akhlesh  Lakhtakia\\
 {\em NanoMM~---~Nanoengineered Metamaterials Group\\ Department of Engineering Science and Mechanics\\
Pennsylvania State University, University Park, PA 16802--6812, USA}

\normalsize

\end{center}

\begin{center}
\vspace{15mm} {\bf Abstract}
\end{center}
The canonical boundary-value problem for  surface-plasmon-polariton (SPP) waves guided by the planar interface of a dielectric material and a plasmonic material was solved for  cases wherein either partnering material could be 
 a uniaxial material with  optic axis  lying in the interface plane.
 Numerical studies revealed that two different SPP waves, with different phase speeds, propagation lengths, and  penetration depths, can propagate in a given direction in the interface plane; in contrast,  the planar interface of isotropic partnering materials supports only one SPP wave for each propagation direction.
Also, for a unique propagation direction in each quadrant of the interface plane, it was demonstrated that a new type of SPP wave~---~called a
 surface-plasmon-polariton--Voigt (SPP--V) wave~---~can exist.
The fields of these  SPP--V waves decay
as the product of a linear and an exponential function of the distance from the interface in the
anisotropic partnering material; in contrast,  the fields of conventional SPP waves  decay only exponentially with distance from the interface. 
Explicit analytic solutions  of the dispersion relation for  SPP--V  waves exist and help establish
 constraints on the constitutive-parameter regimes for the partnering materials that support  SPP--V-wave propagation.

\section{Introduction}

Surface-plasmon-polariton (SPP) waves are guided by the planar interface of a dielectric material and a plasmonic material \c{Pitarke}. While SPP waves cannot be excited by direct illumination, their excitation can be readily achieved indirectly via coupling with a prism \c{Turbadar1964,Otto1968,Kretschmann1968} or surface-relief grating \c{Homola}, for examples. 
SPP waves are of major technological importance: they have been widely exploited for optical sensing \c{Nylander,Homola,AZLe} and microscopy \c{Yeatman87,Rothenhausler88}, and   applications for optical communications \c{Sekhon2011,KimOE08,Berini,Agrahari2018,Agrahari2019} and   harvesting solar energy \c{GL2008,AP2010,Anderson2017}  are on the horizon.

The theory underpinning  SPP-wave propagation is firmly established in the case where the two partnering materials are isotropic  \c{Pitarke}. The case where an isotropic plasmonic material is partnered with an anisotropic dielectric material
has also been considered previously \c{Sprokel1981,WSC1987,PNL07}.
However, SPP-wave propagation in the case where an anisotropic plasmonic material is partnered with an isotropic dielectric material has received scant attention from theorists, even though several  experimental studies on this case have been reported recently
\c{Cao,Feng,Feng_APL,Meng,Srangi,Yin}.

As we demonstrate in this paper, when anisotropic partnering materials are involved,  
some previously unreported SPP-wave characteristics emerge. Most notably, two different SPP waves, with different phase speeds, propagation lengths, and  propagation depths, can propagate in a given direction in the interface plane. Analogously, 
this multiplicity of surface waves can also arise in the case of Dyakonov-wave propagation supported by dissipative anisotropic materials \c{ML_PJ}, and has 
also been reported in the case of SPP-wave propagation supported by periodically nonhomogeneous dielectric materials \c{YJJjnp2011,FLpra2011-2}.

Additionally, we demonstrate that
when anisotropic partnering materials are involved,  
for a unique propagation direction in each quadrant of  the interface plane, a new type of 
SPP wave---with  mixed exponential and linear localization characteristics---can exist. We call this  surface wave a surface-plasmon-polariton--Voigt (SPP--V) wave, because it is closely
related to a singular form of plane waves called Voigt waves that can arise in certain unbounded  anisotropic dielectric mediums \cite{Voigt,Panch,Gerardin}.

A Voigt wave's amplitude  is governed by
the product of an
exponential function of the propagation distance and a linear function of the
propagation distance, in stark contrast to  conventional plane waves that propagate in 
unbounded anisotropic mediums \cite{Chen,EAB}. 
The existence of Voigt waves was established
in
early experimental and theoretical studies  based on
  pleochroic crystals such as andalusite,   iolite, and alexandrite \c{Voigt,Panch,Ranganath}. 
But  greater scope for Voigt-wave  propagation is presented by
more complex mediums \c{Grech,Denis}, such as  bianisotropic \c{Berry} and nonhomogeneous mediums \c{Lakh_helicoidal_bianisotropic_98}.
Furthermore,
 the directions of  Voigt waves  can be selected in  carefully 
engineered materials 
 \c{ML03,ML_WRM,M2011_JOPA,M2014_JNP,Voigt_Pockels}
 A host anisotropic medium that is either dissipative \c{Ranganath,Fedorov} or active \c{ML_EPJ} is a prerequisite for Voigt-wave propagation. However,
  as established in the following,  SPP--V-wave propagation is possible 
  for an anisotropic plasmonic material partnered with a non-dissipative (and non-active) dielectric material.

 In this paper, a unified theory of SPP-wave propagation  and  SPP--V-wave propagation is developed
  by formulating and solving  a canonical boundary-value problem. The cases of 
  \begin{itemize}
  \item[(i)] an anisotropic dielectric material partnered with an isotropic plasmonic material, and
  \item[(ii)]
 an isotropic dielectric material partnered with an anisotropic plasmonic material, 
 \end{itemize}
 are  considered, with emphasis on new combinations of partnering materials.
 Explicit analytic solutions  of the dispersion relation for  SPP--V  waves are derived
 and used to establish
 constraints on the constitutive-parameter regimes for the partnering materials that allow  SPP--V-wave propagation.
Representative numerical results are presented to illustrate the theoretical results.
And some closing remarks are provided at the end. 
 
 In the notation adopted,
double underlining denotes   3$\times$3 dyadics while single underlining denotes 3-vectors;  double underlining and square parenthesis denotes
4$\times$4 matrixes while
 single underlining and square parenthesis denotes
 column 4-vectors. 
 The
identity 3$\times$3  dyadic is written as  $\=I=\ux\ux+\uy\uy+\uz\uz$ \c{Chen}, with the
 triad $\lec \ux, \uy, \uz \ric$ comprising the  Cartesian basis vectors.   The free-space wavenumber is  
 denoted by $\ko = \omega \sqrt{\epso \muo}$, wherein $\omega$ is the angular frequency,
 and the permittivity and permeability of free space are given as $\epso = 8.854 \times 10^{-
12}$~F~m${}^{-1}$
and $\muo = 4 \pi \times 10^{-7}$~H~m${}^{-1}$, respectively.
The free-space wavelength and impedance are written as $\lambdao = 2 \pi / \ko$ and $\etao = \sqrt{\muo/\epso}$, respectively. In addition, $i = \sqrt{-1}$.

\section{Analysis of surface-wave propagation} \label{Cbvp}

\subsection{Matrix ordinary differential equations}

\begin{figure}[!h]
\centering\includegraphics[width=3.in]{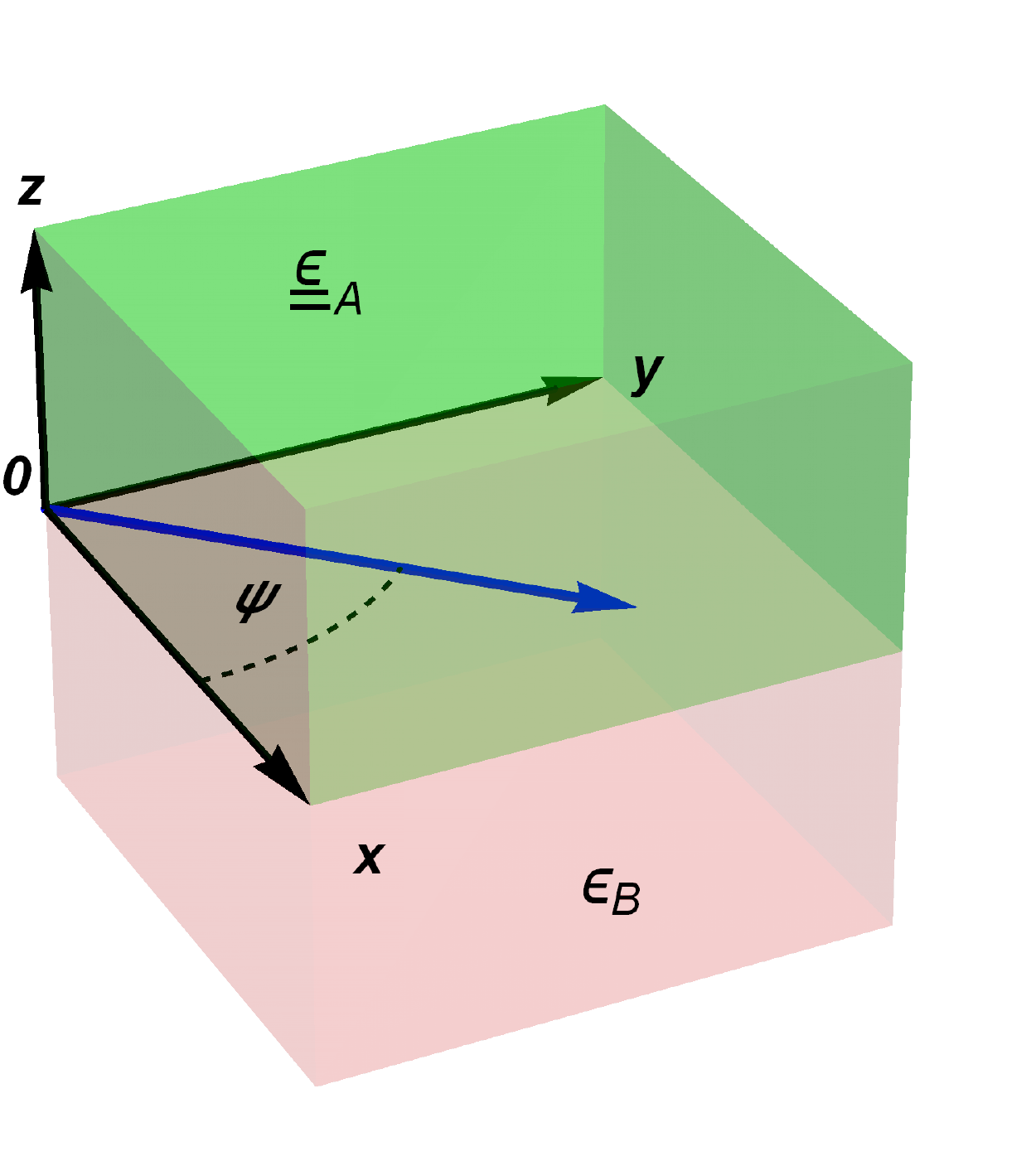}
\caption{A schematic representation of the canonical boundary-value problem. The optic axis of medium $\calA$ is parallel to the $x$ axis. Surface waves propagate parallel to the interface  plane $z=0$ along the direction  at an angle $\psi$ relative to the $x$ axis.}
\label{Fig1}
\end{figure}

A  general formalism for surface-wave propagation \c{ESW_book} is specialized to develop the canonical boundary-value problem for SPP-wave propagation guided by the planar  interface of a uniaxial  material, labeled $\calA$,
and an isotropic  material, labeled $\calB$. The two partnering  materials $\calA$ and $\calB$
are both non-magnetic and non-magnetoelectric \cite{EAB,ODell}.
Material $\mathcal{A}$ occupies
 the half-space $z>0$. As this material is a uniaxial dielectric material, it is characterized by
an
 \textit{ordinary} relative permittivity $\eps_\mathcal{A}^{\rm s}$ and an \textit{extraordinary} relative permittivity
$ \eps_\mathcal{A}^{\rm t}$.  With the unit vector $\ux$  pointing in the direction of the
optic axis,
the relative permittivity
dyadic for material $\calA$ is written  as \c{Chen,EAB}
\begin{equation}
\=\eps_\mathcal{A} = \eps_\mathcal{A}^{\rm s} \=I + \le
\eps_\mathcal{A}^{\rm t} - \eps_\mathcal{A}^{\rm s} \ri \,
\ux \, \ux\,. \l{Ch4_eps_uniaxial}
\end{equation}
Material $\mathcal{B}$ occupies
 the half-space $z<0$ and is characterized by the relative permittivity dyadic
$\=\eps_\mathcal{B}  =\eps_\mathcal{B}\=I$. For SPP waves to be guided by the interface of materials $\calA$ and $\calB$, one of the partnering materials must be a plasmonic material and the other partnering material must be a dielectric material.
 The canonical boundary-value problem is represented  schematically in Fig.~\ref{Fig1}.

The electromagnetic field phasors that characterize a surface wave 
are expressed as
\cite{ESW_book} 
\begin{equation} \label{planewave}
\left.\begin{array}{l}
 \#E (\#r)=  \#e (z) \,\exp\les i q \le x \cos \psi + y \sin \psi \ri \ris \\[4pt]
 \#H  (\#r)=  \#h  (z) \,\exp\les i q \le x \cos \psi + y \sin \psi \ri \ris 
 \end{array}\right\}\,
\end{equation}
 for all $z \in \le - \infty, \infty \ri$. Herein 
 the complex-valued scalar ${q}$ represents the surface  wavenumber;  the angle $\psi\in\left[0,2\pi\right)$
prescribes
 the direction of propagation in the $xy$ plane, relative to  the $x$ axis;
and the auxiliary phasors 
\begin{equation}
\left.
\begin{array}{l}
\#e (z)=e_x(z)\ux + e_y(z)\uy+e_z(z)\uz\\[5pt]
\#h (z)=h_x(z)\ux + h_y(z)\uy+h_z(z)\uz
\end{array}
\ric \,
\end{equation}
have complex-valued components, in general.

  The field phasors \r{planewave} 
  satisfy the source-free, frequency-domain Maxwell curl
postulates \cite{Chen} 
\begin{equation}
\left. \begin{array}{l} \curl \#H (\#r,\omega) + i \omega \epso
\=\eps_{\calA} \. \#E (\#r,\omega) = \#0
\\ \vspace{-6pt}
 \\ \curl \#E (\#r,\omega) - i \omega
\muo \#H (\#r,\omega) =\#0 \end{array} \ric \,, \qquad   z>0\,.
\label{MP_fd_A}
\end{equation}
and
\begin{equation}
\left. \begin{array}{l} 
\curl \#H (\#r,\omega) + i \omega \epso
 \eps_{\calB} \#E(\#r,\omega) = \#0
\\ \vspace{-6pt}
 \\ \curl \#E (\#r,\omega) - i \omega
\muo \#H (\#r,\omega) =\#0 \end{array} \ric \,, \qquad   z<0\,.
\label{MP_fd_B}
\end{equation}
Upon combining with the phasor representations~\r{planewave}, 
with the Maxwell curl postulates \r{MP_fd_A} and \r{MP_fd_B},  respectively,
 the 4$\times$4 matrix ordinary differential
equations \c{Berreman}
\begin{equation}
\label{MODE_A}
\frac{d}{dz}\les\#f  (z)\ris=i \PAmat\.\les\#f  (z)\ris\,,  \qquad   z>0\,,
\end{equation}
and
\begin{equation}
\label{MODE_B}
\frac{d}{dz}\les\#f  (z)\ris=i \PBmat\.\les\#f  (z)\ris\,,  \qquad   z<0\,
\end{equation}
emerge. Herein
the  column 4-vector
\begin{equation}
\les\#f (z)\ris= 
\les
\begin{array}{c}%
e_ x(z)\\[5pt]
e_y(z)\\[5pt]
h_x(z)\\[5pt]
h_y(z)%
\end{array}
\ris \,
\label{f-def}
\end{equation}
contains the $x$-directed and $y$-directed components of the auxiliary phasors, which
are algebraically connected to the   $z$-directed components of the auxiliary phasors  \c{EAB}.
The forms of the 4$\times$4 propagation matrixes  $\les\=P_\calA\ris $
in Eq.~\r{MODE_A} and 
$\les\=P_\calB\ris $
in Eq.~\r{MODE_B} are determined by  the forms
 of $\=\eps_\calA$ and $\=\eps_\calB$, respectively.

 \subsection{Half-space $z>0$}

The matrix on the right side of
Eq.~(\ref{MODE_A}) is given as
\begin{eqnarray}
&&
\les\=P_\calA\ris= \les   
\begin{array}{cccc}
0&0& \displaystyle{ \frac{q^2 \cos \psi \sin \psi}{\omega \epso \eps^s_\calA}} & 
\displaystyle{\frac{\ko^2 \eps^s_\calA- q^2 \cos^2 \psi }{\omega \epso \eps^s_\calA} } \vspace{8pt} \\
0&0& \displaystyle{\frac{-\ko^2 \eps^s_\calA+ q^2 \sin^2 \psi }{\omega \epso \eps^s_\calA} }&
\displaystyle{ -\frac{q^2 \cos \psi \sin \psi}{\omega \epso \eps^s_\calA}} \vspace{8pt}
\\
\displaystyle{ -\frac{q^2 \cos \psi \sin \psi}{\omega \muo}} & 
\displaystyle{\frac{-\ko^2 \eps^s_\calA+ q^2 \cos^2 \psi }{\omega \muo} } &0&0\vspace{8pt} \\
\displaystyle{\frac{\ko^2 \eps^t_\calA- q^2 \sin^2 \psi }{\omega \muo} } &
\displaystyle{ \frac{q^2 \cos \psi \sin \psi}{\omega \muo}}&0&0
\end{array}
\ris,
\label{PA-def}
\end{eqnarray}
and the $z$-directed components of the auxiliary phasors emerge as 
\begin{equation}
\left.
\begin{array}{l}
e_z(z) = \displaystyle{\frac{q \les h_ x(z) \sin \psi - h_ y(z)  \cos \psi  \ris}{\omega \epso \eps^s_\calA}} \vspace{8pt} \\
h_ z(z) = \displaystyle{\frac{q \les e_ y(z) \cos \psi - e_ x(z)  \sin \psi  \ris}{\omega \muo }}
\end{array}
\right\}\,,\qquad z > 0\,.
\end{equation}

\subsubsection{Nonsingular case\label{A-nonsing}}

 The 4$\times$4 matrix  $\les\=P_\calA\ris$ has four distinct eigenvalues, namely
  $\pm \alpha_{\calA 1}$ and $\pm \alpha_{\calA 2}$,
  in the nonsingular case.
Each eigenvalue has  algebraic multiplicity $1$ and geometric multiplicity $1$. The eigenvalues  are given by
\begin{equation} \l{a_decay_const}
\left.
\begin{array}{l}
\alpha_{\calA 1} = i \sqrt{ q^2 - \ko^2 \eps_\calA^s} \vspace{8pt}\\
\alpha_{\calA 2} = \displaystyle{
i\sqrt{\frac{q^2 \beta  - 2 \ko^2 \eps^s_\calA \eps^t_\calA}{2 \eps^s_\calA}}
}
\end{array}
\right\}\,,
\end{equation}
wherein the parameter $\beta =  \le \eps^s_\calA + \eps^t_\calA \ri  - \le \eps^s_\calA - \eps^t_\calA \ri \cos 2 \psi $.
 The signs of the square-root terms in Eqs.~\r{a_decay_const} must be selected such that $\mbox{Im} \lec \alpha_{\calA 1} \ric > 0 $ and $\mbox{Im} \lec \alpha_{\calA 2} \ric > 0 $,
 in order to
 ensure that fields  decay as $z \to +\infty$.
The following pair of eigenvectors of the 4$\times$4
matrix $\les\=P_\calA\ris$   match the eigenvalues $\alpha_{\calA 1}$ and $\alpha_{\calA 2}$,
respectively:
\begin{equation}
\left.
\begin{array}{l}
\#v_{\calA 1} = 
\les \begin{array}{c}
0 \vspace{8pt}\\
\displaystyle{
\frac{\ko \alpha_{\calA 1}}{q^2 \sin \psi \cos \psi}} \vspace{8pt}\\
\displaystyle{\frac{\cot 2 \psi}{\etao} + \frac{\csc 2 \psi}{\etao}  \le 1 - \frac{2 \ko^2 \eps^s_\calA }{q^2} \ri}\vspace{8pt} \\ \etao^{-1}
\end{array}
\ris
\vspace{8pt}
\\
\#v_{\calA 2} = 
\les \begin{array}{c}
\displaystyle{1 - \frac{q^2 \le \cos 2 \psi + 1 \ri}{2\ko^2
 \eps^s_\calA 
}} \vspace{8pt}\\
\displaystyle{-\frac{q^2  \cos  \psi \sin \psi }{ \ko^2 \eps^s_\calA }} \vspace{8pt}\\ 0 \vspace{8pt} \\ 
\displaystyle{\frac{\alpha_{\calA 2}}{\omega \muo}}
\end{array}
\ris
\end{array}
\right\}.
\end{equation}

The general  solution of the matrix differential eq.~\r{MODE_A} for the nonsingular case  is provided by
\begin{equation} \l{D_gen_sol}
\les\#f (z)\ris =C_{\calA 1}  \#v_{\calA 1} \exp \le i \alpha_{\calA 1} z \ri +  C_{\calA 2} \#v_{\calA 2} \exp \le i \alpha_{\calA 2} z \ri 
\end{equation}
for $z>0$.
The constants $C_{\calA 1}$ and $C_{\calA 2}$ herein are determined by the boundary conditions at $z=0$.

\subsubsection{Singular case \label{A-sing}}

In the singular case, the 4$\times$4 matrix
$\les\=P_\calA\ris$ has only two eigenvalues, namely $\pm \alpha_{\calA }$.
Each eigenvalue has algebraic multiplicity $2$ and geometric multiplicity $1$.
This  case arises when 
\begin{equation} \l{q_sol}
q = \sigma \frac{\ko  \sqrt{ \eps^s_{\calA}}}{\cos \psi},
 \end{equation}   
where  the sign parameter $\sigma = +1$ for $\psi \in \le 0, \pi/2 \ri$ and 
$\sigma = -1$ for $\psi \in \le \pi/2, \pi \ri$.
The eigenvalues are given by
\begin{equation} \l{alphaa_sol}
\alpha_\calA = i \sigma \ko \sqrt{\eps^s_{\calA}} \tan \psi,
\end{equation}
wherein the square-root  
term must be 
 selected to have a positive real part in order to achieve $\mbox{Im} \lec \alpha_\calA \ric > 0$, which is required in order that
  fields  decay as $z \to +\infty$
  \cite{ESW_book}. Accordingly,
    SPP--V-wave propagation is not possible for 
    $\psi \in\lec0,\pi\ric$
because
 $\mbox{Im} \lec \alpha_\calA \ric \leq 0$ for $\psi = 0 $ and $\pi$.

The following eigenvector of the 4$\times$4
matrix $\les\=P_\calA\ris$ matches the eigenvalue $\alpha_{\calA}$:
\begin{equation} \l{vA}
\#v_{\calA } = 
\les \begin{array}{c}
0 \vspace{8pt}\\
\displaystyle{\frac{i \sigma  }{\sqrt{ \eps^s_\calA}}} \vspace{8pt}\\
0 \vspace{8pt} \\ \etao^{-1}
\end{array}
\ris.
\end{equation}
Furthermore,
 a  generalized eigenvector that satisfies \c{Boyce}
\begin{equation}
\le \les\=P_\calA\ris - \alpha_{\calA } \=I \ri \. \#w_{\calA } = \#v_{\calA }
\end{equation}
is 
\begin{equation} \l{wA}
\#w_{\calA } = \frac{1}{\ko}
\les \begin{array}{c}
\displaystyle{  \frac{2}{ \eps^t_\calA - \eps^s_\calA  }}
\vspace{8pt}\\
\displaystyle{\frac{\tan \psi}{ \eps^s_\calA} \le \cot^2 \psi - 2\frac{ \eps_\calA^s - \eps^t_\calA \cot^2 \psi }{\eps_\calA^s - \eps^t_\calA}
\ri
} \vspace{8pt}\\
\displaystyle{ \frac{2 i \sigma \sqrt{\eps^s_\calA}}{\etao \le \eps^t_\calA - \eps^s_\calA \ri}}
 \vspace{8pt} \\ 0
\end{array}
\ris.
\end{equation}
The general  solution of the matrix differential eq.~\r{MODE_A}  
for the singular case is provided as
\begin{equation} \l{DV_gen_sol}
\les\#f  (z)\ris = \les C_{\calA 1}  \#v_{\calA }  + C_{\calA 2} \le i  z \, \#v_{\calA}   + \#w_{\calA} \ri  \ris \exp \le i \alpha_{\calA } z \ri
\end{equation}
for $z>0$.
The constants $C_{\calA 1}$ and $C_{\calA 2}$ herein are determined by the boundary conditions at $z=0$.
Notice that the general solution
\r{DV_gen_sol} for the singular case
 contains a term that is linearly proportional to distance from the interface $z$, which is in stark contrast to the general solution
\r{D_gen_sol} for the nonsingular case in Sec.~\ref{A-nonsing}.

\subsection{Half-space $z<0$}

The matrix on the right side of
Eq.~(\ref{MODE_B}) is given as
\cite{Chen,ESW_book}
\begin{eqnarray}
&&
\les\=P_\calB\ris= \les   
\begin{array}{cccc}
0&0& \displaystyle{ \frac{q^2 \cos \psi \sin \psi}{\omega \epso \eps_\calB}} & 
\displaystyle{\frac{\ko^2 \eps_\calB- q^2 \cos^2 \psi }{\omega \epso \eps_\calB} } \vspace{8pt} \\
0&0& \displaystyle{\frac{-\ko^2 \eps_\calB+ q^2 \sin^2 \psi }{\omega \epso \eps_\calB} }&
\displaystyle{ -\frac{q^2 \cos \psi \sin \psi}{\omega \epso \eps_\calB}} \vspace{8pt}
\\
\displaystyle{ -\frac{q^2 \cos \psi \sin \psi}{\omega \muo}} & 
\displaystyle{\frac{-\ko^2 \eps_\calB+ q^2 \cos^2 \psi }{\omega \muo} } &0&0\vspace{8pt} \\
\displaystyle{\frac{\ko^2 \eps_\calB- q^2 \sin^2 \psi }{\omega \muo} } &
\displaystyle{ \frac{q^2 \cos \psi \sin \psi}{\omega \muo}}&0&0
\end{array}
\ris,
\end{eqnarray}
and the $z$-directed components of the auxiliary phasors emerge as 
\begin{equation}
\left.
\begin{array}{l}
e_ z(z) = \displaystyle{\frac{q \les h_ x(z) \sin \psi - h_ y(z)  \cos \psi  \ris}{\omega \epso \eps_\calB}} \vspace{8pt} \\
h_ z(z) = \displaystyle{\frac{q \les e_ y(z) \cos \psi - e_ x(z)  \sin \psi  \ris}{\omega \muo }}
\end{array}
\right\}\,,\qquad z < 0\,.
\end{equation}
The  4$\times$4 matrix $\les\=P_\calB\ris$ has two eigenvalues, namely  $\pm \alpha_{\calB}$.
Each eigenvalue has  algebraic multiplicity $2$ and geometric multiplicity $2$.  The eigenvalues are given by
\begin{equation} \l{b_decay_const}
\alpha_{\calB} =- i \sqrt{q^2 - \ko^2 \eps_\calB},
\end{equation}
wherein the sign of the square-root term must be selected such that
 $\mbox{Im} \lec \alpha_{\calB} \ric < 0 $ to ensure that fields decay as $z \to -\infty$.
 
 The following 
 pair of independent  eigenvectors of the 4$\times$4
matrix $\les\=P_\calB\ris$ match   the eigenvalue $\alpha_{\calB}$:
\begin{equation}
\left.
\begin{array}{l}
\#v_{\calB 1} = 
\les \begin{array}{c}
\displaystyle{1 - \frac{q^2 \cos^2 \psi}{\ko^2 \eps_\calB}} \vspace{8pt}\\
\displaystyle{- \frac{q^2 \cos \psi \sin \psi}{\ko^2 \eps_\calB}} \vspace{8pt}\\ 0 
\vspace{8pt} \\ \displaystyle{\frac{\alpha_\calB}{\omega \muo}}
\end{array}
\ris
\vspace{8pt}
\\
\#v_{\calB 2} = 
\les \begin{array}{c}
\displaystyle{ \frac{q^2 \cos \psi \sin \psi}{\ko^2 \eps_\calB}} \vspace{8pt}\\ 
\displaystyle{-1 + \frac{q^2 \sin^2 \psi}{\ko^2 \eps_\calB}} \vspace{8pt}\\
 \displaystyle{\frac{\alpha_\calB}{\omega \muo}}
\vspace{8pt} \\  0
\end{array}
\ris
\end{array}
\right\}\,.
\end{equation}
The general solution  of  the matrix ordinary differential equation~\r{MODE_B}  is given as
\begin{equation}
\label{2.22-AL}
\les\#f  (z)\ris = \le C_{\calB 1}  \#v_{\calB 1}  +  C_{\calB 2} \#v_{\calB 2} \ri \exp \le i \alpha_{\calB} z \ri 
\end{equation}
for $z < 0$. Herein
 the constants $C_{\calB 1}$ and $C_{\calB 2}$ are determined by the boundary conditions at $z=0$.

\subsection{Canonical boundary-value problem}

\subsubsection{SPP waves} \l{SPP_sec}

 The  tangential  components of the electric and magnetic field
 phasors across the interface $z=0$
 must be continuous \c{Chen}. The
 four algebraic equations that consequently must be satisfied  are compactly expressed as
 \begin{equation}
 \label{2.23-AL}
 \les\#f(0^+)\ris=  \les\#f(0^-)\ris
 \,.
 \end{equation}
 By combining
  Eqs.~\r{D_gen_sol} and \r{2.22-AL} with Eq.~\r{2.23-AL},  the following equation emerges:
\begin{equation}
\les \=M \ris \. \les \begin{array}{c}
 C_{\calA 1} \\
  C_{\calA 2} \\
   C_{\calB 1} \\
    C_{\calB 2}
\end{array}
 \ris =  \les \begin{array}{c}
 0 \\
  0 \\
   0 \\
    0
\end{array}
 \ris.
\end{equation}
The 4$\times$4 characteristic matrix $\les \=M \ris$ herein must be singular for  SPP-wave propagation \c{ESW_book}.
The corresponding 
 dispersion equation $\left\vert \les \=M \ris\right\vert = 0$   is equivalent to the equation
\begin{eqnarray} \l{DE}
 &&\ko^2 \eps^s_\calA  \le \eps^s_\calA \alpha_{\calB}  - \eps_\calB \alpha_{\calA 1}
 \ri \le \alpha_\calB - \alpha_{\calA 2} \ri \tan^2 \psi  \nonumber \\ &&
 = \alpha_{\calA 1}
\le \alpha_\calB - \alpha_{\calA 1} \ri
\le \eps^s_\calA \alpha_\calB \alpha_{\calA 2} - \eps_\calB \alpha^2_{\calA 1} \ri, 
\end{eqnarray}
from which the wavenumber $q$ can be numerically extracted, using the Newton--Raphson method \c{N-R}, for example.
From the symmetry of 
 Eq.~\r{DE} it may be inferred that 
  if a SPP wave  propagates at the orientation specified by  $\psi = \psi^\star$, then
 SPP-wave propagation is also possible for $\psi = - \psi^\star$ and $\psi = \pi \pm \psi^\star$.

\subsubsection{SPP--Voigt  waves} \l{Singular_soln_sec}

As discussed in Sec.~\ref{SPP_sec},   Eq.~\r{2.23-AL} follows from
 the continuity of tangential  components of the electric and magnetic field
 phasors across the interface $z=0$ \c{Chen}.  
 By combining Eqs.~\r{DV_gen_sol} and \r{2.22-AL} with Eq.~\r{2.23-AL}, the following 
 equation emerges:
\begin{equation}
\les \=N \ris \. \les \begin{array}{c}
 C_{\calA 1} \\
  C_{\calA 2} \\
   C_{\calB 1} \\
    C_{\calB 2}
\end{array}
 \ris =  \les \begin{array}{c}
 0 \\
  0 \\
   0 \\
    0
\end{array}
 \ris\,.
\end{equation}
The 4$\times$4 characteristic matrix $\les \=N \ris$ herein must be singular for  surface-wave propagation. And the corresponding
 dispersion equation $\left\vert\les \=N \ris \right\vert= 0$ 
reduces to 
\begin{eqnarray} \l{sing_disp_rel}
&& \les 2 \eps^s_\calA \le \eps_\calB + \eps^s_\calA \ri
+ \le \eps^s_\calA - \eps_\calB \ri \le \eps^s_\calA + \eps^t_\calA \ri \cot^2 \psi \ris 
\nonumber \\ && +
 2  \sqrt{\eps^s_\calA} \le \eps^s_\calA + \eps_\calB \ri
\sqrt { \eps^s_\calA  
  + \le  \eps^s_\calA -\eps_\calB\ri \cot^2 \psi}
= 0.
\end{eqnarray}
The symmetries of Eq.~\r{sing_disp_rel} 
are analogous to
 those of Eq.~\r{DE}. Hence, 
if a  SPP--V wave  propagates at the orientation specified by  $\psi = \psi^\star$, then
  SPP--V waves can also propagate for the orientations  $\psi = - \psi^\star$ and $\psi = \pi \pm \psi^\star$.
Observe that
 Eq.~\r{sing_disp_rel} cannot  be satisfied for $ \eps^s_\calA = \eps_\calB$
 unless $ \eps^s_\calA = \eps_\calB = 0$, 
 but this  eventuality may be dismissed as it   is unphysical.

\subsection{Analytical solutions of the  SPP--V dispersion equation}

Unlike the SPP dispersion equation \r{DE}, the  SPP--V
 dispersion equation \r{sing_disp_rel} 
 yields analytical solutions for the four variables $\eps^s_\calA$, $\eps^t_\calA$, $\eps_\calB$, and $\psi$,  as follows.
\begin{itemize}
\item[(i)] For fixed values of
 $\eps^{t}_\calA$, $\eps_\calB$, and $\psi\in(0,\pi/2)$,  SPP--V-wave
propagation is possible only when
 \begin{eqnarray} \l{epss_sol}
\eps^{s}_\calA&=&\frac{\sec^2 \psi}{12}\Big[ t_1 + \frac{2 t_2}{\le 2 t_3 + 48 \sqrt{6 t_4 t_5} \ri^\frac{1}{3} }   \nonumber \\ && + \le 2 t_3 
+ 48 \sqrt{6 t_4 t_5} \ri^{\frac{1}{3}}\Big],
\end{eqnarray}
 wherein the parameters
 \begin{equation}
 t_1 =10 \eps_\calB - 12 \eps^t_\calA + \le 4 \eps^t_\calA - 6 \eps_\calB \ri \cos 2\psi  ,
\end{equation}
\begin{eqnarray}
t_2&=& 71 \le \eps_\calB \ri^{2} - 126 \eps_\calB \eps^t_\calA + 67  \le \eps^t_\calA \ri^{2} \nonumber \\ && - 4 \Big[ 15  \le \eps_\calB \ri^{2} 
- 34 \eps_\calB \eps^t_\calA + 15 \le \eps^t_\calA \ri^{2} \Big]   \cos 2\psi \nonumber \\ && +\les -3 \le \eps_\calB \ri^{2} + 6 \eps_\calB \eps^t_\calA +  \le \eps^t_\calA \ri^{2}  \ris \cos 4\psi,
\end{eqnarray}
\begin{eqnarray}
 t_3 &=&  2 \Big[ 475 \le \eps_\calB \ri^{3} - 1359 \le \eps_\calB \ri^{2} \eps^t_\calA + 1365  \eps_\calB  \le \eps^t_\calA \ri^{2}
 \nonumber \\ &&
  - 441 \le \eps^t_\calA \ri^{3} \Big] \nonumber  -
3 \Big[ 345 \le \eps_\calB \ri^{3} - 1061 \le \eps_\calB \ri^{2} \eps^t_\calA 
\nonumber \\ &&
+ 1023  \eps_\calB  \le \eps^t_\calA \ri^{2} 
- 347 \le \eps^t_\calA \ri^{3} \Big] \cos 2\psi   +
6 \Big[ 15 \le \eps_\calB \ri^{3} 
\nonumber \\&&
- 51 \le \eps_\calB \ri^{2} \eps^t_\calA + 65  \eps_\calB  \le \eps^t_\calA \ri^{2} - 21 \le \eps^t_\calA \ri^{3} \Big] \cos 4\psi  \nonumber \\&& +
\Big[ 27 \le \eps_\calB \ri^{3} - 63 \le \eps_\calB \ri^{2} \eps^t_\calA + 45  \eps_\calB  \le \eps^t_\calA \ri^{2}
\nonumber \\&&
 -  \le \eps^t_\calA \ri^{3} \Big] \cos 6\psi,
\end{eqnarray}
\begin{eqnarray}
t_4 &=&  2 \Big[ 105 \le \eps_\calB \ri^{4} - 151 \le \eps_\calB \ri^{3} \eps^t_\calA + 17 \le \eps_\calB  \eps^t_\calA \ri ^{2} +6 \le \eps^t_\calA \ri^{4}
\nonumber \\&&
+  67  \eps_\calB  \le \eps^t_\calA \ri^{3}   \Big]  +
 \Big[ 547 \le \eps_\calB \ri^{3} \eps^t_\calA
 -263 \le \eps_\calB \ri^{4} 
 \nonumber \\&&
  -225 \le \eps_\calB  \eps^t_\calA \ri ^{2} +  49  \eps_\calB  \le \eps^t_\calA \ri^{3}
  + 16 \le \eps^t_\calA \ri^{4} \Big] \cos 2\psi  \nonumber \\&& +
2 \Big[ 23 \le \eps_\calB \ri^{4} - 97 \le \eps_\calB \ri^{3} \eps^t_\calA + 135 \le \eps_\calB  \eps^t_\calA \ri ^{2} 
\nonumber \\ &&
- 43  \eps_\calB  \le \eps^t_\calA \ri^{3} + 2 \le \eps^t_\calA \ri^{4} \Big] \cos 4\psi  +
 \Big[ 7 \le \eps_\calB \ri^{4} 
  \nonumber \\&&
 - 19 \le \eps_\calB \ri^{3} \eps^t_\calA + 17 \le \eps_\calB  \eps^t_\calA \ri ^{2} -  \eps_\calB  \le \eps^t_\calA \ri^{3}  \Big] \cos 6\psi ,
\end{eqnarray}
and
\begin{equation}
t_5=-\le \eps_\calB - \eps^t_\calA  \ri^{2} \cos^4 \psi \sin^2 \psi.
\end{equation}

\item[(ii)]
For fixed values of
 $\eps^{s}_\calA$, $\eps_\calB$, and $\psi\in(0,\pi/2)$,  SPP--V-wave
propagation is possible only when
\begin{equation} \l{espt_sol}
\eps^t_\calA = -\eps^s_\calA + \frac{  t_6 \le  \eps^s_\calA + \eps_\calB \ri  \sqrt{\eps^s_\calA } }{\eps_\calB - \eps^s_\calA },
 \end{equation}
 wherein the parameter
 \begin{equation}
 t_6 =  2\,\tan \psi \le  \sqrt{\eps^s_\calA } \tan \psi + 
 \sqrt{\eps^s_\calA \sec^2 \psi - \eps_\calB } \ri.
 \end{equation}
 
 \item[(iii)]
 For fixed values of
 $\eps^{s}_\calA$, $\eps^t_\calA$, and $\psi\in(0,\pi/2)$,  SPP--V-wave
propagation is possible only when
\begin{eqnarray} \l{epsB_1_sol} \eps^{}_\calB &=& \frac{1}{32 \eps^s_\calA}  \Big\{ 4 t_7
- \le  \eps^s_\calA + \eps^t_\calA \ri \csc^2 \psi \nonumber \\ &&
\times \les  4 \le  \eps^s_\calA + \eps^t_\calA \ri 
-
  \sqrt{2 \le t_8 + t_9 \ri}
   \ris \Big\},
 \end{eqnarray} 
 wherein the parameters
 \begin{equation}
 t_7=  \le \eps^t_\calA \ri^2 + 6  \eps^s_\calA  \eps^t_\calA -3 \le  \eps^s_\calA \ri^2 ,
 \end{equation}
 \begin{equation}
 t_8=  \cos 4 \psi \les  \le  \eps^t_\calA \ri^2  + 10   \eps^s_\calA  \eps^t_\calA  - 7 \le  \eps^s_\calA \ri^2
\ris ,
 \end{equation}
 and
  \begin{eqnarray}
 t_9&=&  4 \cos 2 \psi \le   \eps^t_\calA - 3 \eps^s_\calA \ri \le  5  \eps^s_\calA + \eps^t_\calA \ri \nonumber \\ &&
 + 75 \le  \eps^s_\calA  \ri^2 - 2  \eps^s_\calA  \eps^t_\calA + 3 \le   \eps^t_\calA \ri^2.
 \end{eqnarray}
 \item[(iv)]
 For fixed values of
 $\eps^{s}_\calA$,
 $\eps^{t}_\calA$, and $\eps_\calB$,   SPP--V-wave
propagation is possible only when
\begin{equation} \l{psi_sol}
\psi = \arccot \les \frac{2}{\eps^s_\calA + \eps^t_\calA}
\sqrt{\frac{\eps^s_\calA \le \eps_\calB - \eps^t_\calA \ri \le \eps_\calB + \eps^s_\calA \ri }{ \eps^s_\calA - \eps_\calB }} \ris.
\end{equation}
\end{itemize}
In addition, an explicit formula for the 
 surface wavenumber $q$ of a  SPP--V wave is
 provided in Eq.~\r{q_sol}.

\subsection{Constraints on  SPP--V-wave propagation} \l{constraints_sec}

As well as the analytical solutions represented by Eqs.~\r{epss_sol}, \r{espt_sol}, \r{epsB_1_sol}, and \r{psi_sol},  constraints
on the permittivity parameters of the partnering materials for  SPP--V-wave propagation can be developed, as follows. We focus on the $\eps^t_\calA$ solution provided
 in Eq.~\r{espt_sol}. In the following the possibility
 of $\psi = \pi/2$ is discounted, because  the only solution to emerge from the dispersion relation \r{sing_disp_rel}  for this propagation direction is
  $\eps^s_\calA + \eps_\calB = 0$, 
 which is impossible for dissipative materials.
 
 \subsubsection{Anisotropic plasmonic material $\calA$ / isotropic dielectric material $\calB$} \l{Sec_con1}
 
\begin{figure}[!h]
\centering\includegraphics[width=4.2cm]{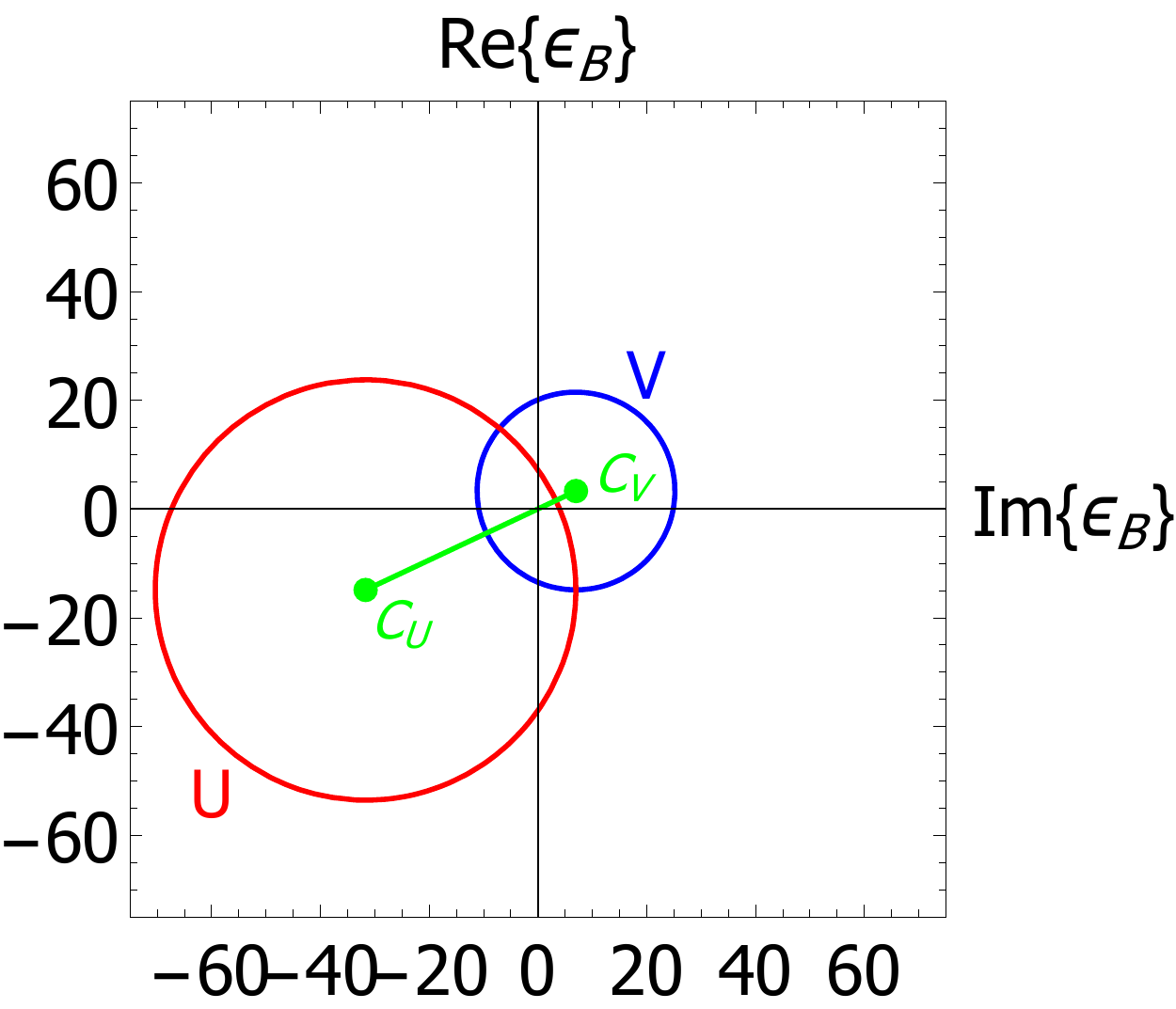} 
\includegraphics[width=4.2cm]{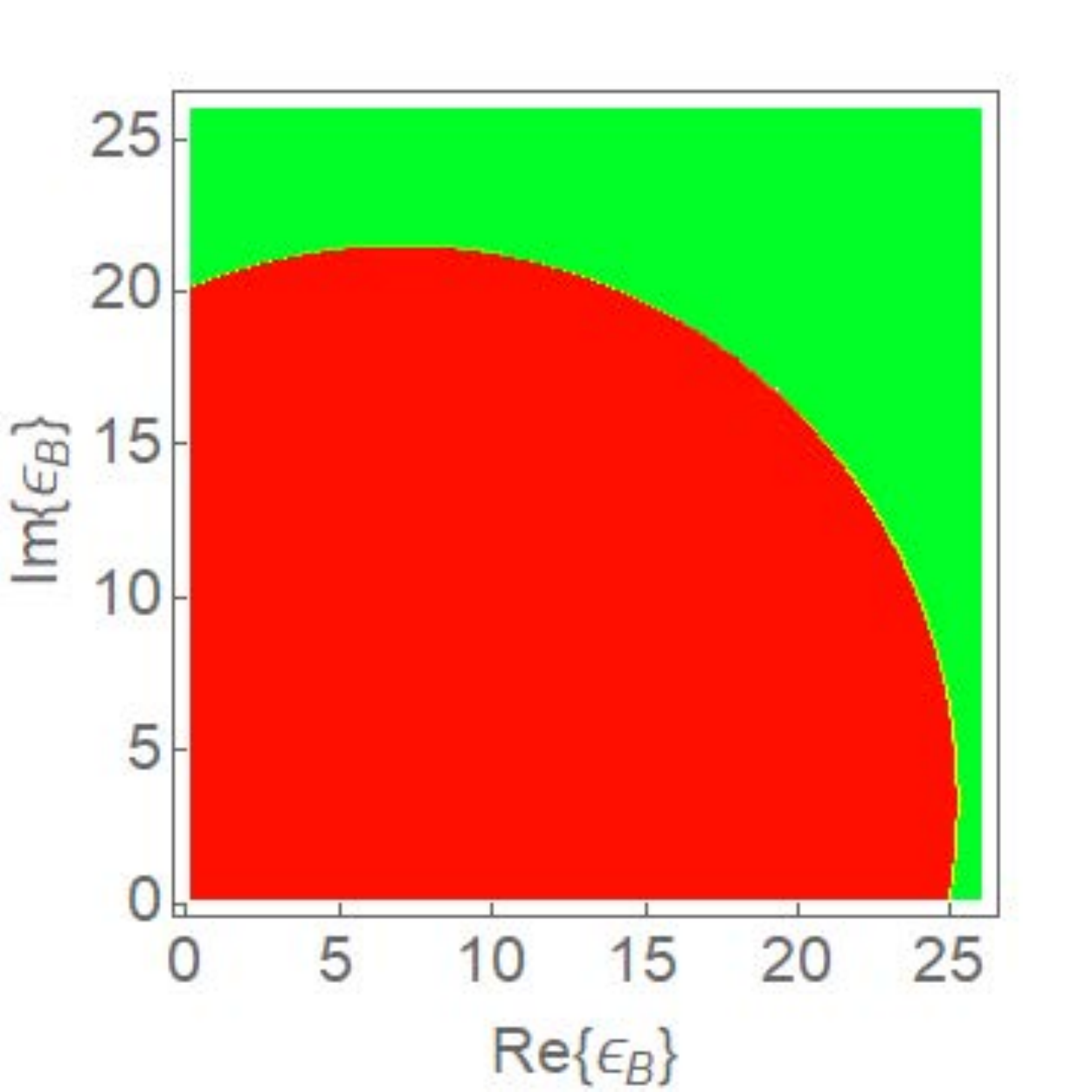}
\caption{Left: Circles $U$ and $V$ in the  complex-$\eps_\calB$ plane. Right: Regions U and V in the first quadrant of the complex-$\eps_\calB$   plane that support  SPP--V-wave propagation (shaded green) and regions in the 
first quadrant of the same plane  that do not support  SPP--V-wave propagation (shaded red). Representative example for    $ \eps^s_\calA = 
 -7.01 + 14.90i$.
}
\label{Fig2}
\end{figure}

 Suppose that material $\calA$ is plasmonic, i.e., $\mbox{Re}\lec \eps^s_\calA \ric < 0$,
$ \mbox{Re}\lec \eps^t_\calA \ric < 0$, $\mbox{Im}\lec \eps^s_\calA \ric > 0$,
and $\mbox{Im}\lec \eps^t_\calA \ric >0$, while material $\calB$ is a
dielectric material that is generally dissipative, i.e.,  $\mbox{Re}\lec \eps_\calB \ric > 0$ and
 $\mbox{Im}\lec \eps_\calB \ric > 0$. We consider values of $\eps^t_\calA$ that support  SPP--V-wave propagation in the direction specified by
 angle $\psi = \le\pi/2\ri - \nu$, wherein
 $ 0 < \nu \ll 1$. Since $\nu$ is taken to be a very small positive parameter, the approximations $\tan \les  \le\pi/2\ri - \nu \ris \approx 1/ \nu$ and $\sec 
 \les  \le\pi/2\ri - \nu \ris \approx 1$ are justified. Accordingly, Eq.~\r{espt_sol} yields
 \begin{equation} \l{epst_approx}
\eps^t_\calA = -\eps^s_\calA 
+ \frac{2}{\nu^2} \les \frac{\eps^s_\calA\le  \eps^s_\calA + \eps_\calB \ri}{\eps_\calB - \eps^s_\calA } \ris, \qquad \qquad  0 < \nu \ll 1 .
 \end{equation}
  For fixed values of $\eps^s_\calA$ and $\eps_\calB$, the absolute value $| \eps^t_\calA |$ becomes increasing large as $\psi$ approaches $\pi/2$, since the possibility  $\eps^s_\calA + \eps_\calB = 0$
is forbidden. In order for $\eps^t_\calA$ to lie in the second quadrant of the complex plane,  the following inequalities
must hold:
\begin{equation} \l{inequal_D}
\left.
\begin{array}{l}
\displaystyle{\mbox{Re} \lec  \frac{\eps^s_\calA\le  \eps^s_\calA + \eps_\calB \ri}{\eps_\calB - \eps^s_\calA } \ric < 0} \vspace{8pt}\\
\displaystyle{\mbox{Im} \lec  \frac{\eps^s_\calA\le  \eps^s_\calA + \eps_\calB \ri}{\eps_\calB - \eps^s_\calA } \ric > 0}
\end{array}
\right\}.
\end{equation}
The inequalities \r{inequal_D} may be conveniently recast in terms of the real and imaginary parts of 
$\eps^s_\calA$ and $\eps_\calB$
 as
\begin{equation} \l{inequal_D2}
\left.
\begin{array}{l}
\displaystyle{ \les \mbox{Re} \lec \eps_\calB \ric  - \frac{\le \mbox{Im} \lec \eps^s_\calA \ric \ri^2}{\mbox{Re} \lec \eps^s_\calA \ric } \ris^2
}
\\
\displaystyle{
+ \le \mbox{Im} \lec \eps_\calB \ric + \mbox{Im} \lec \eps^s_\calA \ric  \ri^2
> \le \frac{\left| \eps^s_\calA  \right|^2}{\mbox{Re} \lec \eps^s_\calA \ric} \ri^2 }
\vspace{8pt} \\
\displaystyle{\le \mbox{Re} \lec \eps_\calB \ric + \mbox{Re} \lec \eps^s_\calA \ric  \ri^2
}
\\
\displaystyle{
+\les \mbox{Im} \lec \eps_\calB \ric  - \frac{\le \mbox{Re} \lec \eps^s_\calA \ric \ri^2}{\mbox{Im} \lec \eps^s_\calA \ric } \ris^2 > 
\le \frac{\left| \eps^s_\calA  \right|^2}{\mbox{Im} \lec \eps^s_\calA \ric} \ri^2 }
\end{array}
\right\},
\end{equation}
which are amenable to a geometrical interpretation.
In the complex-$\eps_\calB$ plane, 
the inequality \r{inequal_D2}${}_1$ prescribes the region outside a circle labeled U, of 
radius $R_U =
\left| \eps^s_\calA  \right|^2 / \left| \mbox{Re} \lec \eps^s_\calA \ric \right|$ and
  centered at the point
$ C_U =\les  \le \mbox{Im} \lec \eps^s_\calA \ric \ri^2 / \mbox{Re} \lec \eps^s_\calA \ric,  - 
\mbox{Im} \lec \eps^s_\calA \ric \ris
$ in the third quadrant, while the
inequality \r{inequal_D2}${}_2$ prescribes the region outside a circle labeled
 V, of 
radius $R_V =
\left| \eps^s_\calA  \right|^2 /  \mbox{Im} \lec \eps^s_\calA \ric $ and
centered at the point
$C_V =\les  -\mbox{Re} \lec \eps^s_\calA \ric,  
\le \mbox{Re} \lec \eps^s_\calA \ric \ri^2 / \mbox{Im} \lec \eps^s_\calA \ric 
\ris
$ in the first quadrant. The straight line connecting the circle centers $C_U$ and $C_V$ passes through the origin.
The distances from the origin to $C_U$ and $C_V$ are 
\begin{equation} \l{DUDV}
\left.
\begin{array}{l}
D_U = \displaystyle{\frac{ \mbox{Im} \lec \eps^s_\calA \ric
\sqrt{ \le \mbox{Re} \lec \eps^s_\calA \ric \ri^2 + \le \mbox{Im} \lec \eps^s_\calA \ric \ri^2}
}{ \left| \mbox{Re} \lec \eps^s_\calA \ric \right|}
} \vspace{8pt} \\
D_V = \displaystyle{\frac{ \left| \mbox{Re} \lec \eps^s_\calA \ric \right|
\sqrt{ \le \mbox{Re} \lec \eps^s_\calA \ric \ri^2 + \le \mbox{Im} \lec \eps^s_\calA \ric \ri^2}
}{  \mbox{Im} \lec \eps^s_\calA \ric }
}
\end{array}
\right\},
\end{equation} 
respectively, while the distance between $C_U$ and $C_V$ is 
\begin{equation} \l{D_UV}
D_{UV} = \frac{\les\le \mbox{Re} \lec \eps^s_\calA \ric \ri^2 + \le \mbox{Im} \lec \eps^s_\calA \ric \ri^2 \ris^{3/2}}{\left|  \mbox{Re} \lec \eps^s_\calA \ric \right| \mbox{Im} \lec \eps^s_\calA \ric}.
\end{equation}
Some manipulation of the expressions \r{DUDV} and \r{D_UV}
 delivers the relations
 \begin{equation}
 \left.
 \begin{array}{l}
 D_U < R_U < D_{UV} \vspace{4pt} \\
 D_V < R_V < D_{UV}
 \end{array}
 \right\}.
 \end{equation}
Consequently, it follows that the region 
in the first quadrant  that lies outside   the
circle centered at $C_V$ is entirely outside   the circle centered at $C_U$. Therefore,  the constraint  \r{inequal_D2}${}_2$ is automatically satisfied provided that the 
constraint  \r{inequal_D2}${}_1$ is satisfied.
The  circles $U$ and $V$, and the parameter space of $\eps_\calB$ that supports  SPP--V-wave propagation, are illustrated in Fig.~\ref{Fig2}, for a representative example.

In summary:
 for the case of a
 plasmonic material $\calA$ and a dissipative dielectric material $\calB$,
   the constraint  \r{inequal_D2}${}_1$ must be satisfied  in order for  SPP--V waves to propagate, in directions close to  $\psi = \pi/2$. In the special case in which material $\calB$ is nondissipative, i.e., $\mbox{Im} \lec \eps_\calB \ric = 0$, the constraint
   \r{inequal_D2}${}_1$ reduces to 
   \begin{equation}
   \eps_\calB > -  \mbox{Re} \lec \eps^s_\calA \ric + \sqrt{2\le  \mbox{Re} \lec \eps^s_\calA \ric \ri^2 + \le  \mbox{Im} \lec \eps^s_\calA \ric \ri^2}.
   \end{equation}
   
\subsubsection{Anisotropic dielectric material $\calA$ / isotropic plasmonic material $\calB$}
 
 Suppose that material $\calA$ is 
 a
dielectric material that is generally dissipative, i.e., $\mbox{Re}\lec \eps^s_\calA \ric > 0$,
$ \mbox{Re}\lec \eps^t_\calA \ric > 0$, $\mbox{Im}\lec \eps^s_\calA \ric > 0$,
and $\mbox{Im}\lec \eps^t_\calA \ric >0$, while material $\calB$ is  plasmonic, i.e.,  $\mbox{Re}\lec \eps_\calB \ric < 0$ and
 $\mbox{Im}\lec \eps_\calB \ric > 0$. 
 As in Sec.~\ref{Sec_con1},
 we consider values of $\eps^t_\calA$ that support  SPP--V propagation in the direction specified by
 angle $\psi = \le\pi/2\ri - \nu$, wherein
 $ 0 < \nu \ll 1$. As the analysis follows in an analogous manner to that given
  in Sec.~\ref{Sec_con1}, the details need not be presented here. In 
  the case of a
 dissipative dielectric  material $\calA$ and a plasmonic material $\calB$,
   the constraint  \r{inequal_D2}${}_2$ must be satisfied  in order for  SPP--V waves to propagate, in directions close to  $\psi = \pi/2$.

   In the special case in which material $\calA$ is nondissipative, i.e., $\mbox{Im} \lec \eps^s_\calA \ric = 0$ and $\mbox{Im} \lec \eps^t_\calA \ric = 0$,  a stronger result can be derived, as follows. 
   Let us introduce the  constant 
   \begin{equation} \l{K_def}
  K = \frac{ \le \eps_\calB - \eps^t_\calA \ri \le \eps_\calB + \eps^s_\calA \ri }{ \eps^s_\calA - \eps_\calB } .
   \end{equation}
   From the analytical solution \r{psi_sol}, $K$ must be real valued and greater than zero  for all values of $\psi \in \le 0, \pi/2 \ri$.
By equating real and imaginary parts, Eq.~\r{K_def} gives rise to the pair of equations
\begin{equation}
\left.
\begin{array}{l}
\le \mbox{Re} \lec \eps_\calB \ric \ri^2 +  \eps^s_\calA \mbox{Re} \lec \eps_\calB \ric
- \le  \mbox{Im} \lec \eps_\calB \ric \ri^2 -
\eps^t_\calA \mbox{Re} \lec \eps_\calB \ric
\\
- \eps^s_\calA \eps^t_\calA - K \le \eps^s_\calA - \mbox{Re} \lec \eps_\calB \ric \ri = 0 \vspace{8pt} \\
\mbox{Im} \lec \eps_\calB \ric
\les
  \eps^s_\calA  - \eps^t_\calA +
  2 \, \mbox{Re} \lec \eps_\calB \ric
  + K \ris = 0
\end{array}
\right\}.
\end{equation}
The inequality $K > 0 $ thus yields the twin inequalities
 \begin{equation}
 \left.
 \begin{array}{l}
  \mbox{Re} \lec \eps_\calB \ric \le  \mbox{Re} \lec \eps_\calB \ric + \eps^s_\calA - \eps^t_\calA \ri > \le  \mbox{Im} \lec \eps_\calB \ric \ri^2 + \eps^s_\calA \eps^t_\calA  \vspace{4pt} \\
  2 \mbox{Re} \lec \eps_\calB \ric + \eps^s_\calA - \eps^t_\calA < 0
 \end{array}
 \right\},
 \end{equation}
 which together imply the impossible result $- \le  \mbox{Re} \lec \eps_\calB \ric \ri^2 > 0$. Therefore, 
 if material $\calA$ is a nondissipative dielectric material and material $\calB$ is plasmonic, then  SPP--V-wave propagation is impossible for any value of $\psi$.

\section{Numerical studies: SPP-wave propagation}
\label{SPP-num}

In order to use realistic relative 
permittivity parameters that can be conveniently varied, a homogenized composite material (HCM) is introduced to play the role of uniaxial material $\calA$.
The HCM 
arises from a mixture of identically oriented needle-shaped particles  of two component materials labeled  $a$ and   $b$. Particles of both component materials are oriented with their long axes parallel to $\ux$. The volume fraction of  component material $a$  is denoted by $f_a\in[0,1]$ whereas that of component material $b$ is $f_b = 1-f_a$.  For the numerical results presented here,
$\eps_a = -11.63 + 17.45 i$ (cobalt at $\lambdao = 600$ nm \c{J_Co}) and $\eps_b= 5$ (a generic non-dissipative dielectric material).

Provided that the component particles are small in linear dimensions relative to the electromagnetic wavelength(s) involved, the composite material may be regarded as an effectively homogeneous uniaxial material, whose relative permittivity dyadic has the form given in Eq.~\r{Ch4_eps_uniaxial} \c{MAEH}. The relative permittivity parameters of material $\calA$ are estimated by the Bruggeman homogenization formalism as \c{M_PNFA}
\begin{equation} \l{eps_na}
\left.
\begin{array}{l}
\eps_\mathcal{A}^{\rm s} = \frac{1}{2} \Big[ \le f_b-f_a \ri \le \eps_b - \eps_a \ri 
\\ \hspace{24pt} + \sqrt{\les \le f_b-f_a \ri \le \eps_b - \eps_a \ri \ris^2 + 4 \eps_a \eps_b} \, \Big] \vspace{8pt} \\
\eps_\mathcal{A}^{\rm t} = f_a \eps_a + f_b \eps_b
\end{array}
\right\},
\end{equation}
wherein  the square-root term in the expression for $\eps_\mathcal{A}^{\rm s}$ must be taken
to have a positive-valued imaginary part. 

\begin{figure}[!htb]
\centering
\includegraphics[width=4.2cm]{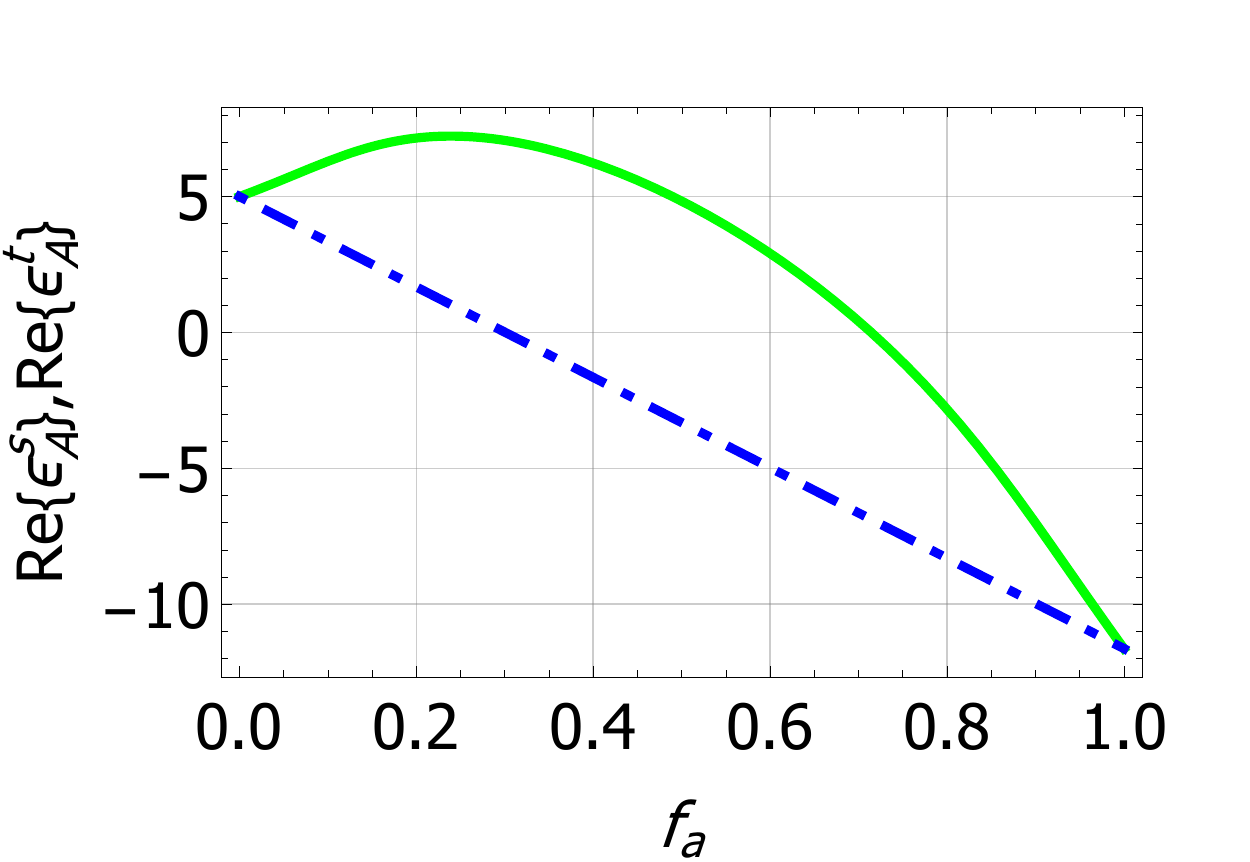}   
\includegraphics[width=4.2cm]{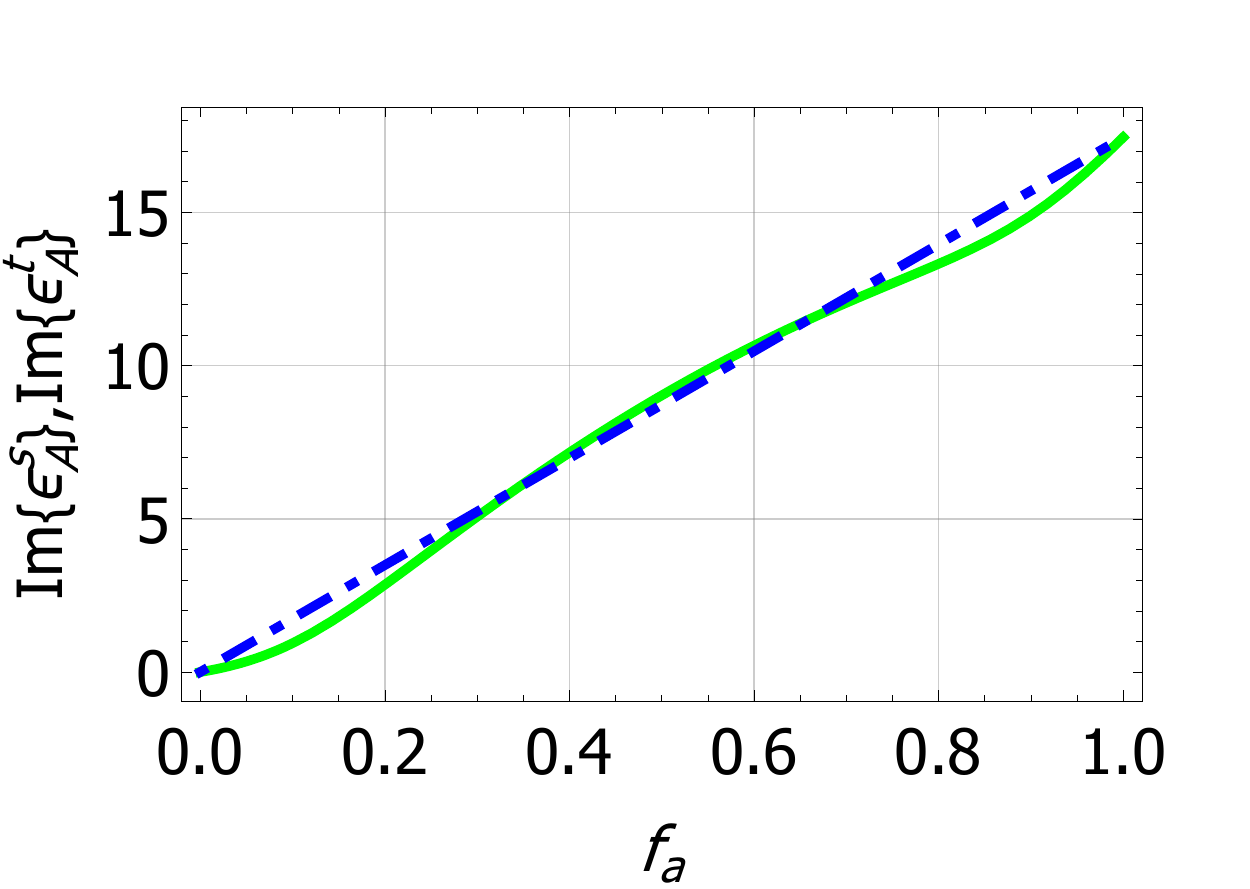}  
  \caption{\label{Fig3}   
  Real and imaginary parts of  $\eps^s_\calA$  (green solid  curves) and $\eps^t_\calA$ (blue broken-dashed curves), as provided by Eqs.~\r{eps_na}, plotted against $f_a \in \les 0,1 \ris$ for $\eps_a = -11.63 + 17.45 i$ and $\eps_b= 5$.  }
\end{figure}

The real and imaginary parts of
$\eps_\mathcal{A}^{\rm s}$ and $\eps_\mathcal{A}^{\rm t}$, as provided by Eqs.~\r{eps_na},  are plotted against $f_a \in \les 0, 1 \ris$ in Fig.~\ref{Fig3}.
For $f_a < 0.31$, material $\calA$ is a dissipative dielectric material with
  $\mbox{Re} \lec
\eps_\mathcal{A}^{\rm s}
 \ric>0$ and $\mbox{Re} \lec
\eps_\mathcal{A}^{\rm t}
 \ric > 0$.
 Specifically, 
 $\eps_\mathcal{A}^{\rm s} = 6.30 + 0.94 i$ and $\eps_\mathcal{A}^{\rm t} = 3.34 + 1.75 i$ for  $f_a = 0.1$; 
 $\eps_\mathcal{A}^{\rm s} = 7.15 + 2.85 i$ and $\eps_\mathcal{A}^{\rm t} = 1.67 + 3.49 i$ for $f_a = 0.2$; and 
  $\eps_\mathcal{A}^{\rm s} = 7.07 + 5.07 i$ and $\eps_\mathcal{A}^{\rm t} = 0.011 + 5.24 i$ for $f_a = 0.3$. 
  For $f_a > 0.71$, material $\calA$ is a uniaxial plasmonic material since  $\mbox{Re} \lec
\eps_\mathcal{A}^{\rm s}
 \ric <0$ and  $\mbox{Re} \lec
\eps_\mathcal{A}^{\rm t}
 \ric < 0$.
  Specifically, 
 $\eps_\mathcal{A}^{\rm s} = -0.17 + 12.31 i$ and $\eps_\mathcal{A}^{\rm t} = -6.97 + 12.56 i$ for  $f_a = 0.72$; 
 $\eps_\mathcal{A}^{\rm s} =-2.82 + 13.32 i$ and $\eps_\mathcal{A}^{\rm t} = -8.30 + 13.96 i$ for $f_a = 0.80$; and 
  $\eps_\mathcal{A}^{\rm s} = -7.01 + 14.90 i$ and $\eps_\mathcal{A}^{\rm t} =  -9.97 + 15.71 i$ for $f_a = 0.9$. 
In the regime $0.31 < f_a < 0.71$, material $\calA$ is classified as a hyperbolic material
 since $\mbox{Re} \lec
\eps_\mathcal{A}^{\rm s}
 \ric \mbox{Re} \lec
\eps_\mathcal{A}^{\rm t}
 \ric < 0$ \c{ML_JO}. The hyperbolic regime is not  considered in the 
 following numerical studies, but
 this may be an interesting regime to investigate in the future~---~especially 
since hyperbolic partnering materials have recently been found to support surface waves with negative phase velocity \c{ML_JO}.

\subsection{Anisotropic plasmonic material $\calA$ / isotropic dielectric material $\calB$}

Consider the case where  material $\calA$ is a plasmonic material, specified by the relative permittivity parameters \r{eps_na} with $f_a > 0.71$. Material $\calB$ is taken to be a generic non-dissipative dielectric material with relative permittivity $\eps_\calB = 5$.

\begin{figure}[!htb]
\centering
\includegraphics[width=4.2cm]{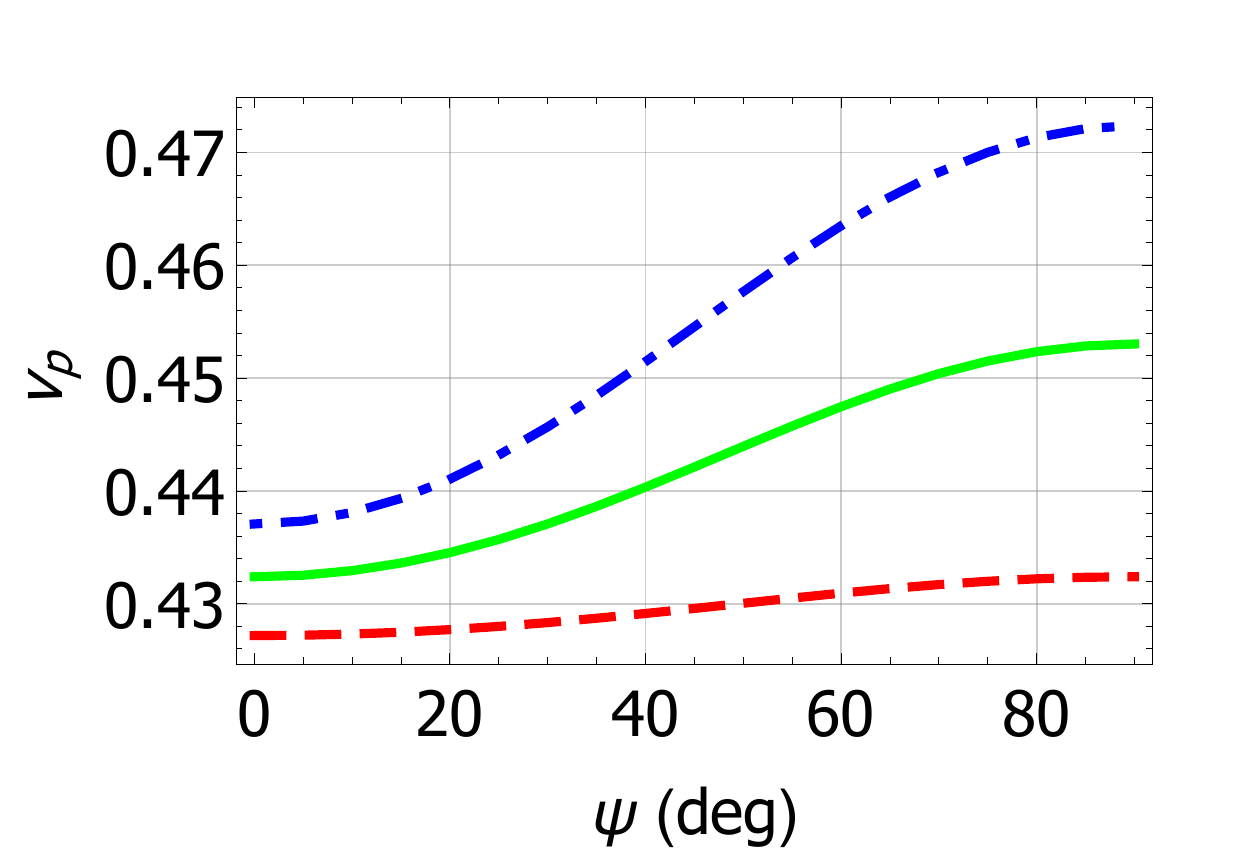}  \includegraphics[width=4.2cm]{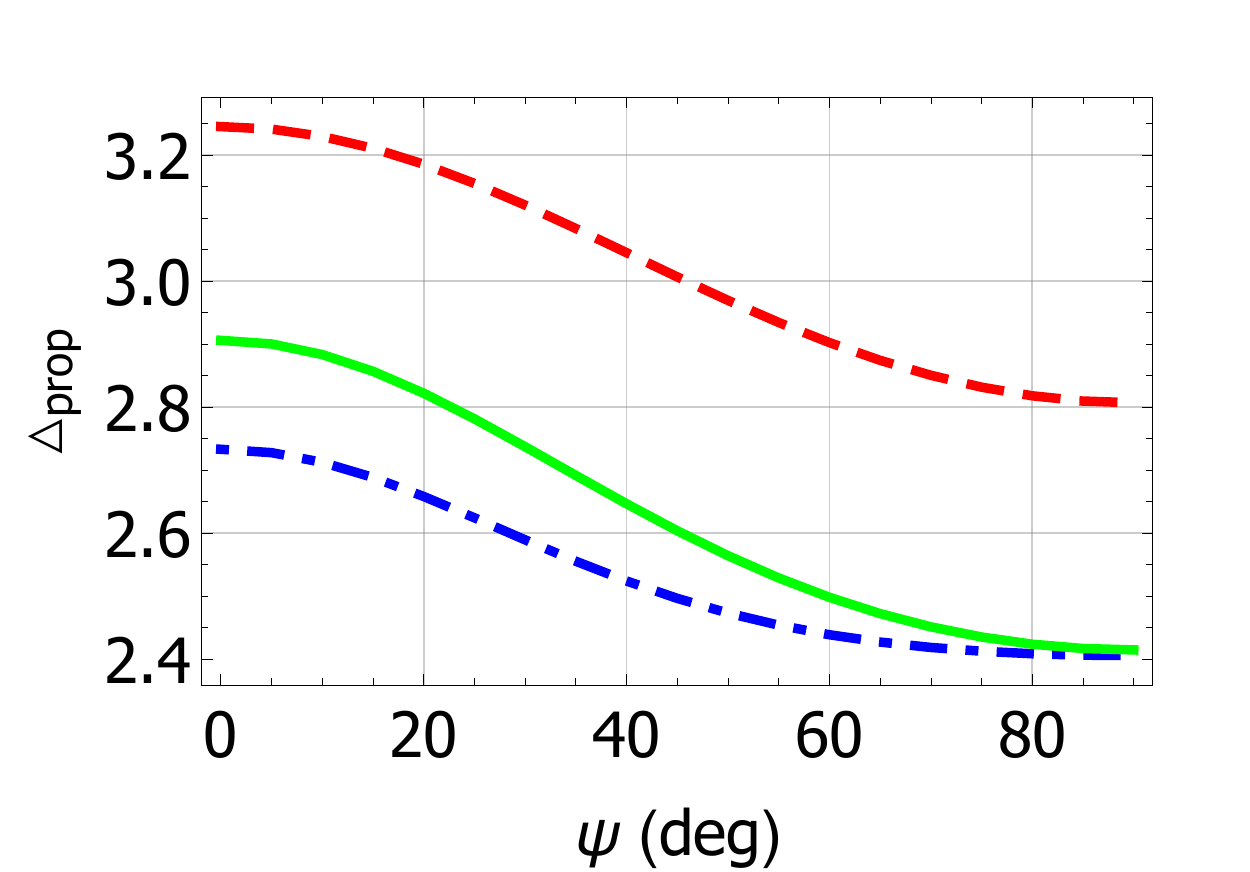}\\ 
\includegraphics[width=4.2cm]{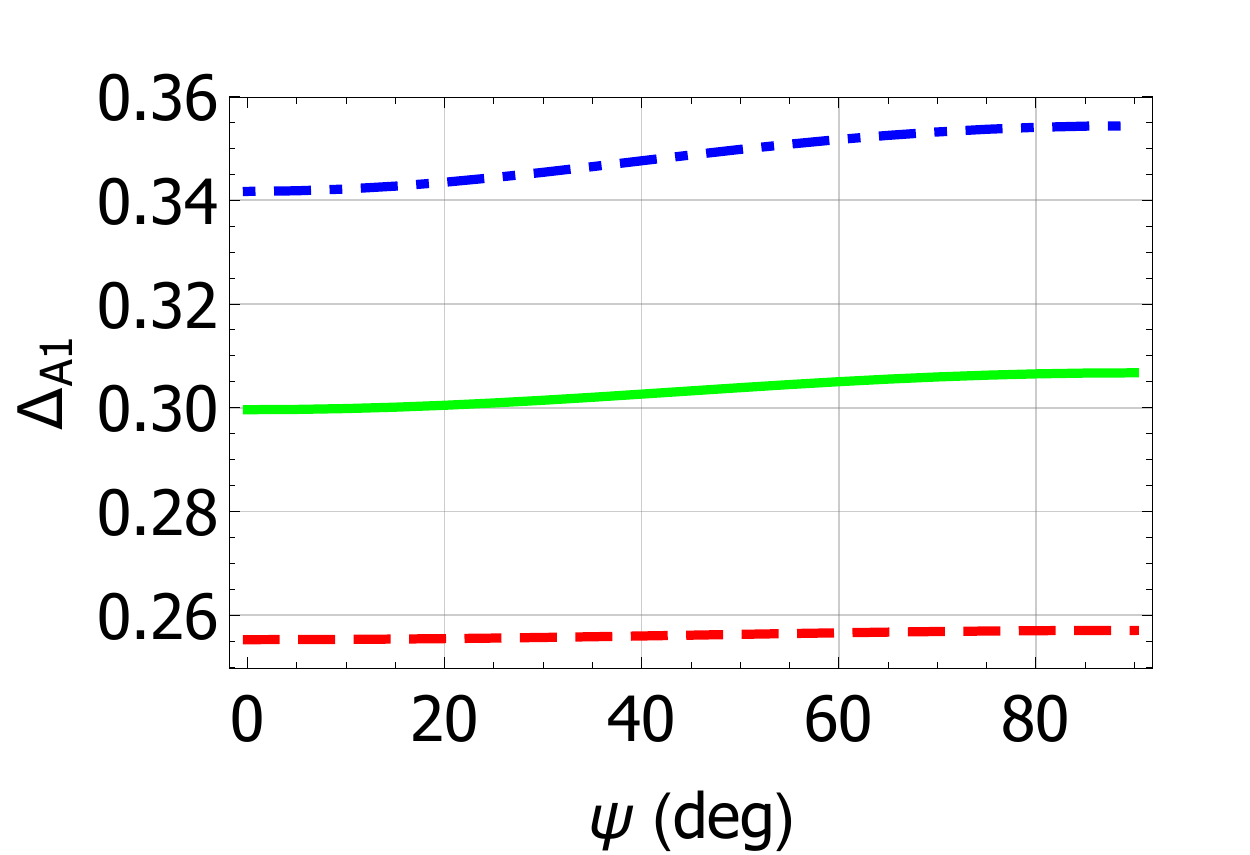}  \includegraphics[width=4.2cm]{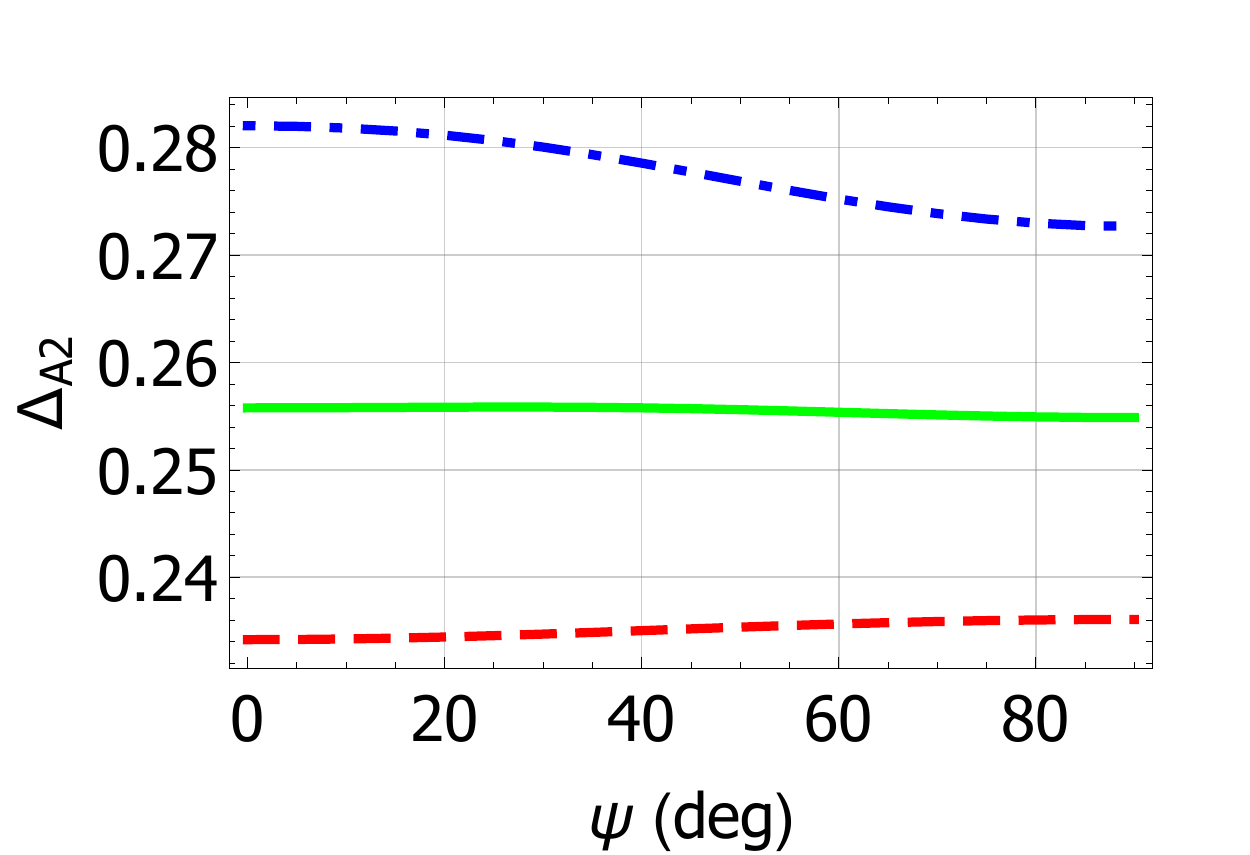}\\
\includegraphics[width=4.2cm]{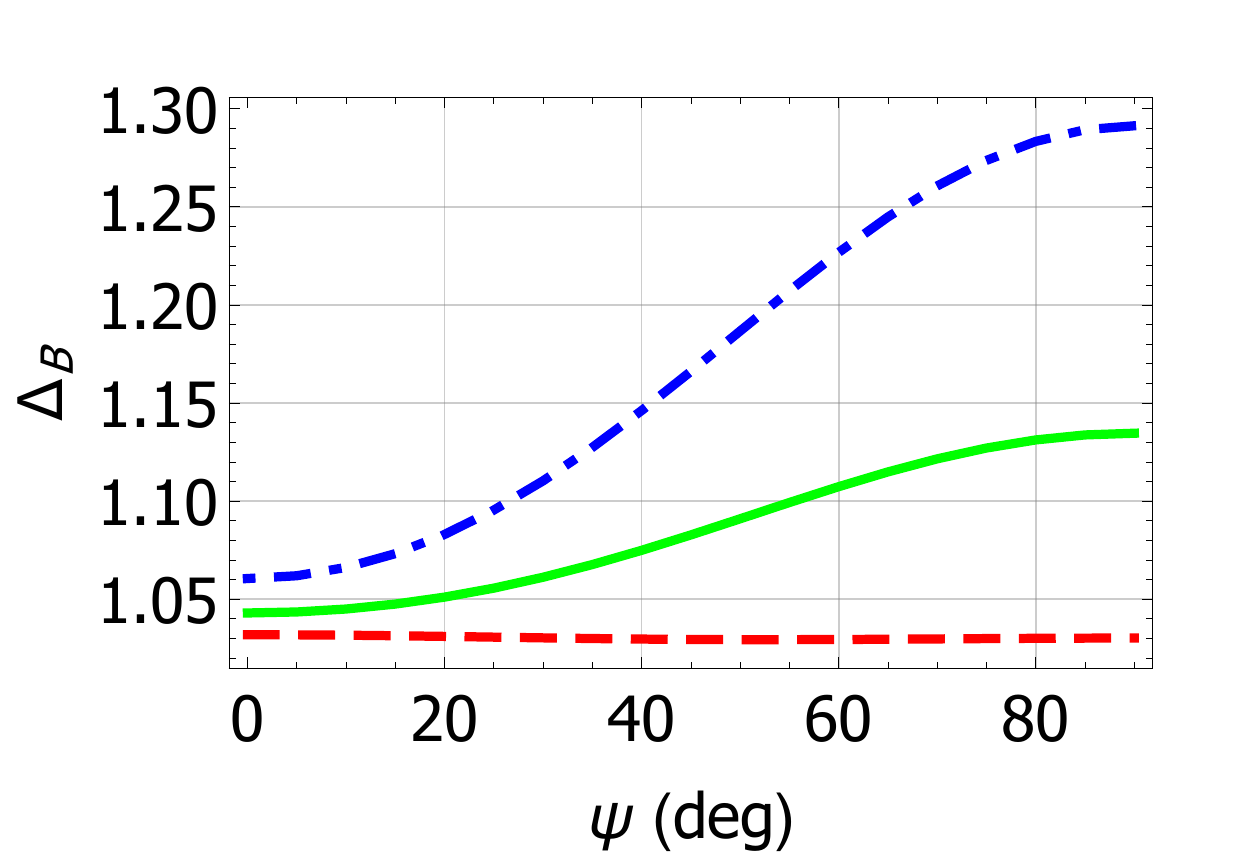}
  \caption{\label{Fig4}    SPP waves:
Plots of the normalized phase speed $v_\text{p}$ and normalized  propagation length $\Delta_{\text{prop}}$, computed using values of   $q $ extracted numerically from Eq.~\r{DE}, 
and the normalized penetration depths $\Delta_{\calA1}$,
$\Delta_{\calA2}$, and $\Delta_\calB$,   as calculated from Eqs.~\r{a_decay_const} and \r{b_decay_const}, 
versus 
$\psi \in \le 0, \pi/2 \ri$ 
for $f_a =  0.72$ (blue broken-dashed curves),  $0.80$  (green solid  curves), and  $0.90$ (red dashed curves).
    }
\end{figure}

Plots of the 
normalized phase speed
\begin{equation}
v_\text{p} = \frac{\ko}{\mbox{Re} \lec q \ric}
\end{equation}
and normalized propagation length
\begin{equation}
\Delta_{\text{prop}} = \frac{\ko}{\mbox{Im} \lec q \ric},
\end{equation} as
 computed using values of 
 $q $ extracted numerically from Eq.~\r{DE}, versus 
$\psi \in \le 0, \pi/2 \ri$ are provided in Fig.~\ref{Fig4}
for $f_a \in \lec 0.72, 0.80, 0.90 \ric$. Also provided in Fig.~\ref{Fig4} are corresponding plots of the
 normalized penetration depths 
 \begin{equation} \l{pd_s}
\left.
\begin{array}{l}
\Delta_{\calA \ell} = \displaystyle{\frac{\ko}{\mbox{Im} \lec \alpha_{\calA \ell} \ric}}, \qquad (\ell = 1, 2) \vspace{8pt}\\
\Delta_\calB = \displaystyle{\frac{\ko}{- \mbox{Im} \lec \alpha_\calB \ric}}
\end{array}
\right\},
\end{equation}
as calculated from Eqs.~\r{a_decay_const} and \r{b_decay_const}, respectively.
In Fig.~\ref{Fig4},
the normalized phase speed, as well as the normalized  penetration depths in both partnering materials, vary more as $\psi$ increases for smaller values of $f_a$.
Also, the penetration depths 
in material $\calB$ are substantially greater than the penetration depths in material $\calA$. This observation is in line with what would be expected for a plasmonic/dielectric interface, regardless of anisotropy of the partnering material $\calA$.

\begin{figure}[!htb]
\centering
\includegraphics[width=4.2cm]{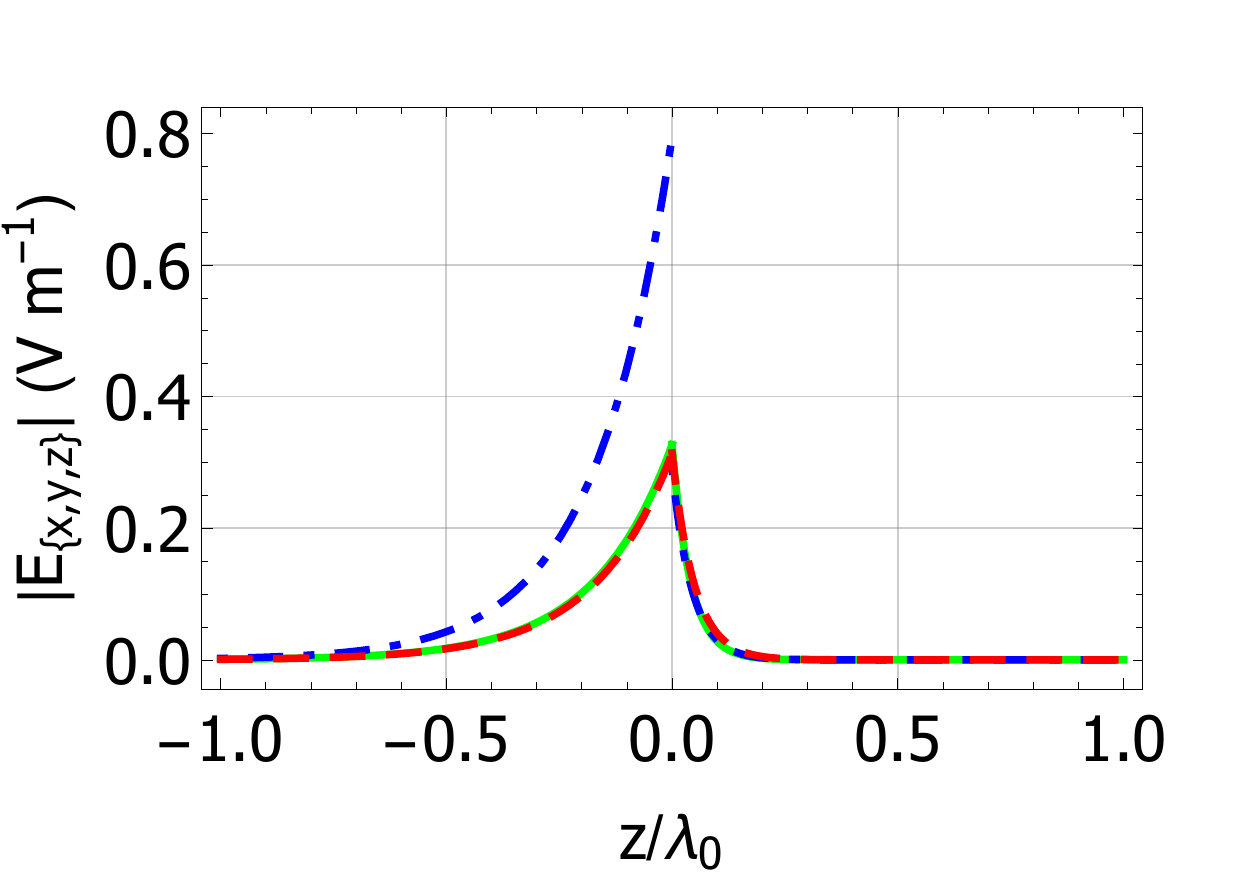} 
 \includegraphics[width=4.2cm]{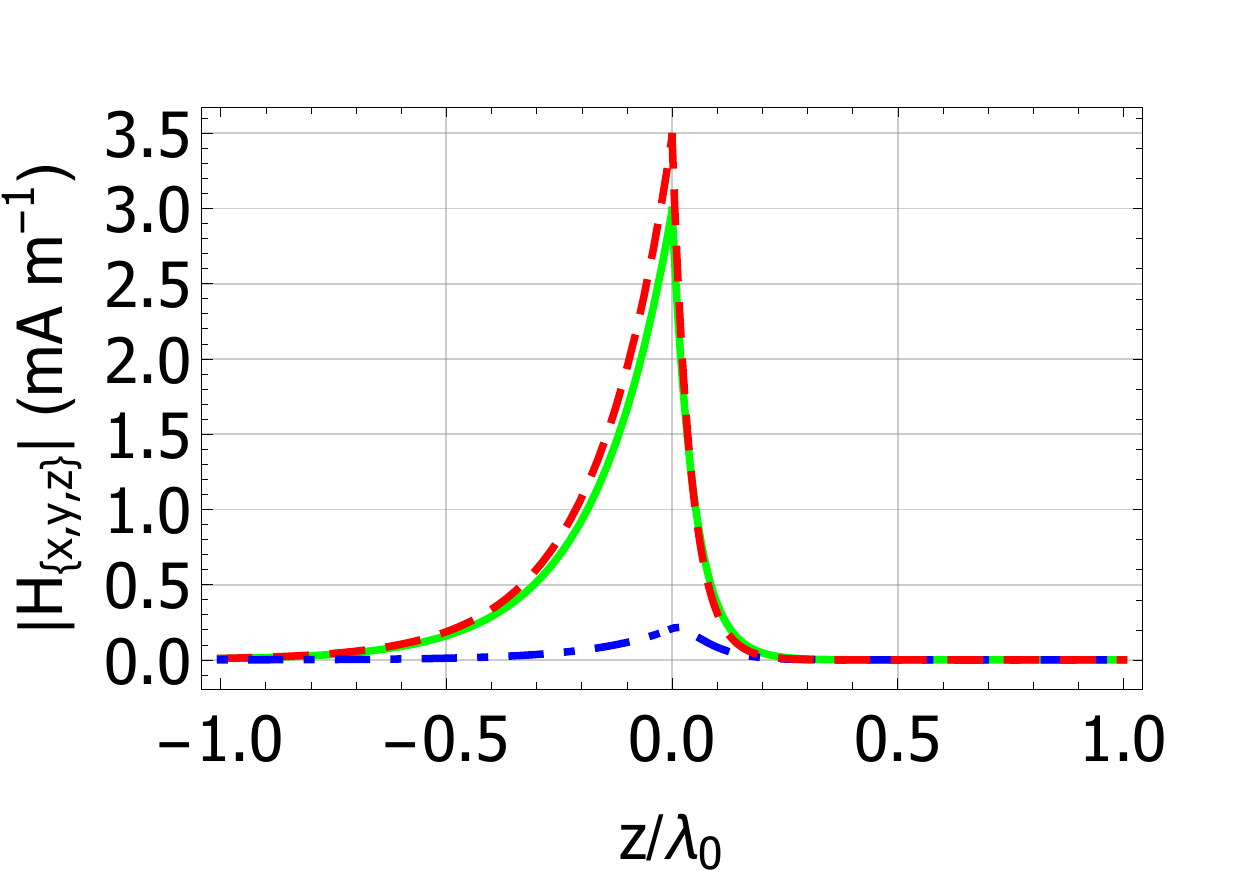}\\
  \includegraphics[width=4.2cm]{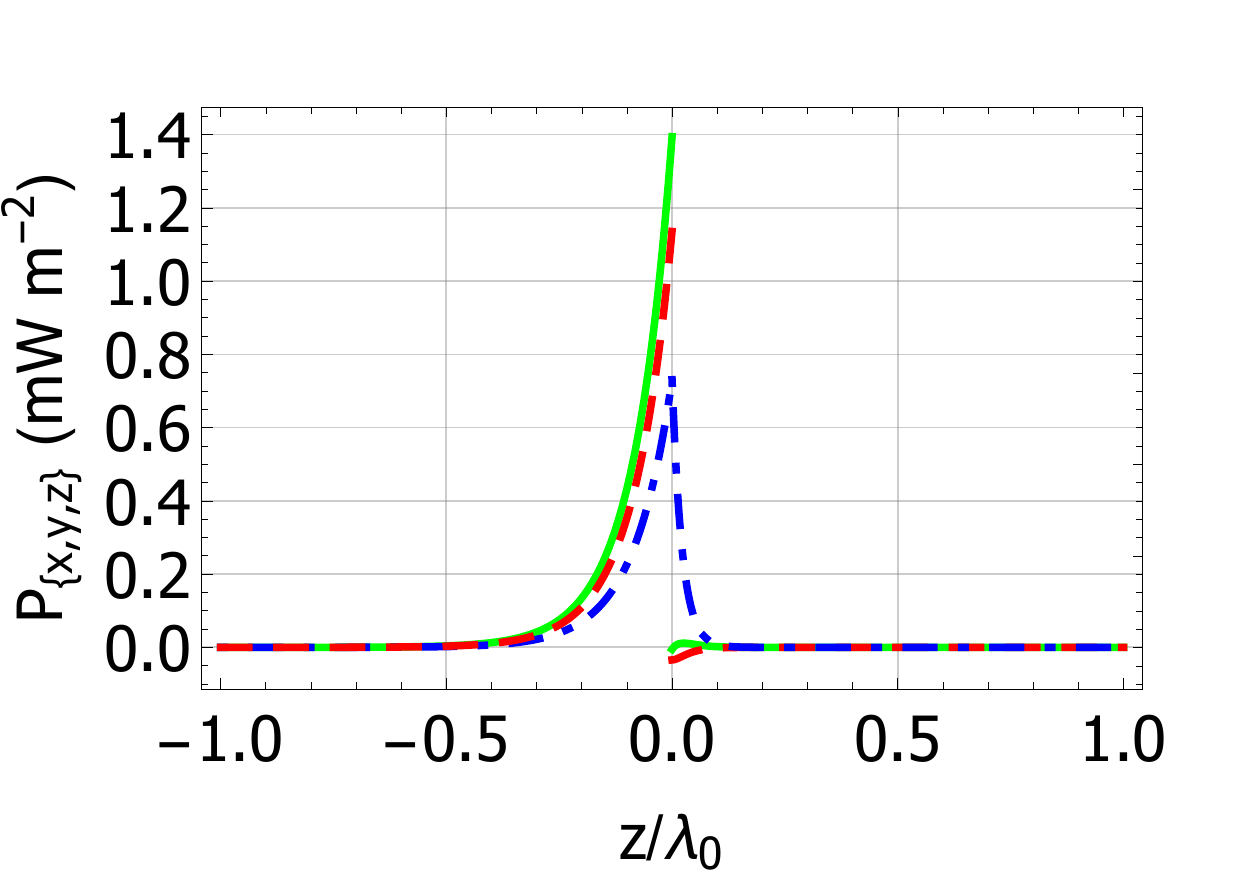}  
  \caption{\label{Fig5}   
 SPP waves:  
$\vert{E_{\lec x,y,z\ric}(z\uz)}\vert$,  $\vert{H_{\lec x,y,z\ric}(z\uz)}\vert$, and
$P_{\lec x,y,z\ric}(z\uz)$ plotted versus $z/\lambdao$, 
 for
 $f_a = 0.80$  and $\psi = 40^\circ$,
 with  $C_{\mathcal{B}1} = 1$ V m${}^{-1}$.
  Key:   $x$-directed components: green solid curves; 
 $y$-directed components: red dashed  curves; $z$-directed components: blue  broken-dashed curves.
}
\end{figure}

 The nature of the SPP waves represented in  Fig.~\ref{Fig4} is further illuminated 
in Fig.~\ref{Fig5} wherein
$\vert{E_{\lec x,y,z\ric}(z\uz)}\vert$ and  $\vert{H_{\lec x,y,z\ric}(z\uz)}\vert$
 are plotted versus $z/\lambdao$ for the case $f_a = 0.80$ with $\psi = 40^\circ$.
Also plotted are ${P_{\lec x,y,z\ric}(z\uz)}$ which represent
the Cartesian components  of the time-averaged Poynting vector
\begin{equation}
\underline{P}  (\#r) =
 \frac{1}{2} \mbox{Re} \les \, \underline{E}  (\#r)  \times 
\underline{H}^*  (\#r) \, \ris, 
\end{equation}
where the asterisk denotes the complex conjugate.
For these computations, we fixed
   $C_{\mathcal{B}1} = 1$ V m${}^{-1}$. The  localization of the SPP wave to the interface $z=0$  is clearly evident, with the degree of localization being substantially greater in the half-space $z>0$  than in the half-space $z<0$, as would be expected from the plots of the penetration depths in Fig.~\ref{Fig4}.

\subsection{Anisotropic dielectric material $\calA$ / isotropic plasmonic material $\calB$}

Consider the case where  material $\calA$ is a dissipative dielectric material, specified by the relative permittivity parameters \r{eps_na} with $f_a < 0.31 $. Material $\calB$ is taken to be a plasmonic material with relative permittivity $\eps_\calB = -11.63 + 17.45 i$.

\begin{figure}[!htb]
\centering
\includegraphics[width=4.2cm]{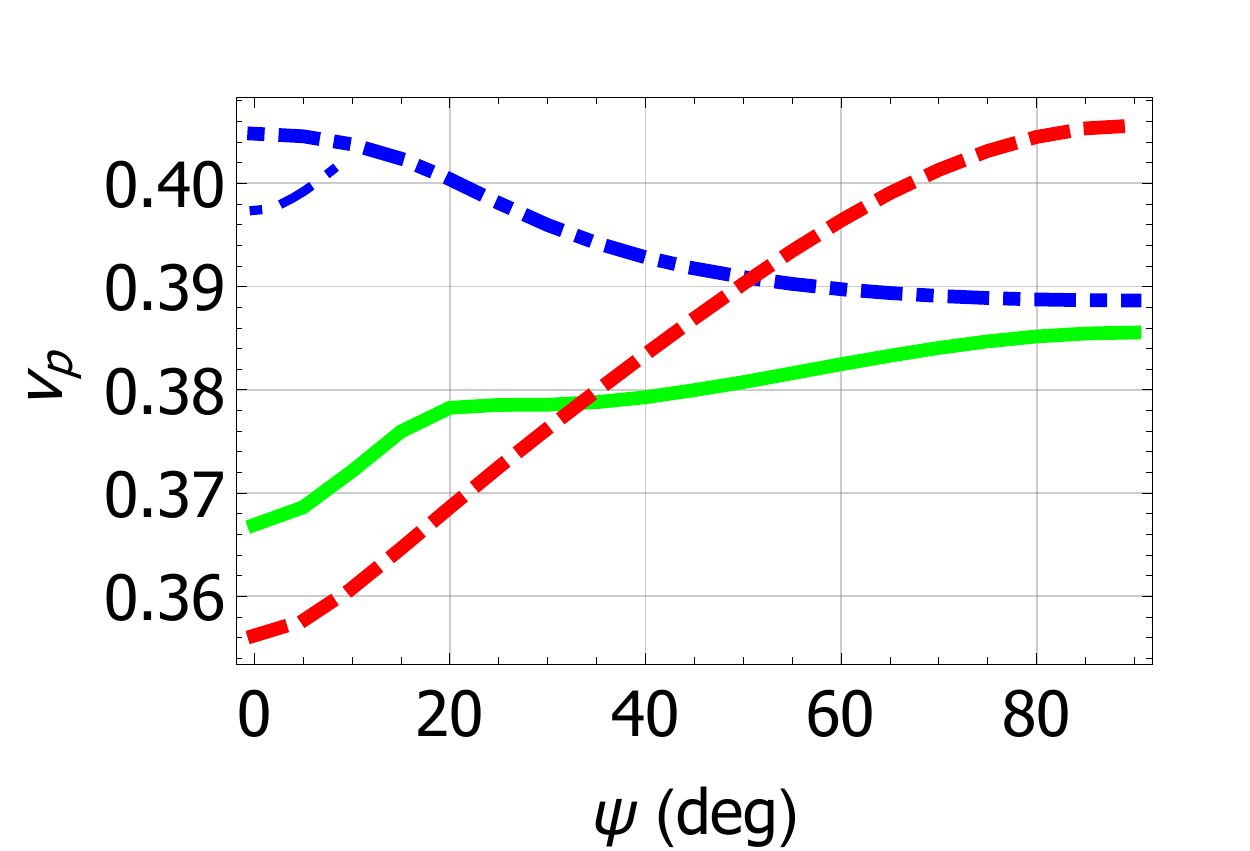}  \includegraphics[width=4.2cm]{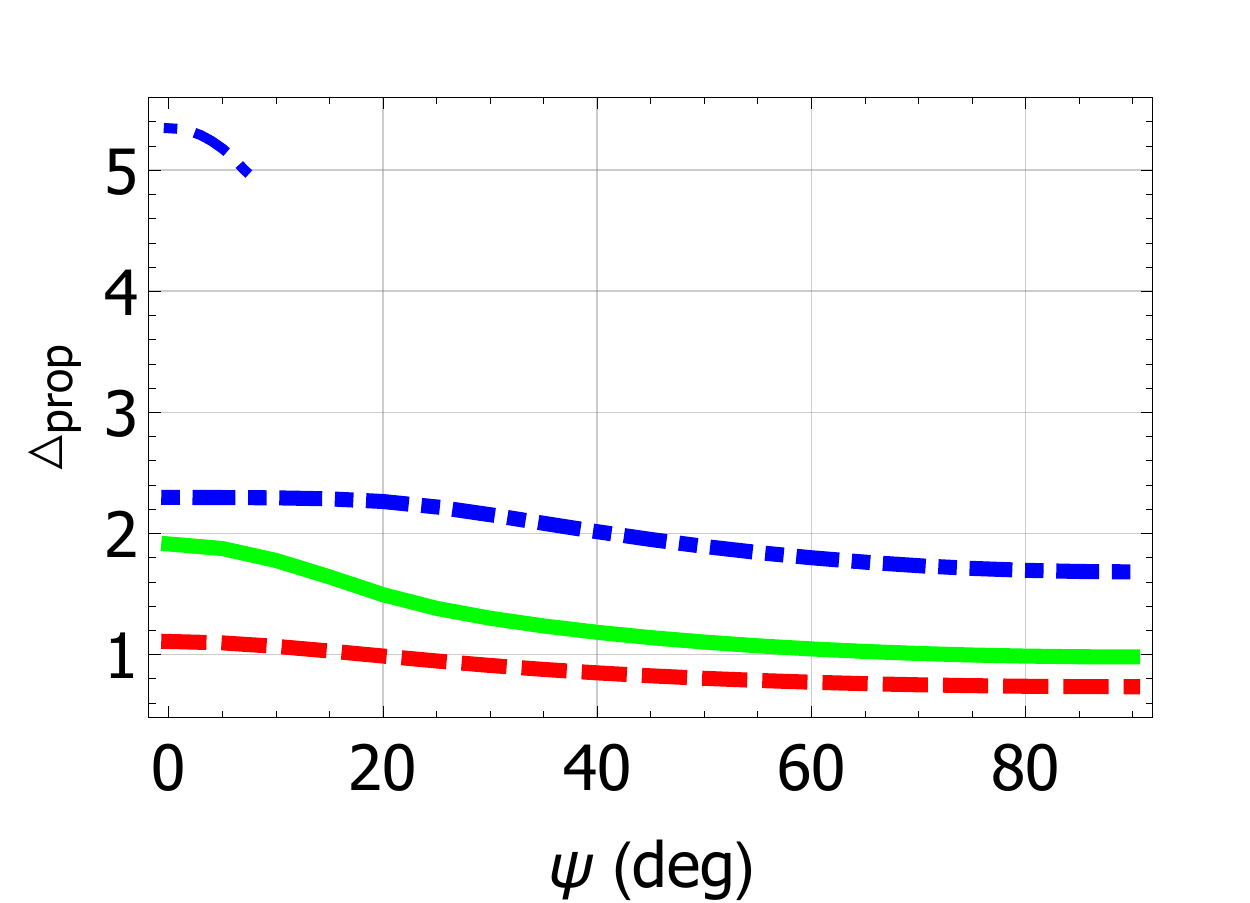}\\ 
\includegraphics[width=4.2cm]{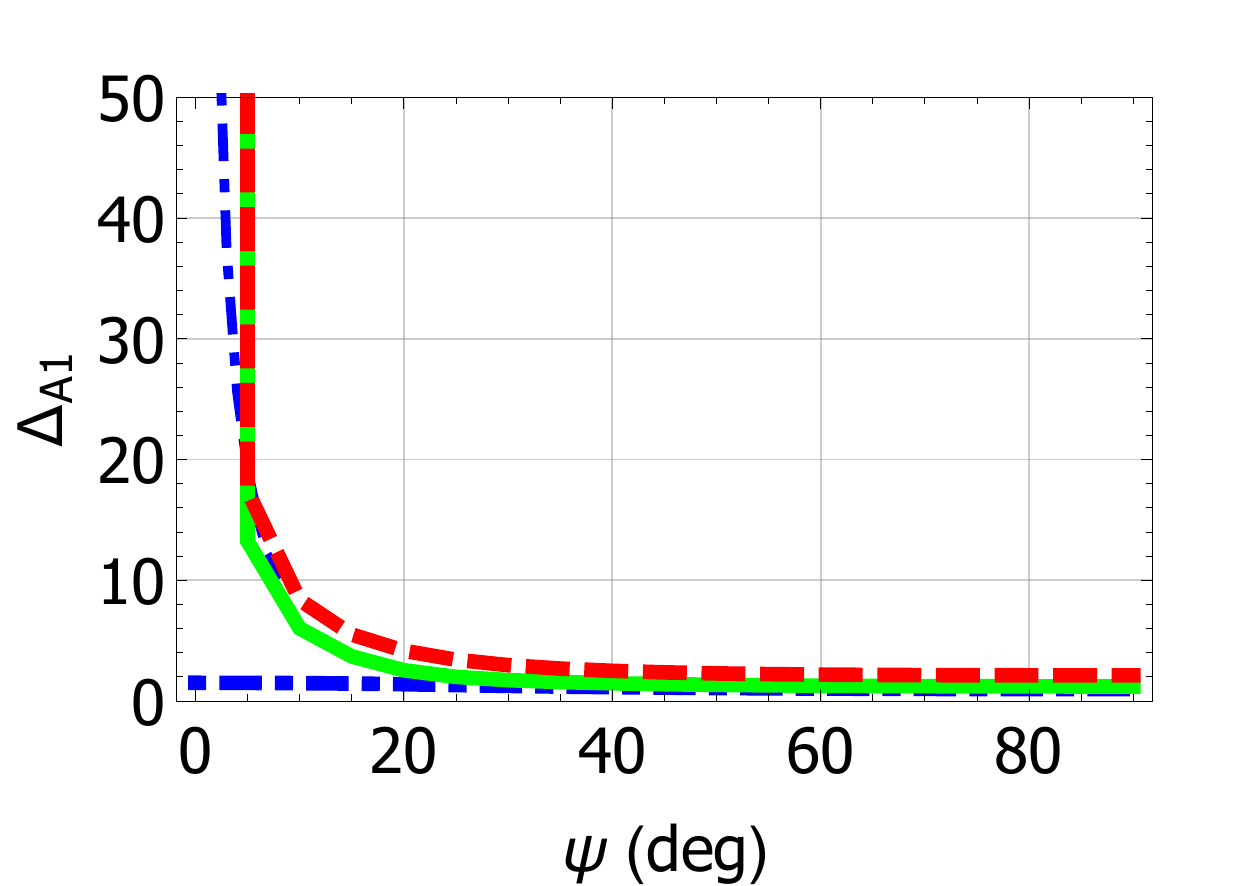}  \includegraphics[width=4.2cm]{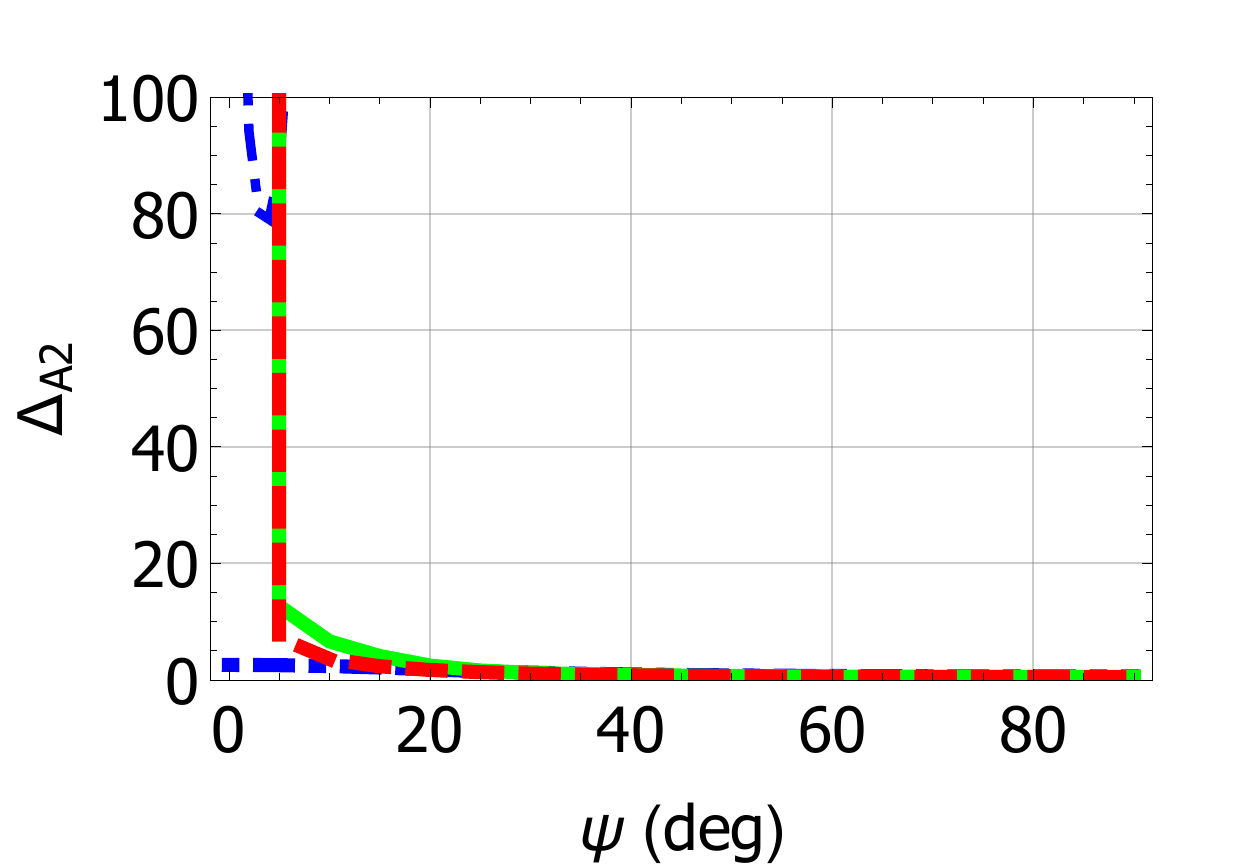}\\
\includegraphics[width=4.2cm]{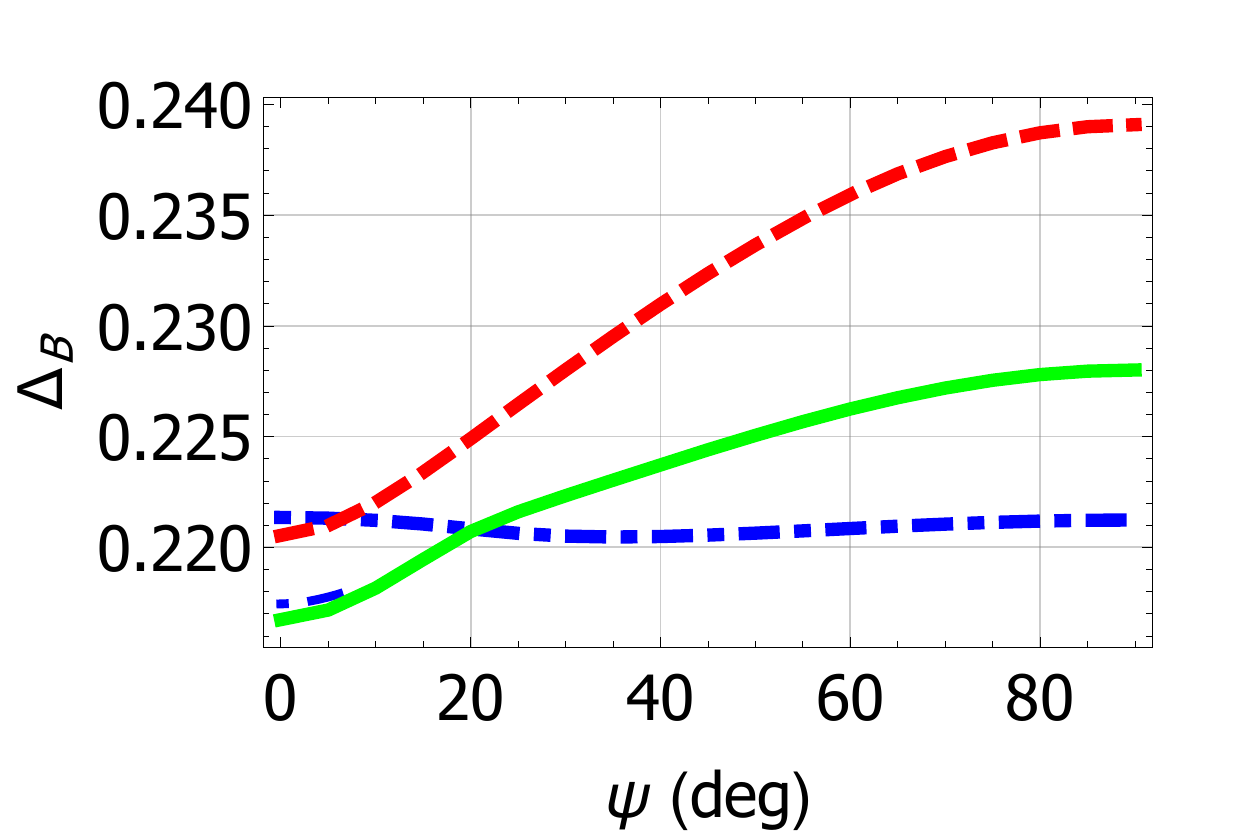}
  \caption{\label{Fig6}    SPP waves: As Fig.~\ref{Fig4} but for $f_a =  0.10$ (blue broken-dashed curves),  $0.20$  (green solid  curves), and  $0.30$ (red dashed curves). The logarithms of $\Delta_{\calA 1}$ and $\Delta_{\calA 2}$ are plotted instead of $\Delta_{\calA 1}$ and $\Delta_{\calA 2}$.
    }
\end{figure}

Plots of the normalized phase speed $v_\text{p}$ and 
normalized 
propagation length $\Delta_{\text{prop}}$, computed using values of   $q $ extracted numerically from Eq.~\r{DE}, versus 
$\psi \in \le 0, \pi/2 \ri$ are provided in Fig.~\ref{Fig6}
for $f_a \in \lec 0.10, 0.20, 0.30 \ric$.  Also provided in Fig.~\ref{Fig6} are corresponding plots of the logarithms of the
 normalized penetration depths $\Delta_{\calA 1}$ and $\Delta_{\calA 2}$, and 
 the  normalized penetration depth
 $\Delta_{\calB}$, as calculated from Eqs.~\r{a_decay_const} and \r{b_decay_const}, respectively.  The SPP-wave solutions represented in Fig.~\ref{Fig6} are both qualitatively and quantitatively different to those represented in Fig.~\ref{Fig4}.
 Most obviously,  two solution branches exist for the case $f_a = 0.10$: the first branch  exists for all $\psi \in \le 0 , \pi/2 \ri$ whereas the second branch   exists only for $0^\circ < \psi < 8.07^\circ$. In contrast, only one solution arises  for
  $f_a \in \lec  0.20, 0.30 \ric$ and it exists
 for all values of $\psi \in \le 0 , \pi/2 \ri$.  Also only one
 solution exists 
 at each value of $\psi$ considered in Fig.~\ref{Fig4}.
 Furthermore, the SPP waves on the  branch  that exists for $0^\circ < \psi < 8.07^\circ$
at $f_a = 0.10$ penetrate much further into material $\calA$ than do the SPP waves on the branch  that  exists for $0^\circ < \psi < 90^\circ$
at $f_a = 0.10$, but the penetration depths into material $\calB$ solutions are similar for solutions on both branches. In addition, the penetration depths into material $\calA$ for the  SPP-wave solutions for
  $f_a \in \lec  0.20, 0.30 \ric$ are much greater for $0^\circ < \psi \lessapprox 10^\circ$ than they are for $10^\circ \lessapprox \psi < 90^\circ$, but this is not the case for the penetration depths into material $\calB$.

\begin{figure}[!htb]
\centering
\includegraphics[width=4.2cm]{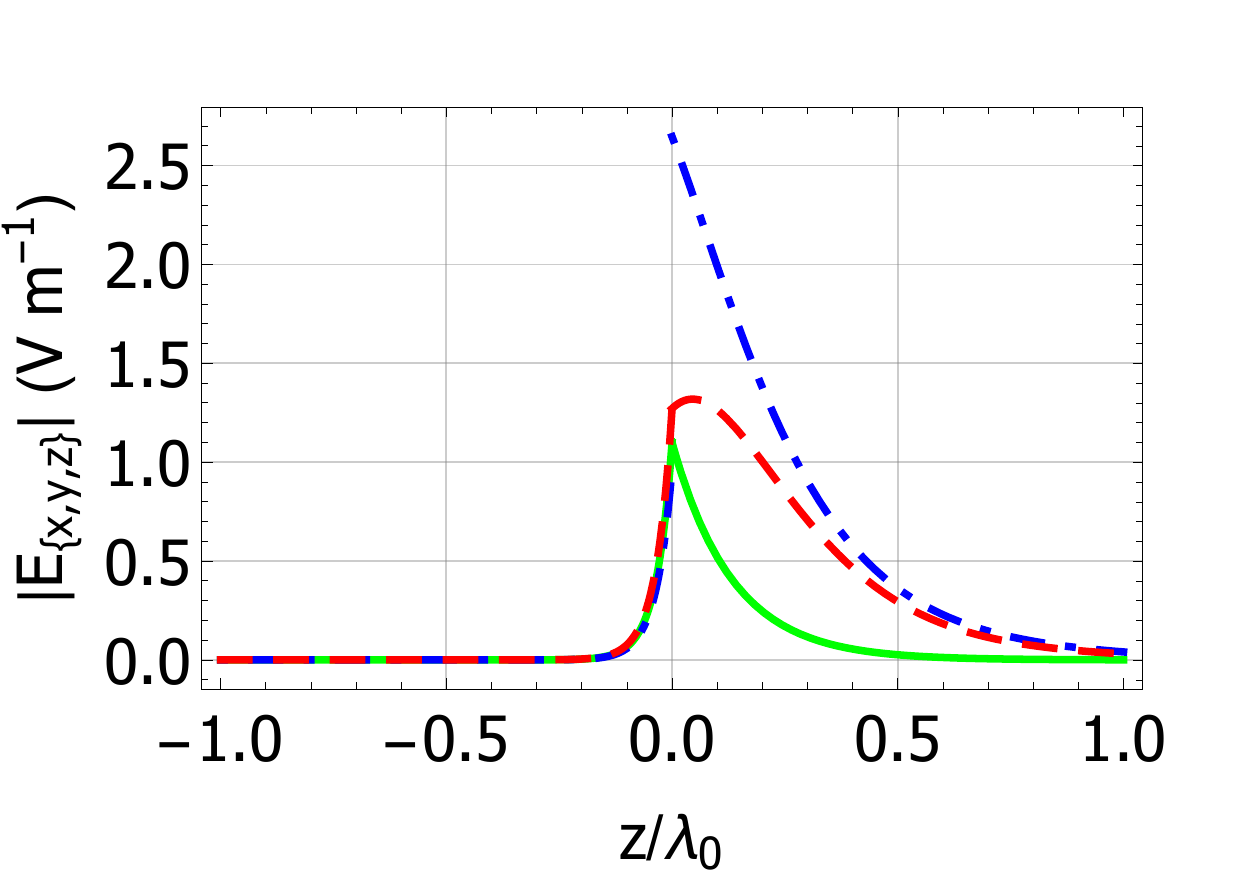} 
 \includegraphics[width=4.2cm]{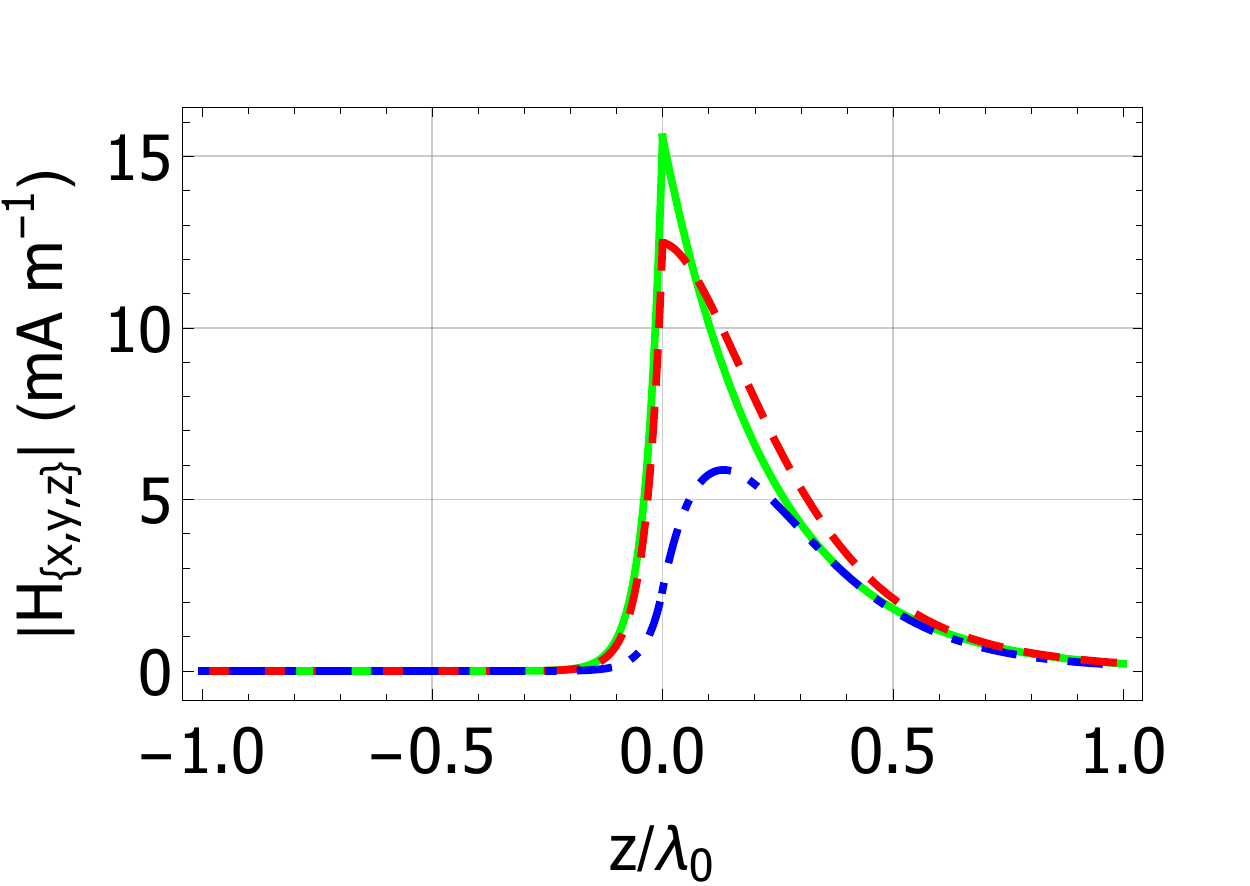}\\
  \includegraphics[width=4.2cm]{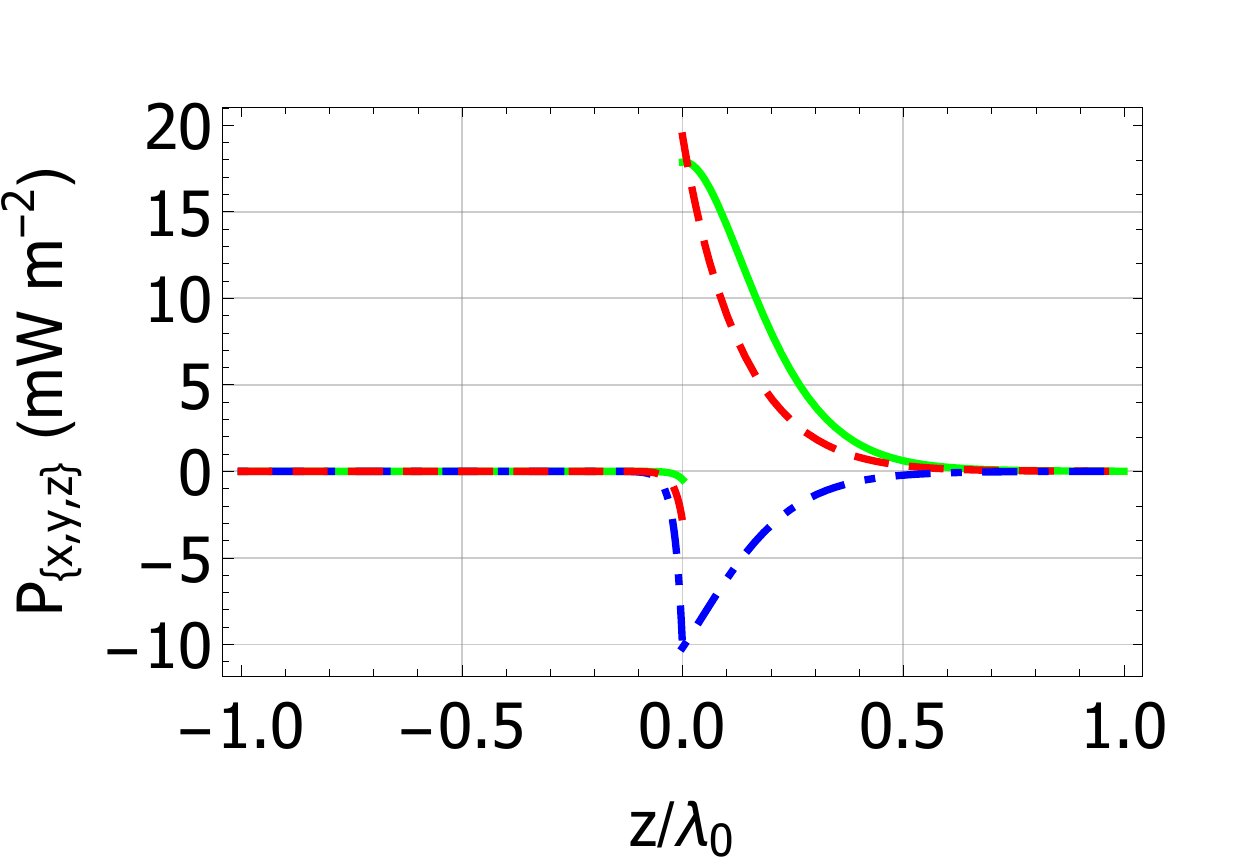}  
  \caption{\label{Fig7}   
 SPP waves: 
  As Fig.~\ref{Fig5} but for $f_a = 0.20$.
}
\end{figure}

 Further light  is  shed onto the nature of the  SPP waves represented in  Fig.~\ref{Fig6} by considering the field profiles in the direction normal to the interface $z=0$.
  In Fig.~\ref{Fig7},
$\vert{E_{\lec x,y,z\ric}(z\uz)}\vert$,  $\vert{H_{\lec x,y,z\ric}(z\uz)}\vert$, and
${P_{\lec x,y,z\ric}(z\uz)}$ are plotted versus $z/\lambdao$ for the case $f_a = 0.20$ with $\psi = 40^\circ$,
and
   $C_{\mathcal{B}1} = 1$ V m${}^{-1}$.
   Unlike the case presented in Fig.~\ref{Fig5}, the degree of localization of the SPP wave to the interface $z=0$   in Fig.~\ref{Fig7}
is substantially greater in the half-space $z<0$  than it is in the half-space $z>0$.

\section{Numerical studies:  SPP--V-wave propagation}
\label{SPPV-num}

\subsection{Anisotropic plasmonic material $\calA$ / isotropic dielectric material $\calB$}

Next, material $\calA$ is taken to be a plasmonic material while  material $\calB$ is taken to be a non-dissipative dielectric material.

\begin{figure}[!htb]
\centering
\includegraphics[width=4.2cm]{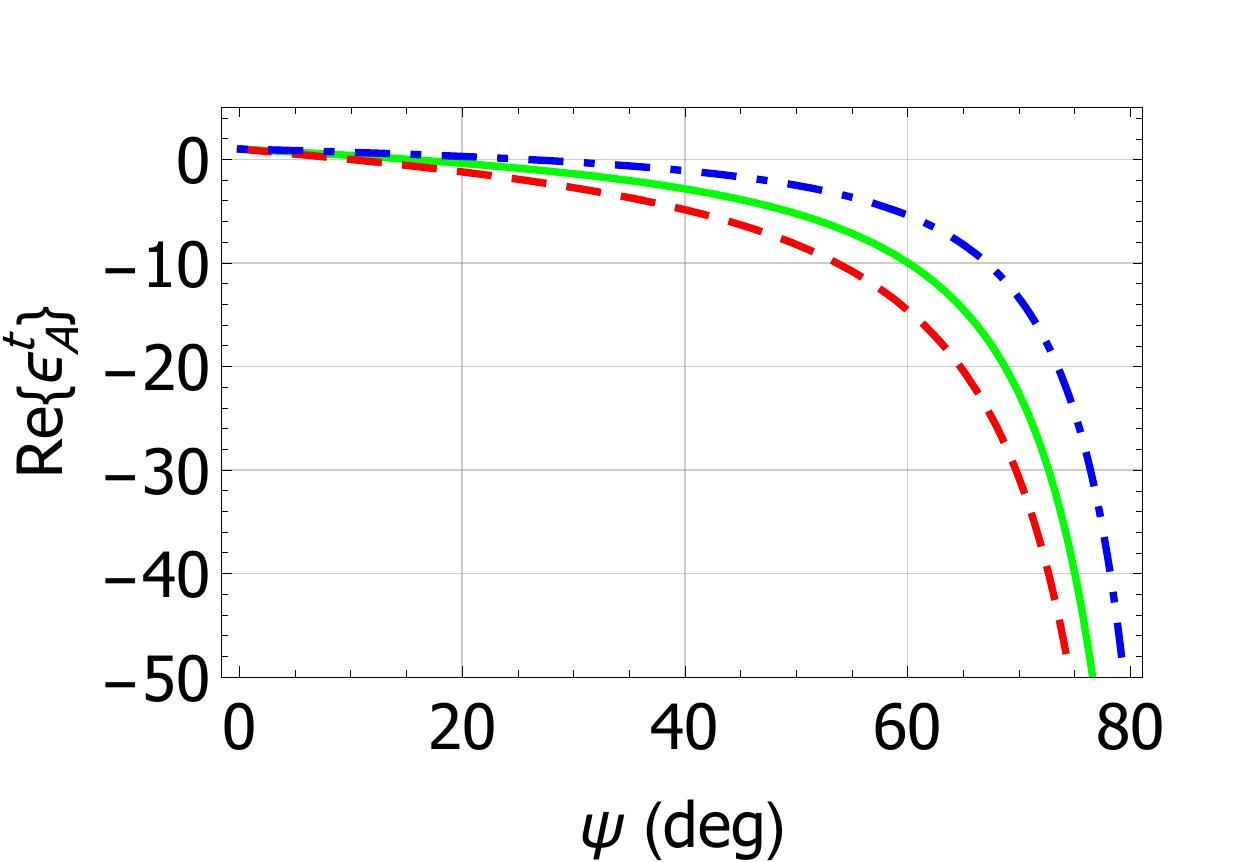}  \includegraphics[width=4.2cm]{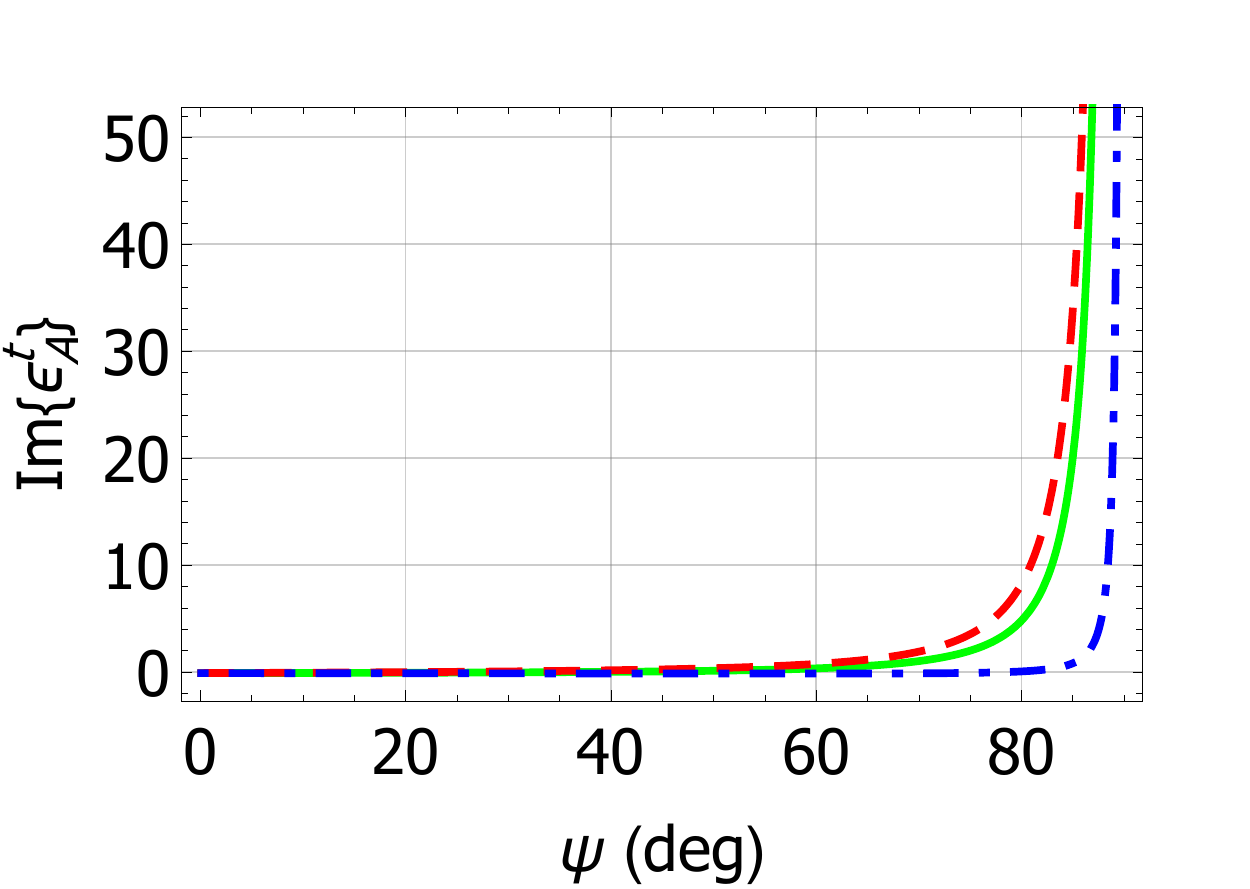}\\ 
\includegraphics[width=4.2cm]{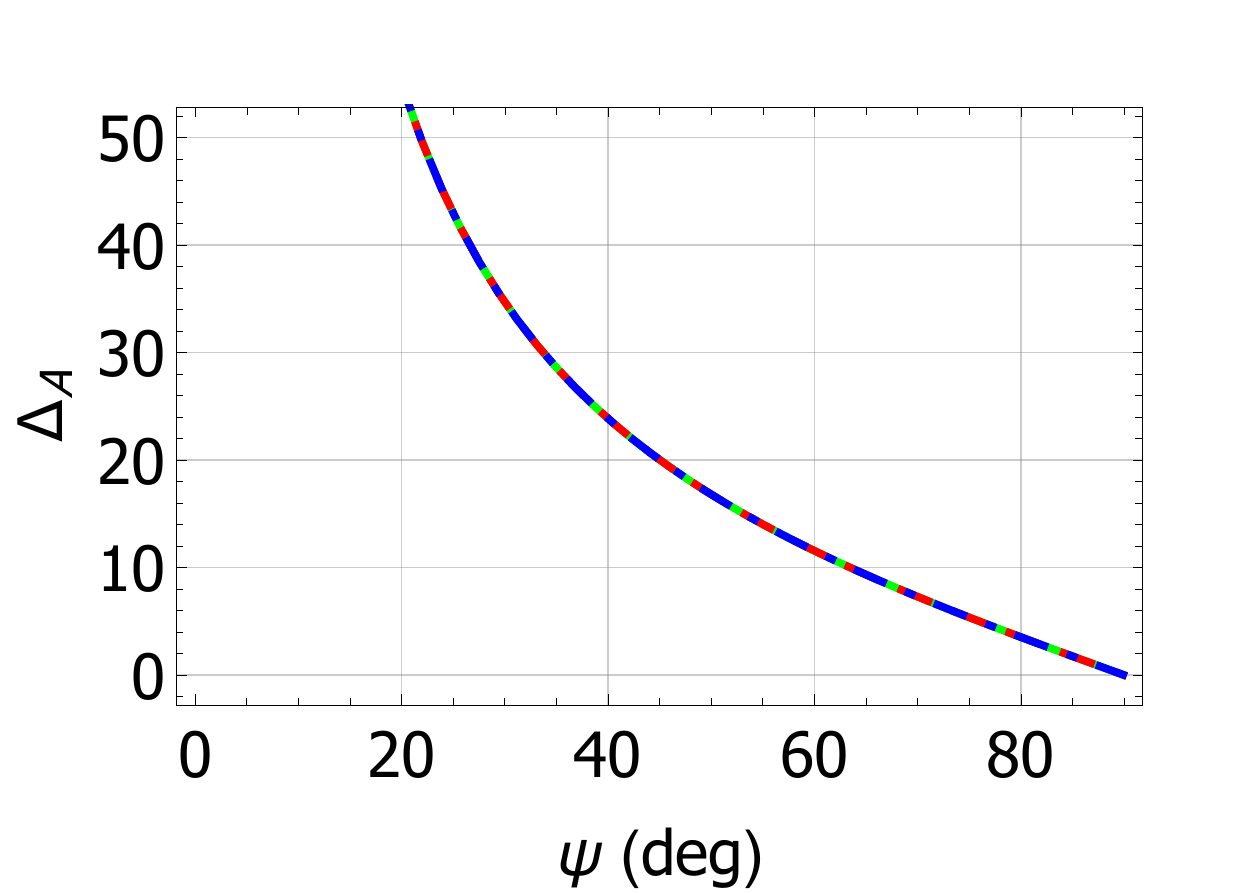}  \includegraphics[width=4.2cm]{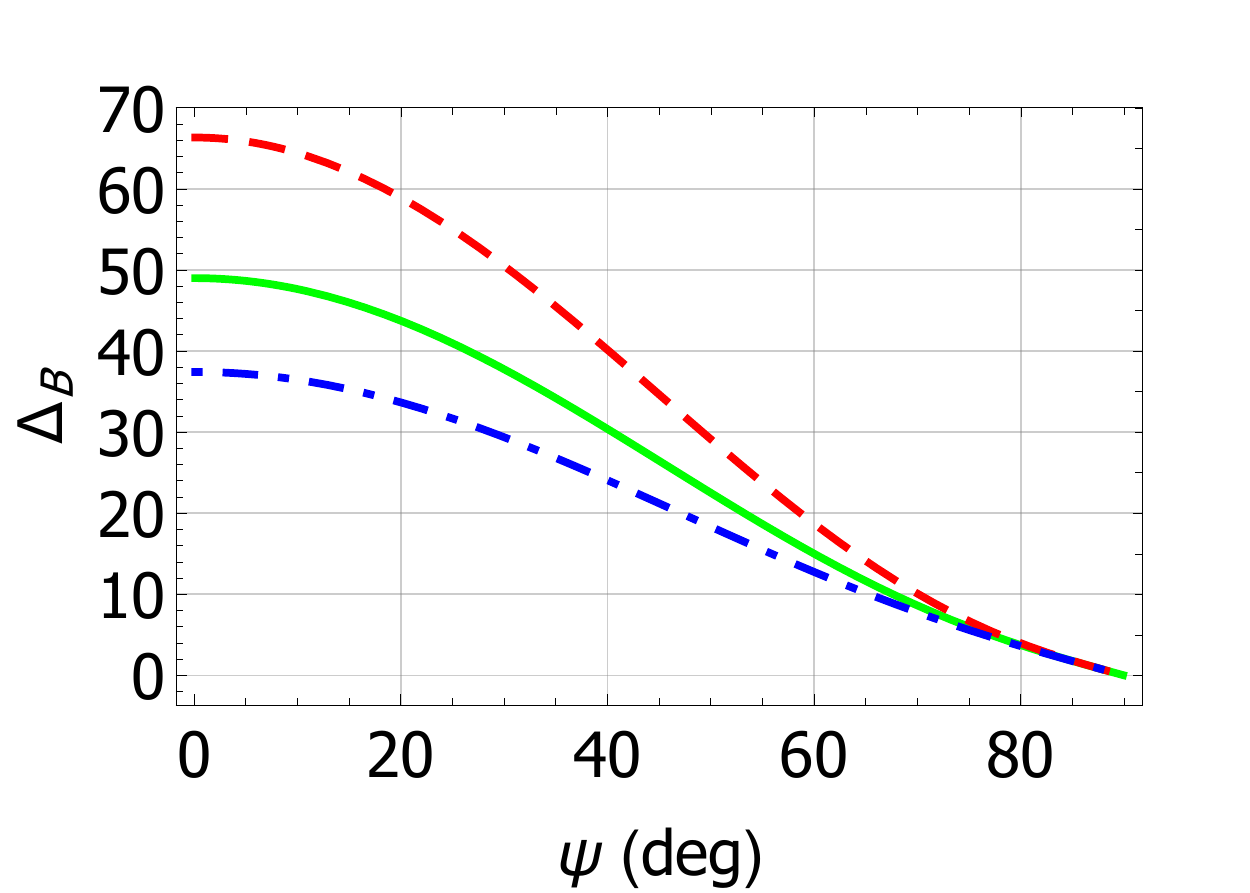}
  \caption{\label{Fig8}   
 SPP--V waves:    Plots of the real and imaginary parts of  $\eps^t_\calA$, 
and the normalized penetration depths $\Delta_\calA$ and $\Delta_\calB$, 
  versus $\psi \in \le 0, \pi/2 \ri$ for $\eps^s_\calA =- 1 + 0.1 i$
 with  $\eps_\calB = 2.5$  (blue broken-dashed curves),  $5$  (green solid  curves), and $10$  (red dashed curves).
    }
\end{figure}

Let  $\eps^s_\calA = -1 + 0.1i$. The real and imaginary  parts of $\eps^t_\calA$ that support  SPP--V-wave propagation,
as calculated from Eq.~\r{espt_sol}, are plotted
 versus $\psi \in \le 0, \pi/2 \ri$ in Fig.~\ref{Fig8} for $\eps_\calB \in \lec 2.5, 5, 10 \ric$.
 The plots of  $\mbox{Re} \lec \eps^t_\calA \ric$ converge to  $- \mbox{Re} \lec \eps^s_\calA \ric$, and 
 the plots of  $\mbox{Im} \lec \eps^t_\calA \ric$ converge to  $- \mbox{Im} \lec \eps^s_\calA \ric$, as $\psi$ approaches zero.
 On the other hand,  the plots of $\mbox{Re} \lec \eps^t_\calA \ric$ and $\mbox{Im} \lec \eps^t_\calA \ric$ both become unbounded 
 as $\psi$ approaches $\pi/2$.
 
 Also plotted in Fig.~\ref{Fig8}
are the 
 corresponding plots of the normalized penetration depths in materials $\calA$ and $\calB$, namely $\Delta_\calA$ and $\Delta_\calB$, 
 as defined in Eqs.~\r{pd_s} but with the symbol $\calA \ell$ therein replaced by $\calA$, and
 as calculated from Eqs.~\r{alphaa_sol} and \r{b_decay_const}, respectively.
Both penetration depths $\Delta_\calA$ and $\Delta_\calB$ converge to zero  as $\psi$ approaches $\pi/2$. On the other hand,  
 $\Delta_\calA$ becomes unbounded as $\psi$ approaches zero whereas $\Delta_\calB$ does not. Also, the plotted values of
  $\Delta_\calA$ are almost independent of $\eps_\calB$ whereas the plotted values of
$\Delta_\calB$ are  greater for larger values of  $\eps_\calB$, especially so at smaller values of $\psi$.

\begin{figure}[!htb]
\centering
\includegraphics[width=4.2cm]{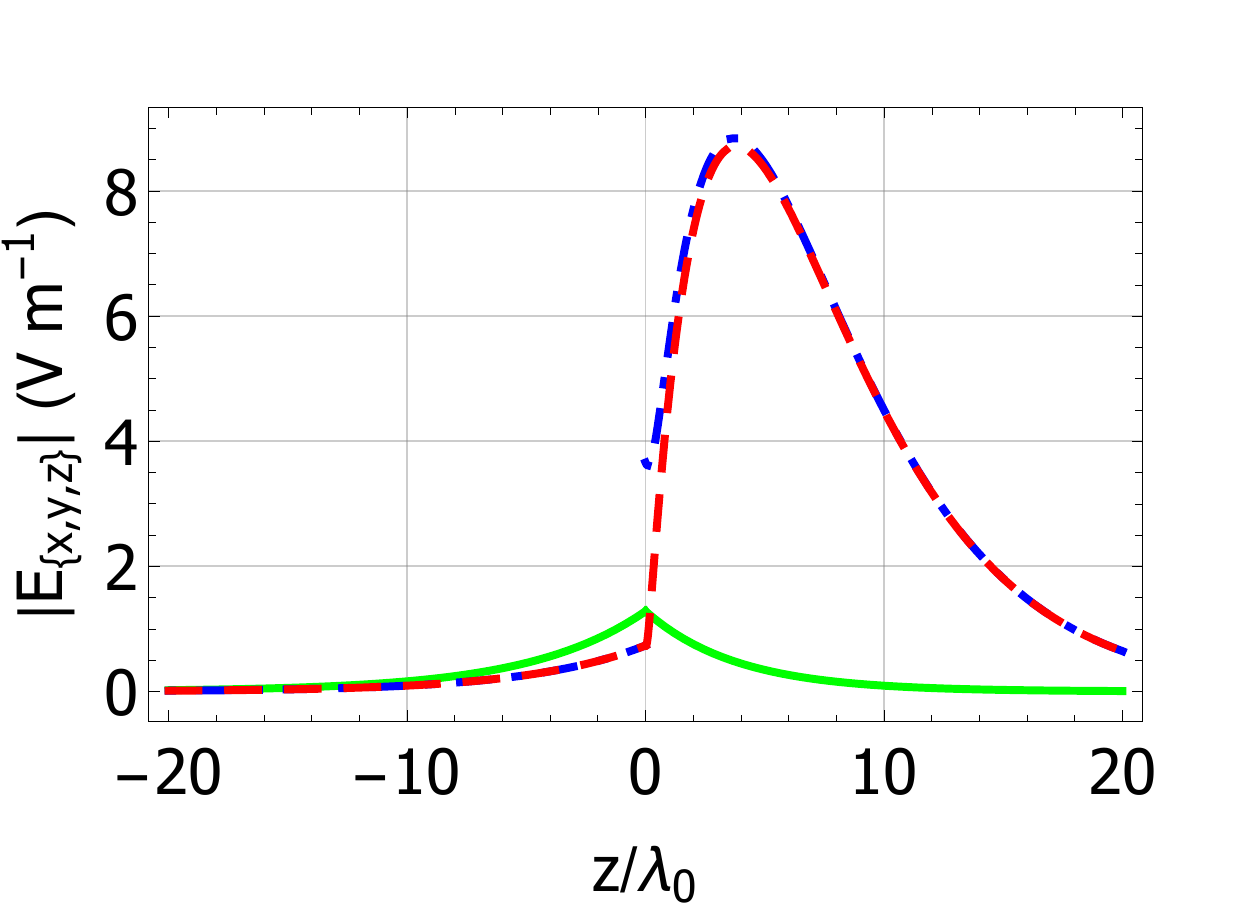} 
 \includegraphics[width=4.2cm]{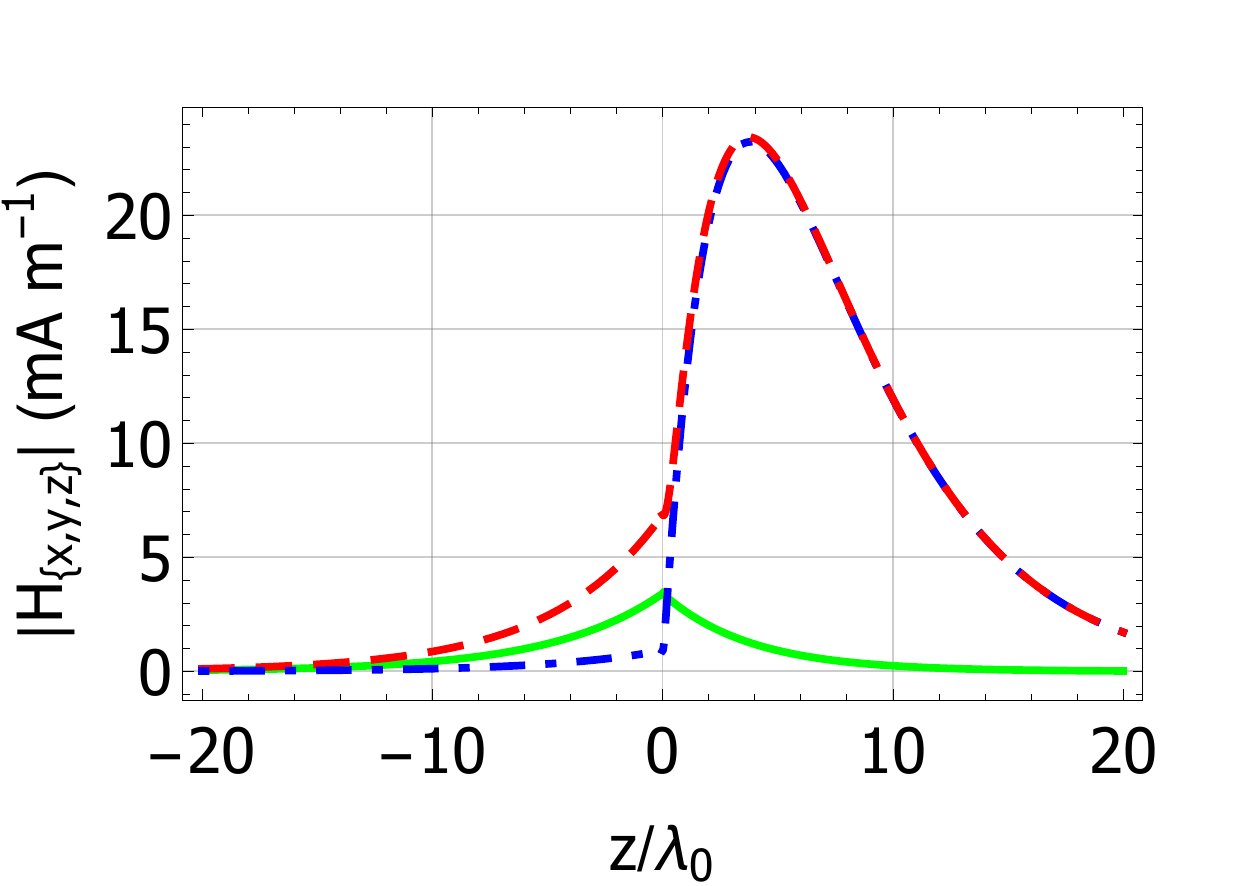}\\
  \includegraphics[width=4.2cm]{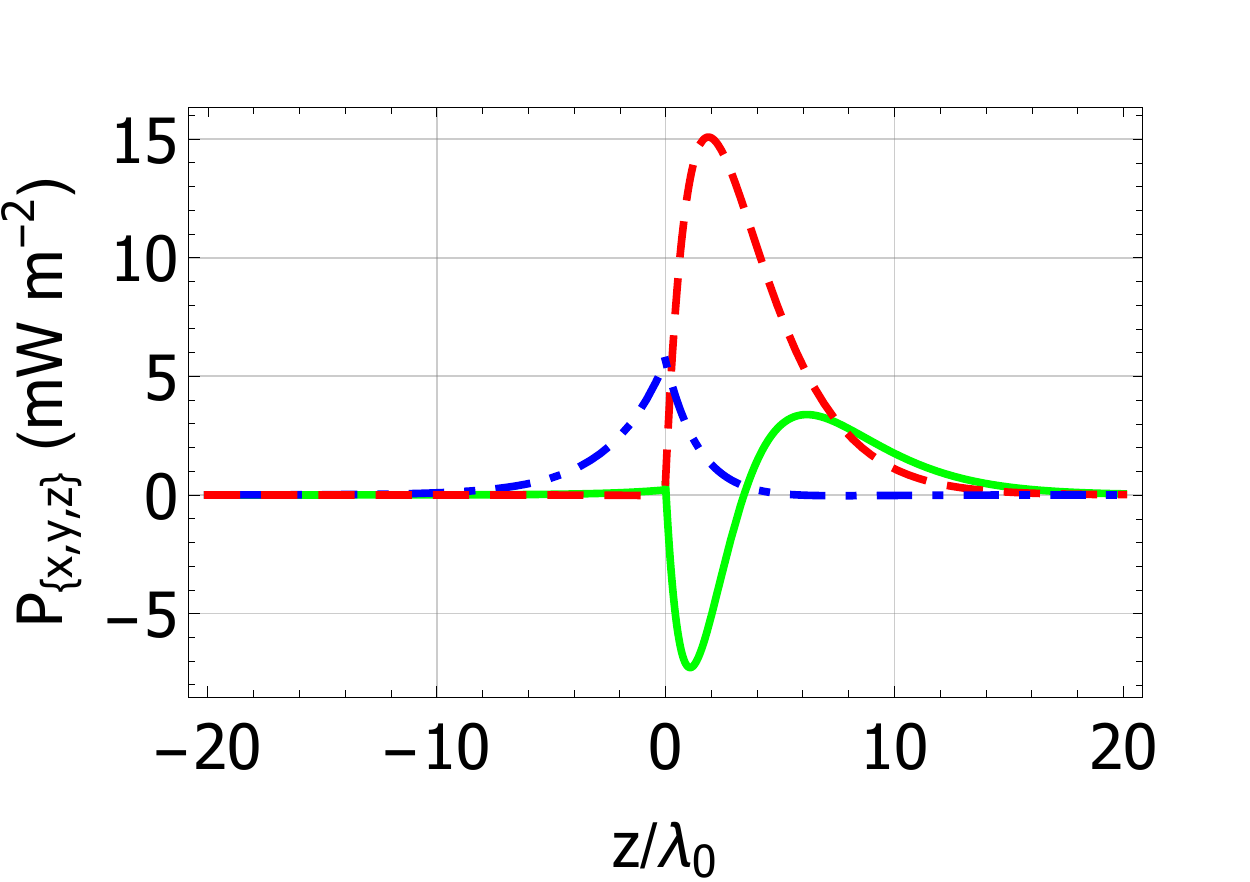}  
  \caption{\label{Fig9}   
  SPP--V waves:   
$\vert{E_{\lec x,y,z\ric}(z\uz)}\vert$,  $\vert{H_{\lec x,y,z\ric}(z\uz)}\vert$, and
$P_{\lec x,y,z\ric}(z\uz)$ plotted versus $z/\lambdao$, 
 for same parameter values as Fig.~\ref{Fig8} with
  $\eps_\mathcal{B} =5 $,  $\psi = 40^\circ$, and
   $C_{\mathcal{B}1} = 1$ V m${}^{-1}$.
  Key: as for Fig.~\ref{Fig5}.
}
\end{figure}

 The nature of  the SPP--V waves represented in  Fig.~\ref{Fig8} is further illuminated 
in Fig.~\ref{Fig9} wherein
$\vert{E_{\lec x,y,z\ric}(z\uz)}\vert$,  $\vert{H_{\lec x,y,z\ric}(z\uz)}\vert$, and
${P_{\lec x,y,z\ric}(z\uz)}$ are plotted versus $z/\lambdao$ 
 for
  $\eps_\mathcal{B} =5$, $\psi = 40^\circ$ (which corresponds to $\eps^t_\calA = -2.85+0.015 i $), and
   $C_{\mathcal{B}1} = 1$ V m${}^{-1}$. The  localization of the SPP--V wave to the interface $z=0$   is clearly demonstrated, with the degree of localization being substantially greater in the half-space $z<0$  than in the half-space $z>0$.

\begin{figure}[!htb]
\centering
\includegraphics[width=4.2cm]{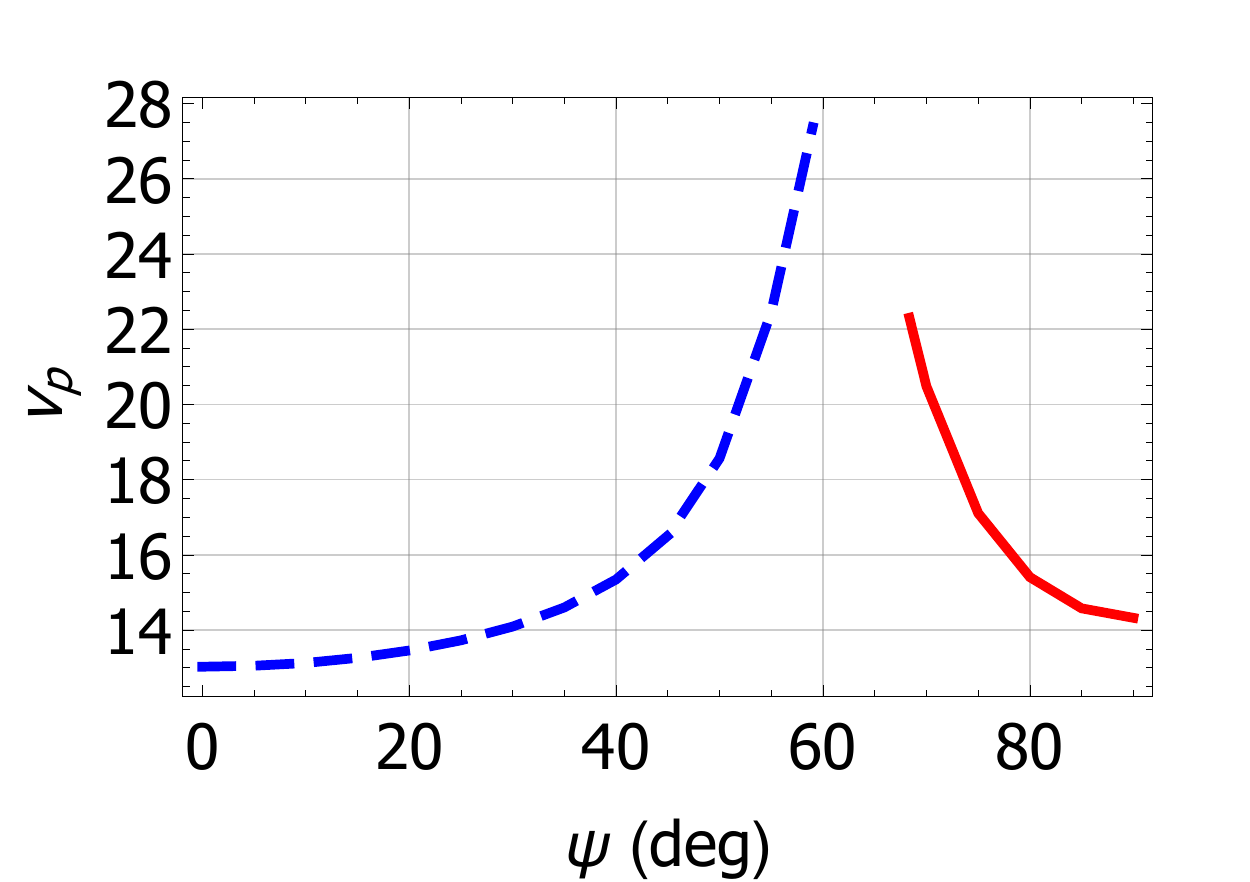}  \includegraphics[width=4.2cm]{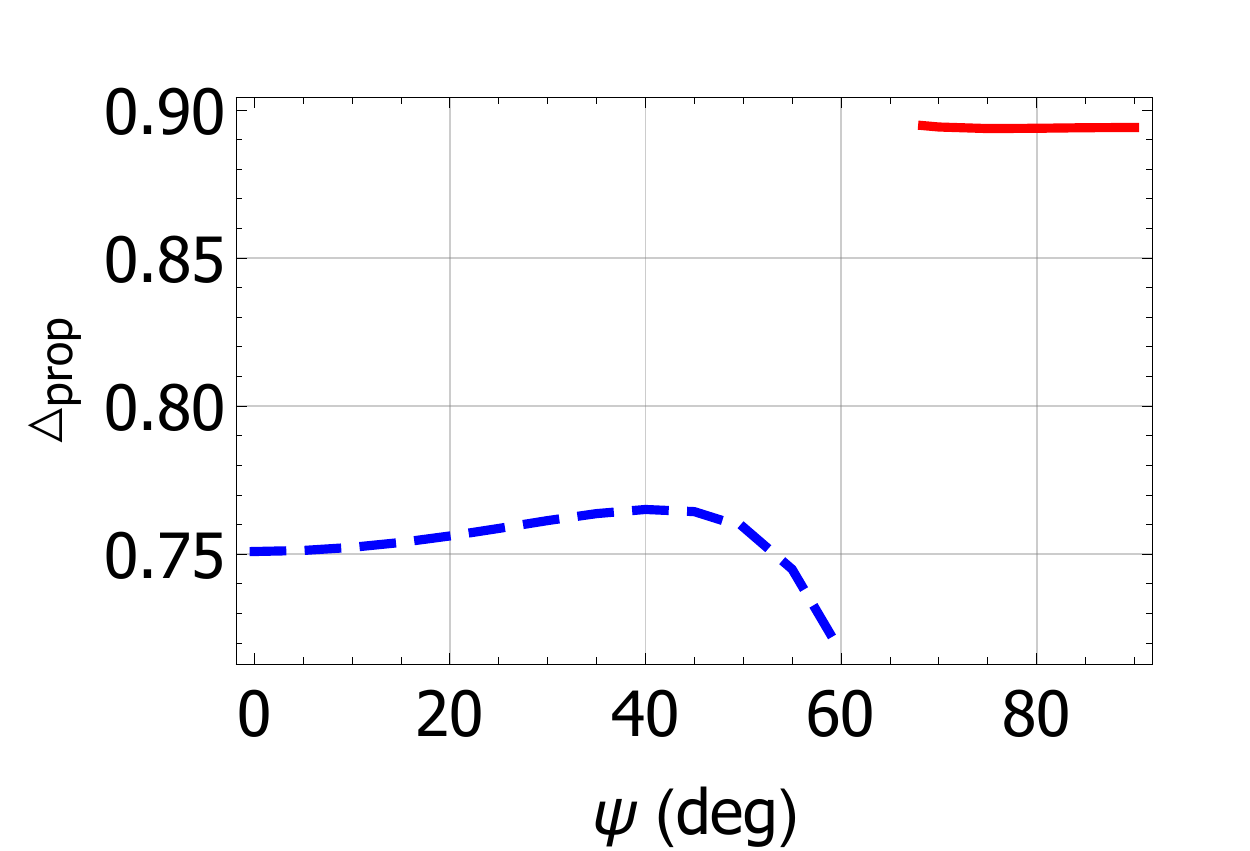}\\ 
\includegraphics[width=4.2cm]{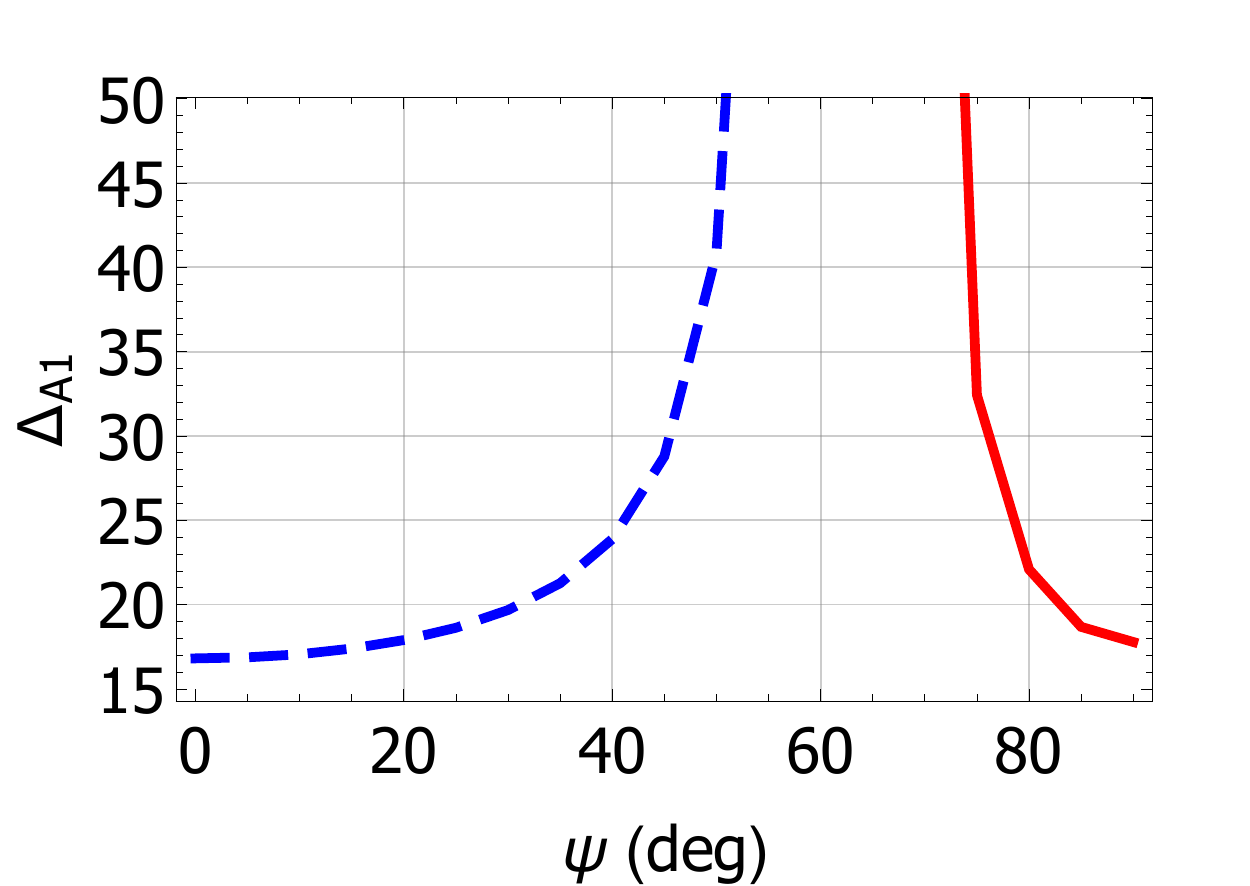}  \includegraphics[width=4.2cm]{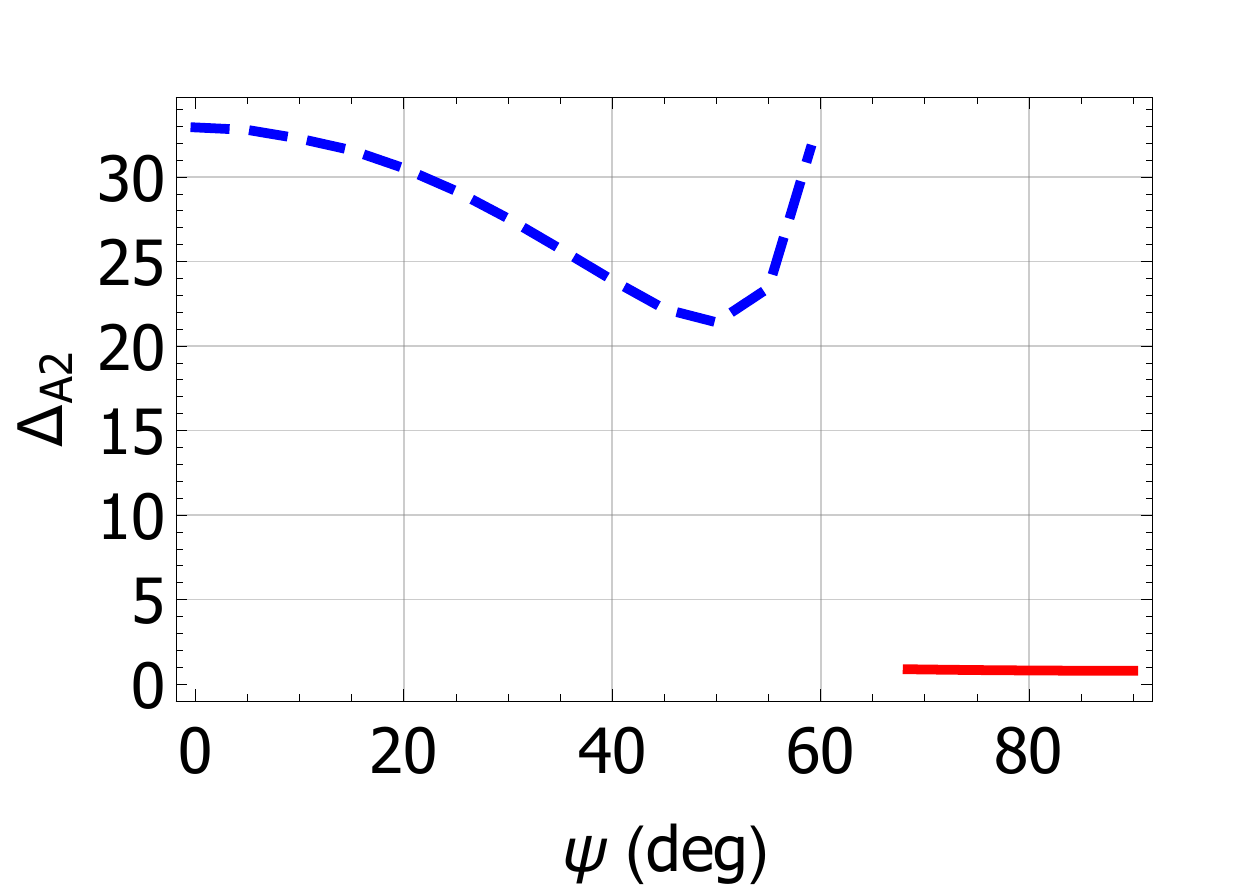}\\
\includegraphics[width=4.2cm]{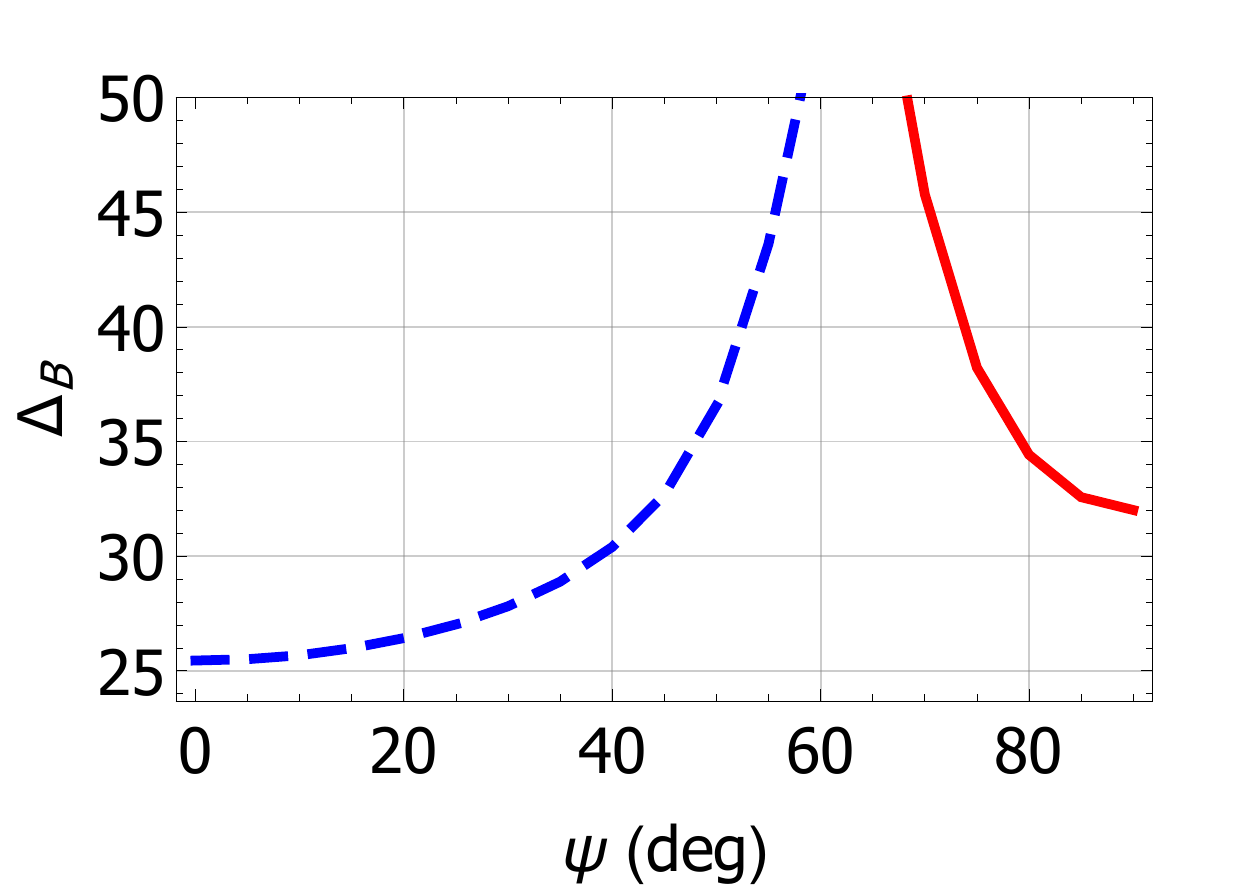} 
  \caption{\label{Fig10}    SPP waves ( SPP--V at $\psi = 40^\circ$):
Plots of the normalized phase speed $v_\text{p}$ and normalized propagation length $\Delta_{\text{prop}}$, computed using values of   $q $ extracted numerically from Eq.~\r{DE}, 
and the normalized penetration depths $\Delta_{\calA1}$,
$\Delta_{\calA2}$, and $\Delta_\calB$, as calculated from Eqs.~\r{a_decay_const} and \r{b_decay_const}, 
versus 
$\psi \in \le 0, \pi/2 \ri$ using the same relative permittivity parameters as were used for Fig.~\ref{Fig8}
with
 $\eps_\mathcal{B} =5$ at $\psi = 40^\circ$.
    }
\end{figure}

To allow a  comparison between SPP and  SPP--V waves in the same neighborhood of relative permittivity parameter values,
 in Fig.~\ref{Fig10} the 
  normalized phase speed $v_\text{p}$ and normalized propagation length $\Delta_{\text{prop}}$, computed using values of   $q $ extracted numerically from Eq.~\r{DE}, 
 are plotted versus 
$\psi \in \le 0, \pi/2 \ri$ using the same relative permittivity parameters  as were used for Fig.~\ref{Fig8}, i.e., $\eps^s_\calA =- 1 + 0.1 i$
and
 $\eps_\mathcal{B} =5$. The value  $\eps^t_\calA = -2.85+0.015 i $ was taken, which corresponds to  $\psi = 40^\circ$ in Fig.~\ref{Fig8}.
 
Also provided in Fig.~\ref{Fig10} are corresponding plots of the
 normalized penetration depths $\Delta_{\calA 1}$, $\Delta_{\calA 2}$, and $\Delta_{\calB}$, as calculated from Eqs.~\r{a_decay_const} and \r{b_decay_const}, respectively.
 The SPP-wave solution represented in Fig.~\ref{Fig10} only exists for two disjoint $\psi$ intervals: $0^\circ < \psi < 59.29^\circ$ and 
 $68.30^\circ < \psi < 90^\circ$. The penetration depths $\Delta_{\calA 1}$ and $\Delta_{\calB}$ become unbounded as $\psi$ approaches $59.29^\circ$ from below and 
 as $\psi$ 
 approaches $68.30^\circ$ from above, whereas the penetration  depth $\Delta_{\calA 1}$ remains bounded for all values of $\psi$.
 Notice that for Fig.~\ref{Fig8}, with $\eps^s_\calA =- 1 + 0.1 i$ and $\psi = 40^\circ$, the corresponding value of 
  $q / \ko$ is $0.065 + 1.307i$, as delivered by Eq.~\r{q_sol}, and this value agrees with the value of $q/\ko$ plotted in Fig.~\ref{Fig10} at $\psi = 40^\circ$.
Also, at $\psi = 40^\circ$ the penetration depths $\Delta_{\calA 1}$ and $\Delta_{\calA 2}$ 
in Fig.~\ref{Fig10}
coincide and these depths agree with $\Delta_{\calA}$ at $\psi = 40^\circ$ in Fig.~\ref{Fig8}. And also at $\psi = 40^\circ$ the penetration depths $\Delta_{\calB}$ in Figs.~\ref{Fig8} and \ref{Fig10} coincide.
Thus, the solution presented in Fig.~\ref{Fig10} represents a SPP wave for $\psi \neq 40^\circ$, but it represents a  SPP--V wave at the singular orientation $\psi = 40^\circ$.

\subsection{Anisotropic dielectric material $\calA$ / isotropic plasmonic material $\calB$}

As discussed in Sec.~\ref{constraints_sec},  SPP--V-wave propagation is not supported if material $\calA$ is a nondissipative dielectric  material and material $\calB$ is a plasmonic material. Accordingly, 
material $\calA$ is now taken to be a dissipative dielectric material  while  material $\calB$ is taken to be a plasmonic material.

\begin{figure}[!htb]
\centering
\includegraphics[width=4.2cm]{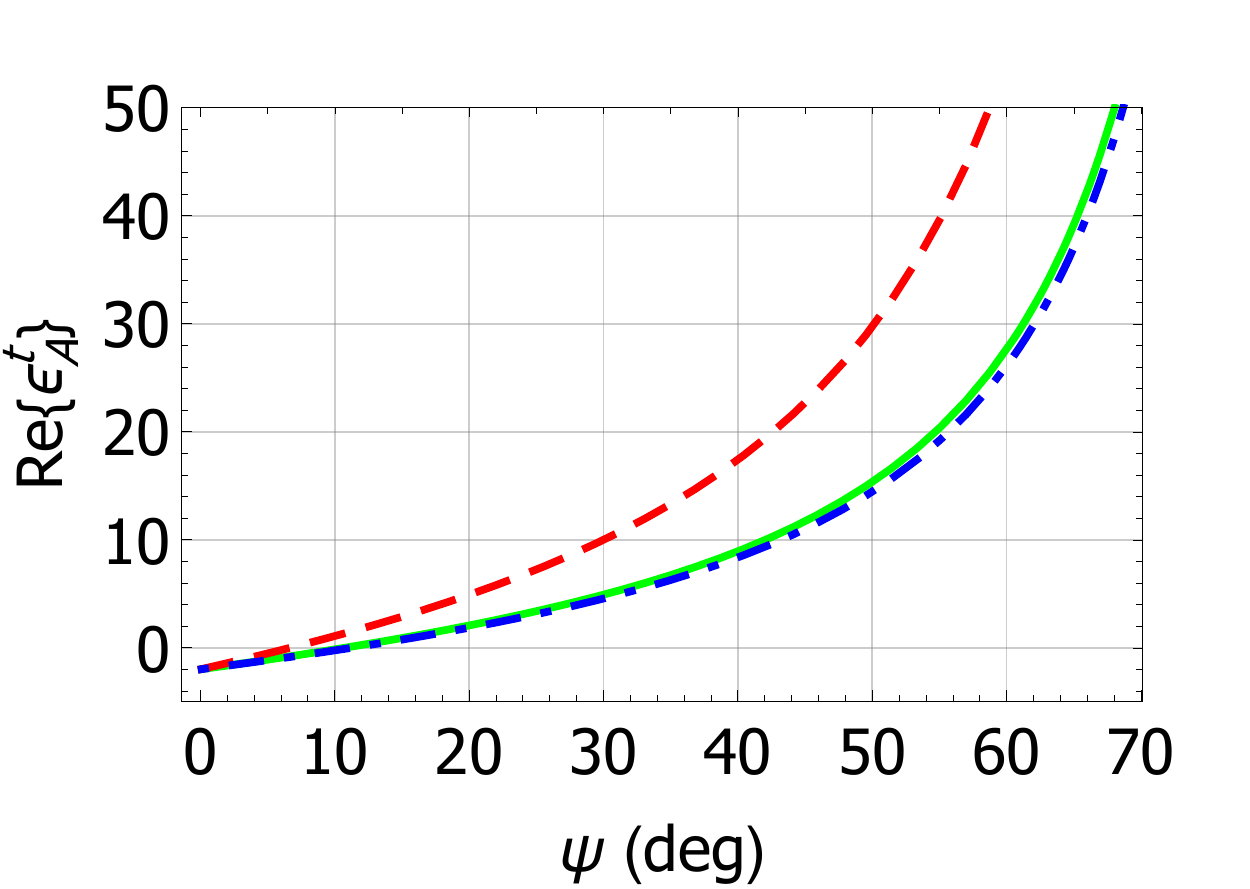}  \includegraphics[width=4.2cm]{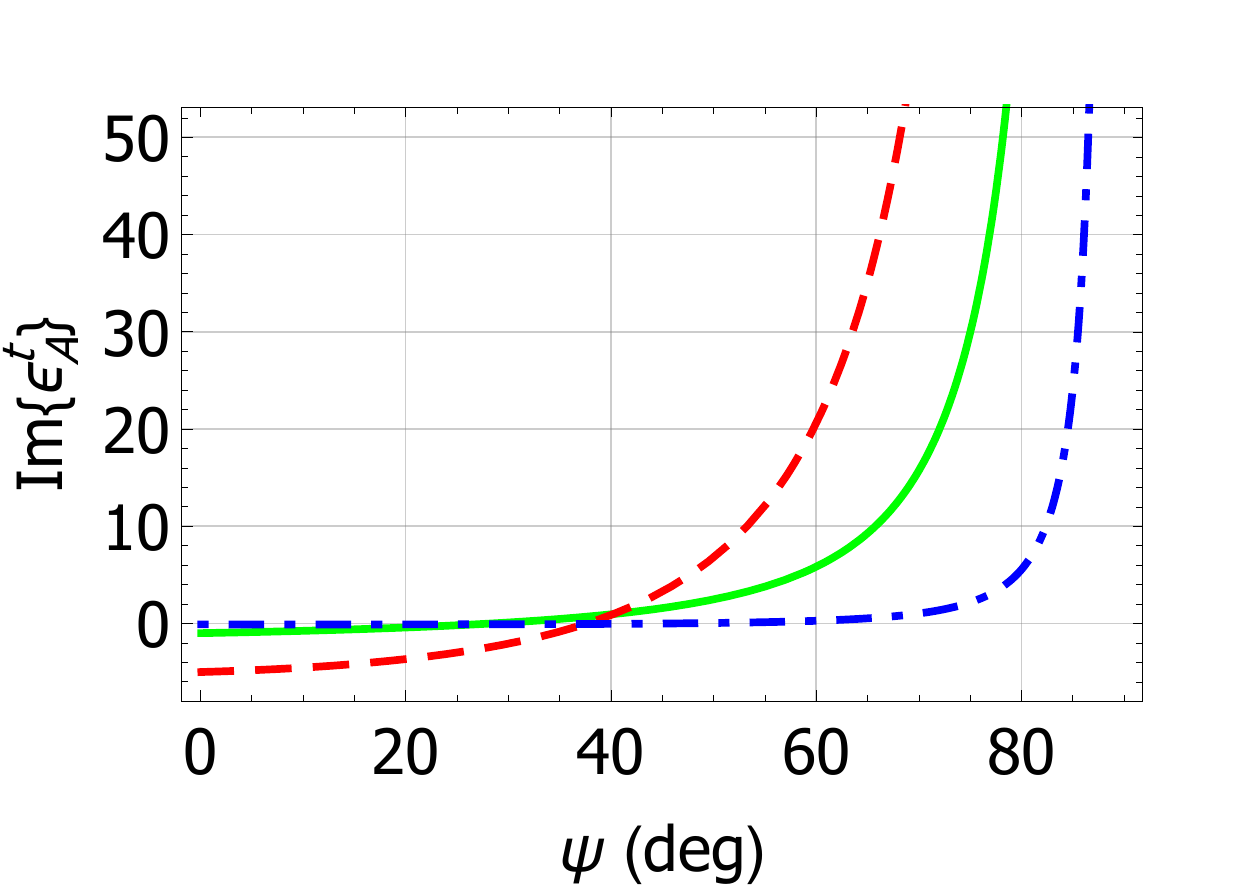}\\ 
\includegraphics[width=4.2cm]{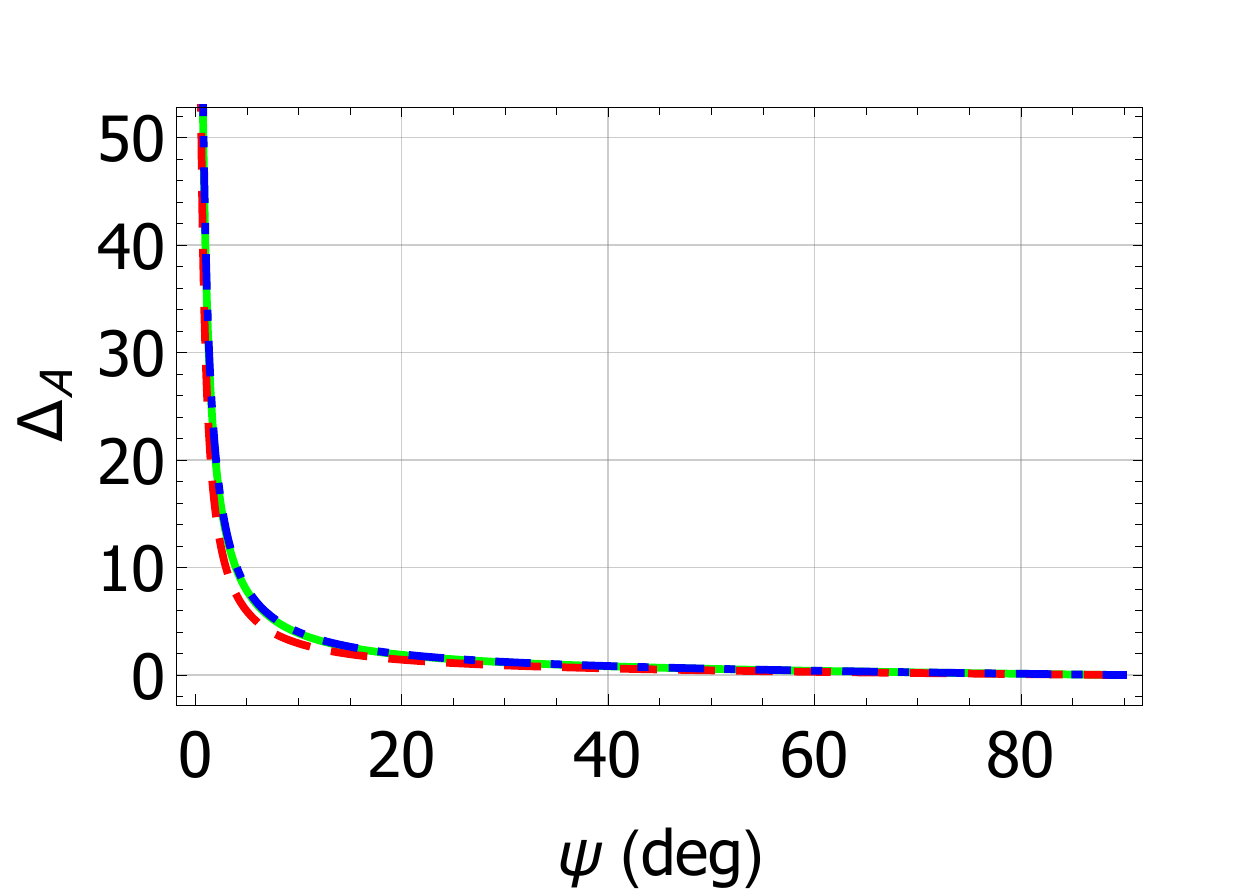}  \includegraphics[width=4.2cm]{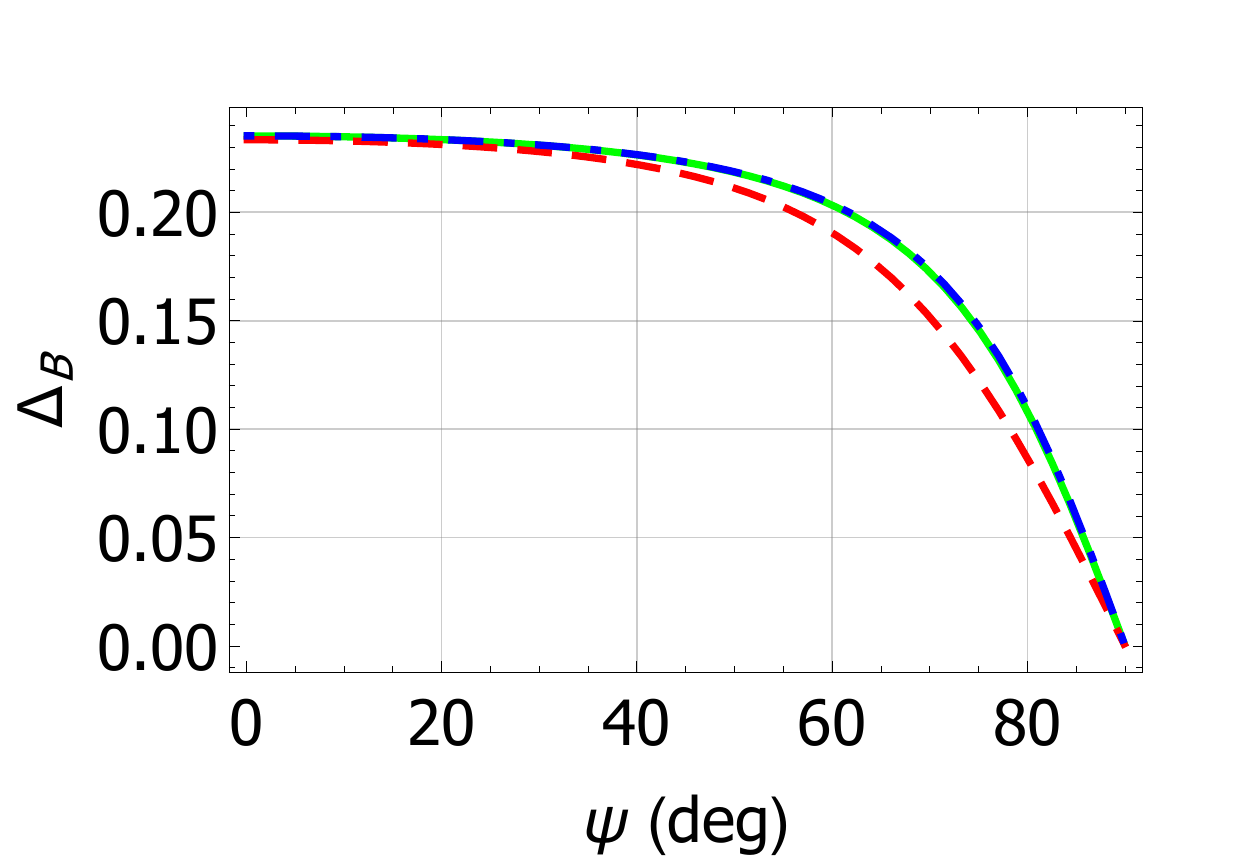}
  \caption{\label{Fig11}   
 SPP--V waves:    
  As Fig.~\ref{Fig8} but for
   $\eps_\calB = -16.07 + 0.44 i$  with $\eps^s_\calA = 2+0.1i $ (blue broken-dashed curves),  $2+i $  (green solid  curves), and $2+5i$   (red dashed curves).
    }
\end{figure}

Let  $\eps_\calB = -16.07 + 0.44 i$ (silver at $\lambdao = 600$ nm \c{Johnson}).  The real and imaginary parts  of $\eps^t_\calA$ that support  SPP--V waves are plotted
 versus $\psi \in \le 0, \pi/2 \ri$ in Fig.~\ref{Fig11} for $\eps^s_\calA \in \lec 2+0.1i,  2+i,  2+5i  \ric$.
 As in Fig.~\ref{Fig8}, the plots of  $\mbox{Re} \lec \eps^t_\calA \ric$ in Fig.~\ref{Fig11} converge to  $- \mbox{Re} \lec \eps^s_\calA \ric$, and 
 the plots of  $\mbox{Im} \lec \eps^t_\calA \ric$ in Fig.~\ref{Fig11} converge to  $- \mbox{Im} \lec \eps^s_\calA \ric$, as $\psi$ approaches zero.
  As $\psi$ approaches $\pi/2$,  the plots of $\mbox{Re} \lec \eps^t_\calA \ric$ and $\mbox{Im} \lec \eps^t_\calA \ric$ both become unbounded.
  
 Also presented in Fig.~\ref{Fig11}
are the 
 corresponding plots of the normalized penetration depths in materials $\calA$ and $\calB$, namely $\Delta_\calA$ and $\Delta_\calB$, 
 as defined in Eqs.~\r{pd_s} but with the symbol $\calA \ell$ therein replaced by $\calA$, and
 as calculated from Eqs.~\r{alphaa_sol} and \r{b_decay_const}, respectively.
As in Fig.~\ref{Fig8},
both penetration depths $\Delta_\calA$ and $\Delta_\calB$ 
 in Fig.~\ref{Fig11}
converge to zero  as $\psi$ approaches $\pi/2$.
As $\psi$ approaches zero,
 $\Delta_\calA$ becomes unbounded  whereas $\Delta_\calB$ does not. Also, the plotted values of
  $\Delta_\calA$ and $\Delta_\calB$ are almost independent of $\eps_\calB$.
 
\begin{figure}[!htb]
\centering
\includegraphics[width=4.2cm]{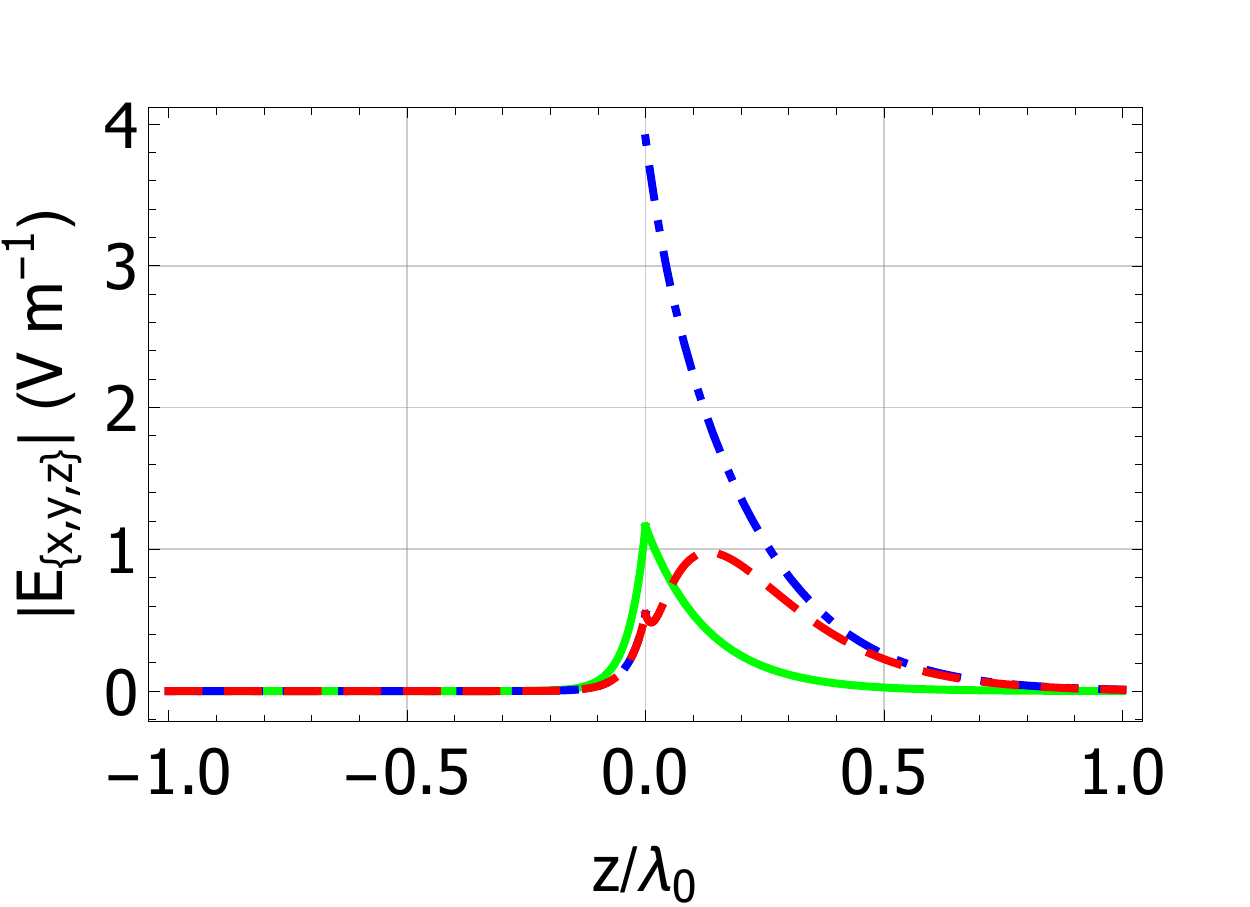} 
 \includegraphics[width=4.2cm]{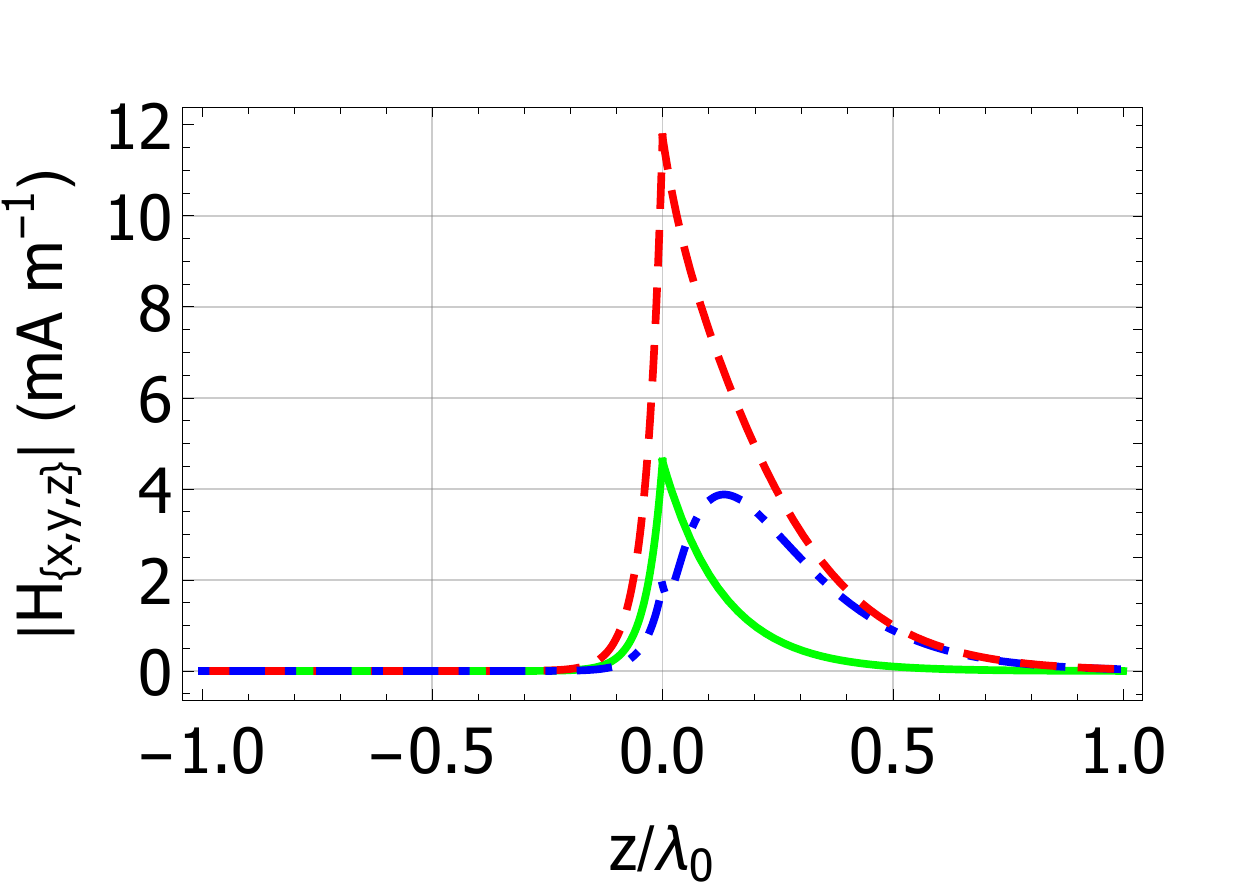} \\
  \includegraphics[width=4.2cm]{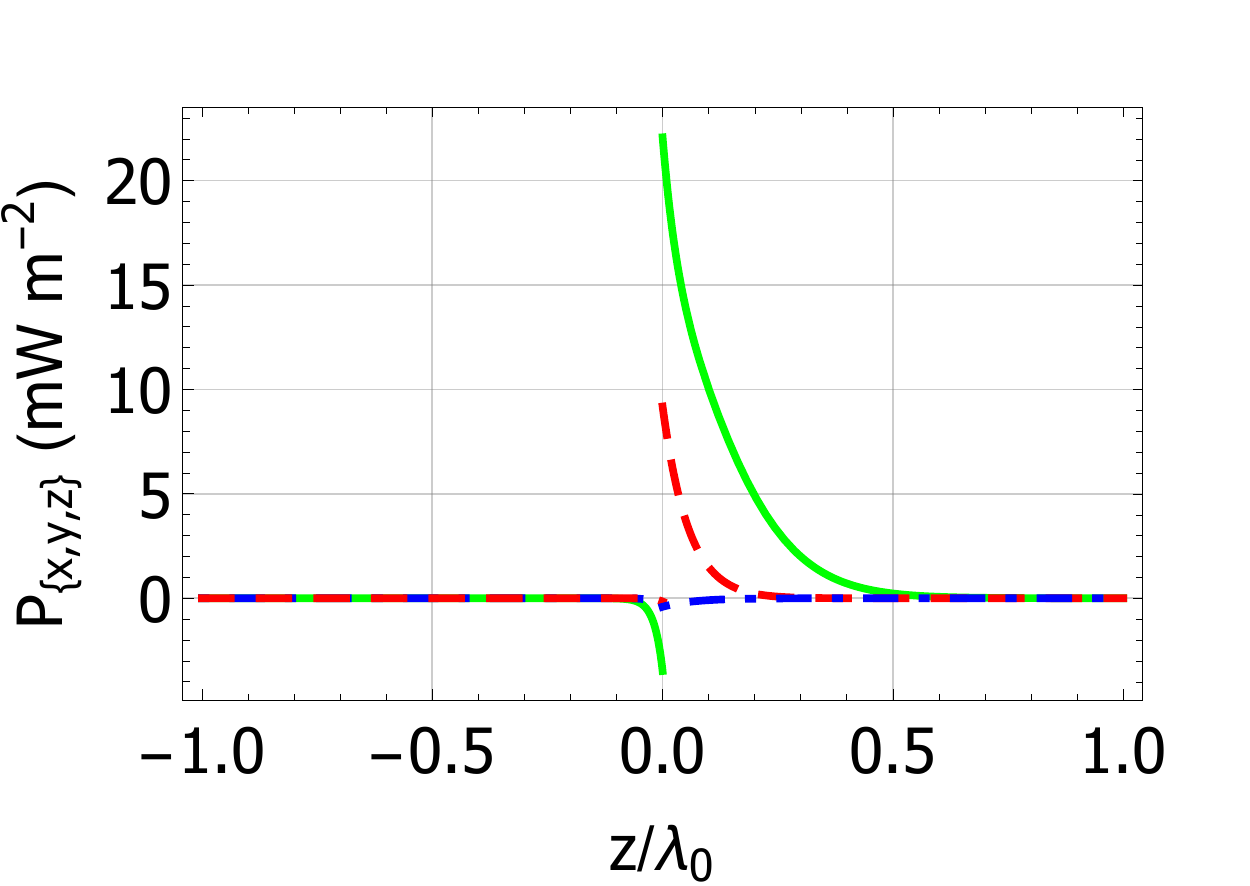}  
  \caption{\label{Fig12}   
 SPP--V waves:  
$\vert{E_{\lec x,y,z\ric}(z\uz)}\vert$,  $\vert{H_{\lec x,y,z\ric}(z\uz)}\vert$, and
${P_{\lec x,y,z\ric}(z\uz)}$ plotted versus $z/\lambdao$, 
for same parameter values as Fig.~\ref{Fig11} with
   $\eps^s_\calA = 2+ i$ and $\psi = 40^\circ$.
Also  $C_{\mathcal{B}1} = 1$ V m${}^{-1}$.
  Key: as for Fig.~\ref{Fig5}.
}
\end{figure}

  The nature of the  SPP--V waves represented in  Fig.~\ref{Fig11} is further illuminated 
in Fig.~\ref{Fig12} wherein
$\vert{E_{\lec x,y,z\ric}(z\uz)}\vert$,  $\vert{H_{\lec x,y,z\ric}(z\uz)}\vert$, and
${P_{\lec x,y,z\ric}(z\uz)}$ are plotted versus $z/\lambdao$ 
 for  $\eps^s_\calA = 2+ i$
   and $\psi = 40^\circ$ (which corresponds to $\eps^t_\calA =  8.93+0.94 i $).
For these computations, we fixed
   $C_{\mathcal{B}1} = 1$ V m${}^{-1}$. The  localization of the SPP--V wave to the interface $z=0$  is easy to see, with the degree of localization being substantially greater in the half-space $z<0$  than in the half-space $z>0$. Also, the  SPP--V wave represented in Fig.~\ref{Fig12} is localized to the interface $z=0$ to a substantially greater
   degree than the  SPP--V wave represented in Fig.~\ref{Fig9}.

\begin{figure}[!htb]
\centering
\includegraphics[width=4.2cm]{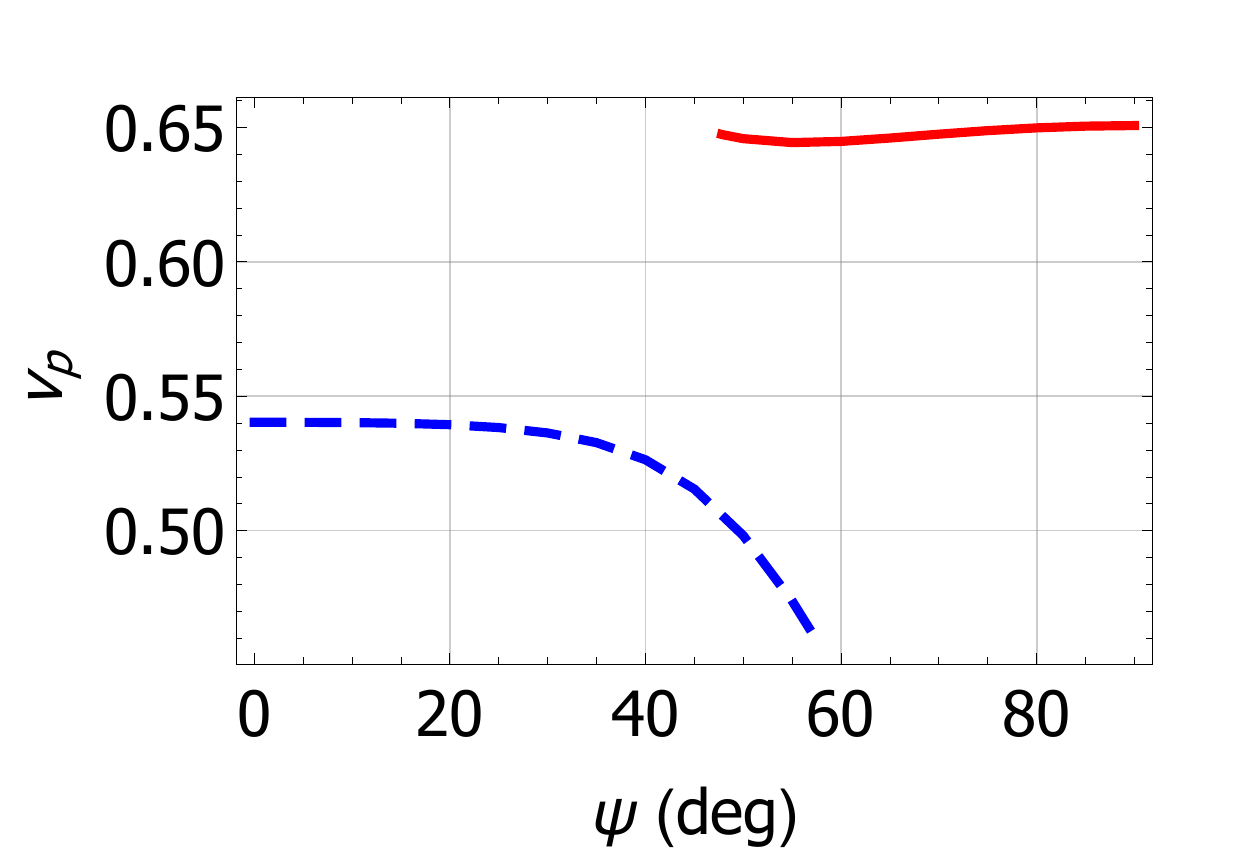}  \includegraphics[width=4.2cm]{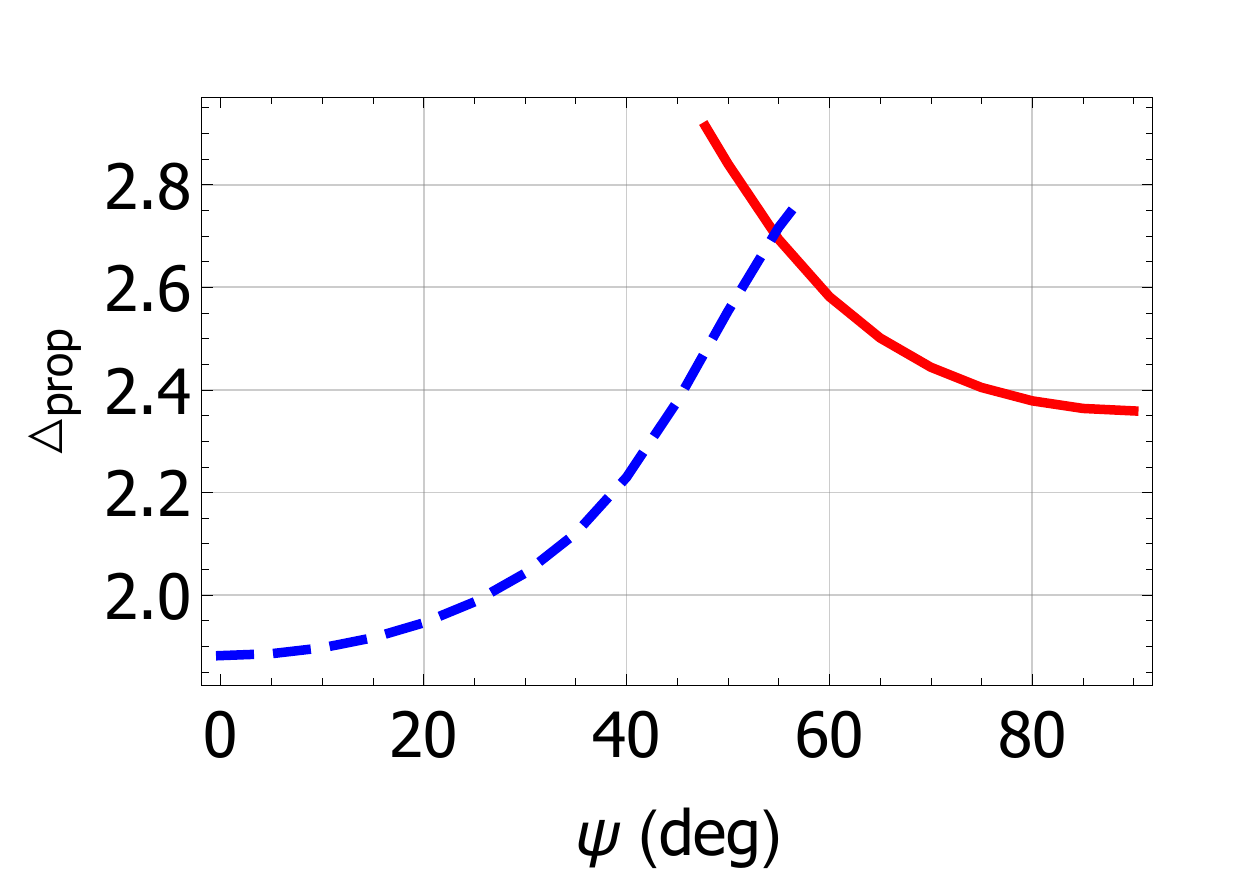}\\ 
\includegraphics[width=4.2cm]{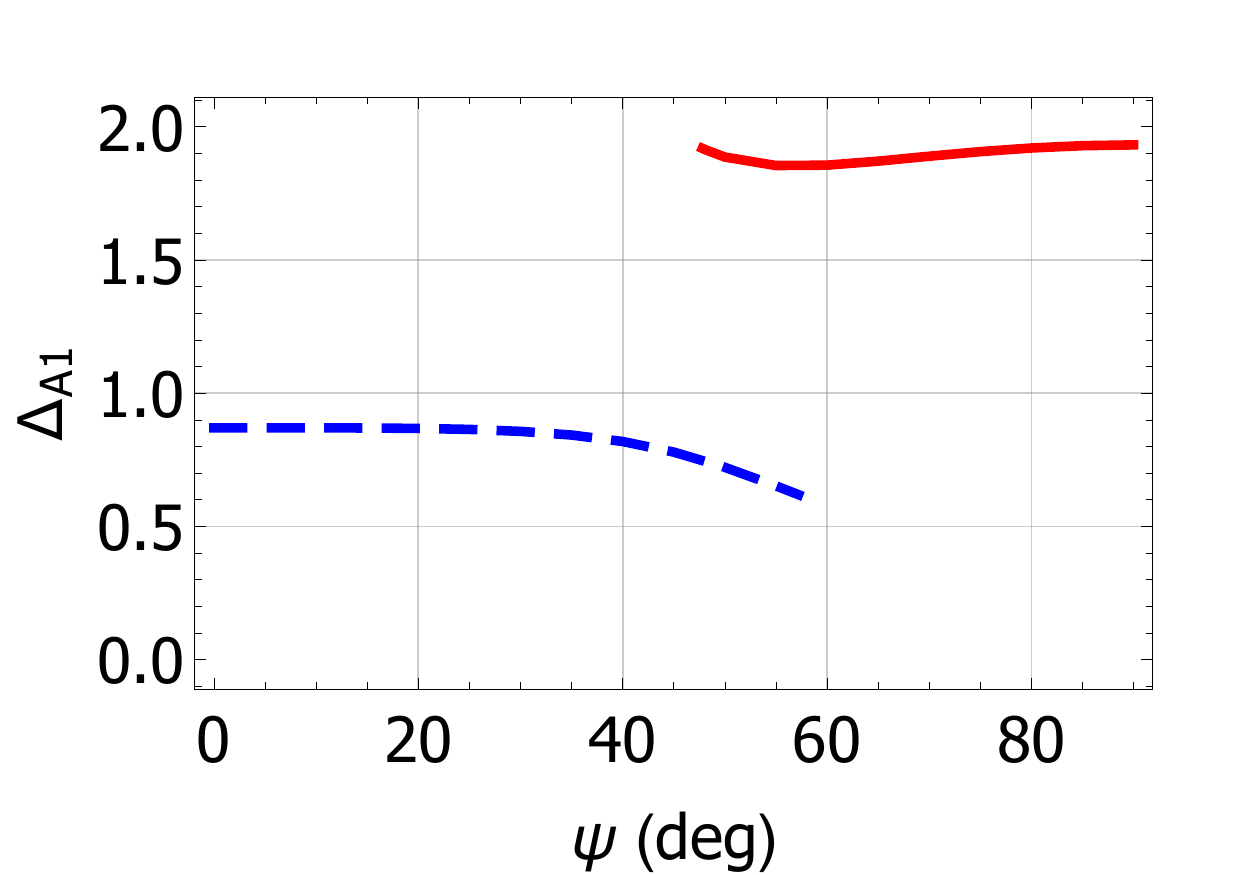}  \includegraphics[width=4.2cm]{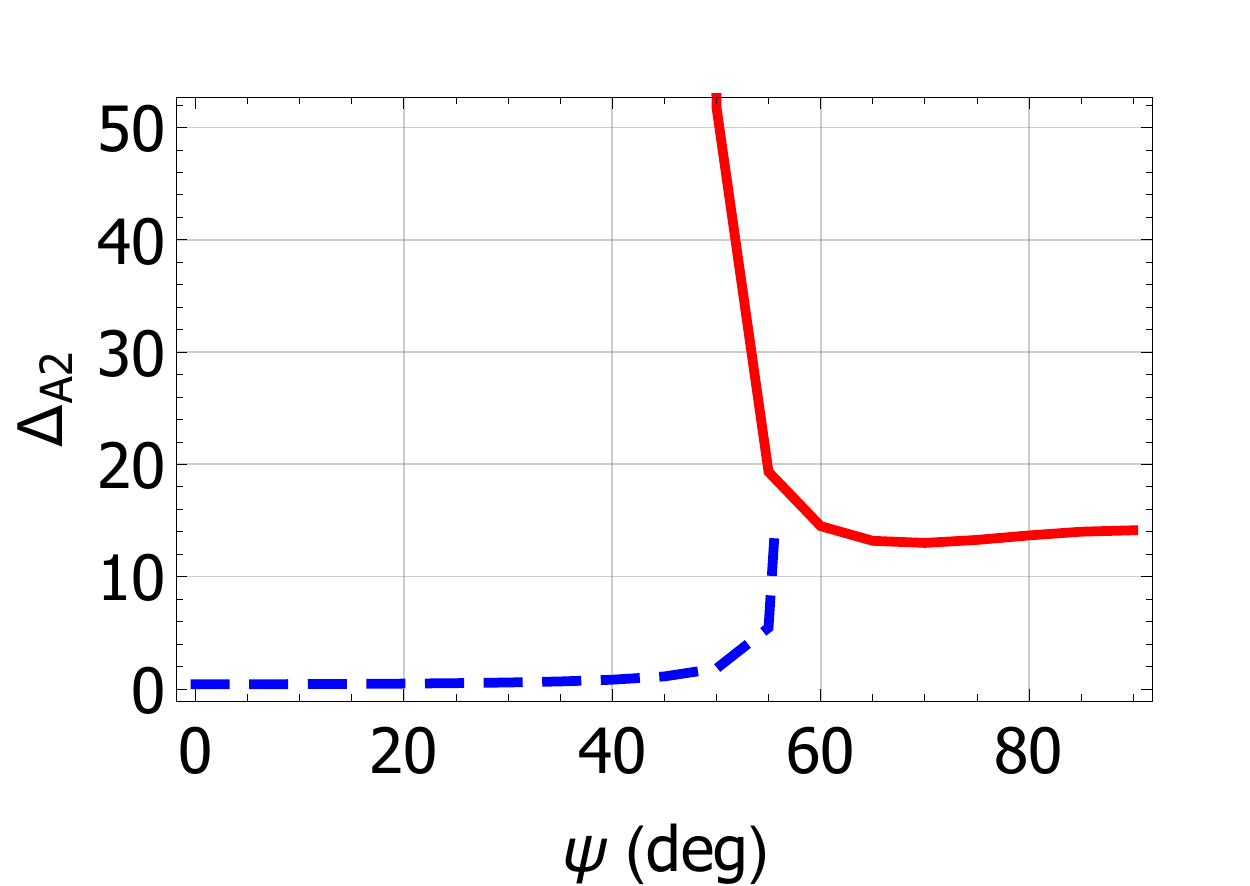}\\
\includegraphics[width=4.2cm]{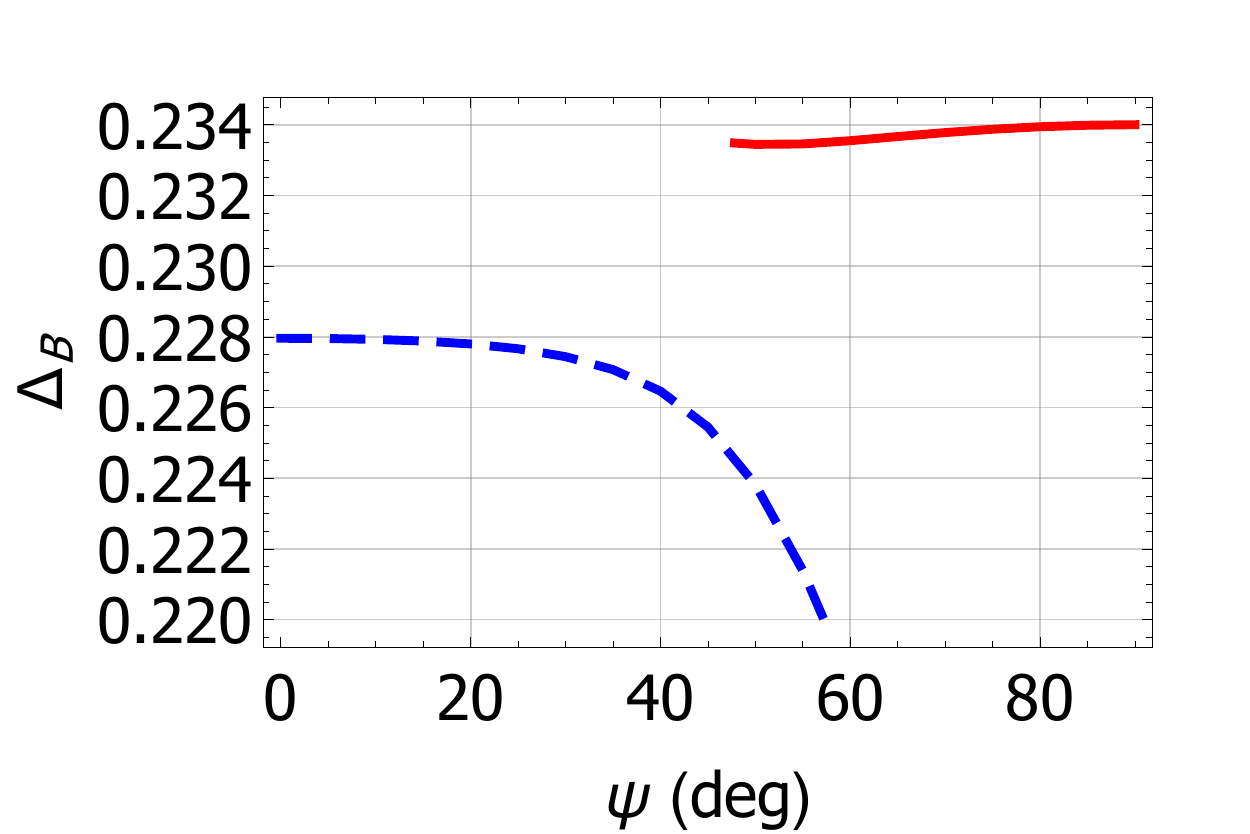} 
  \caption{\label{Fig13}    SPP waves (SPP--V wave at $\psi = 40^\circ$):
  Plots of the normalized phase speed $v_\text{p}$ and normalized propagation length $\Delta_{\text{prop}}$, computed using values of   $q $ extracted numerically from Eq.~\r{DE}, 
and the normalized penetration depths $\Delta_{\calA1}$,
$\Delta_{\calA2}$, and $\Delta_\calB$, as calculated from Eqs.~\r{a_decay_const} and \r{b_decay_const}, 
versus 
$\psi \in \le 0, \pi/2 \ri$ using the same relative permittivity parameters as were used for Fig.~\ref{Fig11}
with $\eps^s_\calA = 2+ i$ at $\psi = 40^\circ$.
    }
\end{figure}

To allow a  comparison between SPP and  SPP--V waves in the same neighborhood of relative permittivity parameter values, in Fig.~\ref{Fig13} the 
 normalized phase speed $v_\text{p}$ and normalized propagation length $\Delta_{\text{prop}}$, computed using values of   $q $ extracted numerically from Eq.~\r{DE}, 
 are plotted versus 
$\psi \in \le 0, \pi/2 \ri$ using the same relative permittivity parameters  as were used for Fig.~\ref{Fig11}, i.e., 
$\eps_\calB = -16.07 + 0.44 i$
with
$\eps^s_\calA = 2+ i$.
The value  $\eps^t_\calA = 8.93+0.94 i $ was taken, which corresponds to  $\psi = 40^\circ$ in Fig.~\ref{Fig11}.

Also provided in Fig.~\ref{Fig13} are corresponding plots of the
 normalized penetration depths $\Delta_{\calA 1}$, $\Delta_{\calA 2}$, and $\Delta_{\calB}$, as calculated from Eqs.~\r{a_decay_const} and \r{b_decay_const}, respectively.
 The SPP-wave solutions
  represented in Fig.~\ref{Fig13}
  are organized in two overlapping  branches: the first   branch exists for  $0^\circ < \psi < 57.37^\circ$
  while the second   branch exists for  $47.79^\circ < \psi < 90^\circ$.
  At  each orientation
  in the overlapping interval $47.79^\circ < \psi < 57.37^\circ$,  two SPP waves can exist.
   The penetration depths $\Delta_{\calA 2}$ and $\Delta_{\calB}$ for the solution on the first branch  become unbounded as $\psi$ approaches $57.37^\circ$ from below,
    and  $\Delta_{\calA 2}$ for the solution on the second branch  becomes unbounded as $\psi$
 approaches $47.79^\circ$ from above. In contrast, the penetration  depth $\Delta_{\calA 1}$ remains bounded for all values of $\psi$, for both solution branches.
 Notice that for Fig.~\ref{Fig11}, with $\eps^s_\calA = 2+ i$ and $\psi = 40^\circ$, the corresponding value of 
  $q / \ko$ is $1.900 + 0.448i$, as delivered by Eq.~\r{q_sol}, and this value agrees with the value of $q/\ko$ plotted in Fig.~\ref{Fig13} at $\psi = 40^\circ$ (first-branch solution).
Also, at $\psi = 40^\circ$ the penetration depths $\Delta_{\calA 1}$ and $\Delta_{\calA 2}$
in Fig.~\ref{Fig13}
coincide and these depths agree with $\Delta_{\calA}$ at $\psi = 40^\circ$ in Fig.~\ref{Fig11}. And also at $\psi = 40^\circ$ the penetration depths $\Delta_{\calB}$ in Figs.~\ref{Fig11} and \ref{Fig13} coincide.
Thus, the solution branches presented in Fig.~\ref{Fig13} represent SPP waves for $\psi \neq 40^\circ$, but
the presented solution
 represents a  SPP--V wave at the singular orientation $\psi = 40^\circ$.

\section{Closing discussion}

The theoretical underpinnings of  SPP-wave propagation supported by isotropic partnering materials are comprehensively described in the literature \c{Pitarke,ESW_book}. And the 
   case where an isotropic plasmonic material is partnered with an anisotropic dielectric material has also been considered previously \c{Sprokel1981,WSC1987,PNL07}.
   However,
 the same is not true for  SPP-wave propagation supported by anisotropic  plasmonic materials.  
 
 The matter of anisotropic partnering materials
 is  addressed in the preceding sections.
 In particular, our theoretical and numerical studies have revealed several 
  characteristics of SPP-wave propagation that do not arise for isotropic partnering materials and are therefore
 attributable to the anisotropy of the partnering materials. These characteristics are as follows:
 
 First, the phase speeds, propagation lengths, and penetration depths for SPP waves supported by anisotropic materials vary with propagation direction.
 Furthermore,  SPP-wave propagation is   not necessarily possible for all directions in the interface plane. That is, the
  angular existence domains of these SPP waves may be less than $360^\circ$, as is illustrated in Fig.~\ref{Fig10}. 
  
  Second, for certain relative permittivity-parameter regimes of the partnering materials and for certain propagation directions, the propagation of two distinct SPP waves is supported. As 
 is illustrated in Fig.~\ref{Fig13} for example, these two SPP waves have different  phase speeds, propagation lengths, and  penetration depths.

Third,  for a unique direction in each quadrant of the interface plane and for certain relative permittivity-parameter regimes of the partnering materials, the propagation of  SPP--V waves is possible. These  SPP--V waves are fundamentally different from the conventional SPP waves insofar as 
the 
 fields of   SPP--V waves decay
as the product of a linear and an exponential function of the distance from the interface in the
anisotropic partnering material; in contrast,  the fields of conventional SPP waves  decay only exponentially
with  distance from the interface.
A Voigt wave emerges in unbounded anisotropic materials when 
two planewave modes
 coalesce to form a singular wave whose amplitude varies with propagation distance \c{Panch,Ranganath}.
 An analogous physical interpretation may be extended to the emergence of SPP--V waves.

 The preceding numerical studies were based on realistic values for the relative
permittivity parameters of the partnering materials $\calA$ and $\calB$. 
With these realistic values, the requirements  for
multiple SPP-wave propagation (i.e., a  dielectric partnering material
that is both  anisotropic  and dissipative) and constraints for
 SPP--V wave propagation (as established in Sec.~IIF) could be satisfied. For the purposes  of flexibility of presentation, the anisotropic partnering material was taken  to be 
a homogenized composite material whose  relative
permittivity parameters could be conveniently varied. However, there is no reason to suspect that the requirements for
multiple SPP-wave propagation and constraints for  
SPP--V wave propagation could not satisfied by an anisotropic partnering  material with a simpler microstructure.
Numerical results (not presented in this paper) that are qualitatively similar to those presented  in Secs.~\ref{SPP-num} and \ref{SPPV-num} were obtained when the relative permittivity parameters were varied by modest amounts.
Parenthetically, where an homogenized composite material is used 
as the anisotropic plasmonic partnering material, care must be exercised to choose
a plasmonic component material that exhibits a moderately high degree of dissipation; we chose cobalt which has a relative permittivity of  $-11.63 + 17.45 i$  at $\lambdao = 600$ nm  \c{J_Co}. This is because conventional  homogenization formalisms that can be used to estimate the constitutive parameters of such homogenized composite materials, such as the Bruggeman formalism \c{MAEH} adopted herein, 
can give unphysical estimates if the plasmonic component material is only weakly dissipative \c{ML_Br}.

In closing, we note that the
existence of multiple SPP waves 
(for homogeneous partnering materials), and
the existence
 of SPP--V waves with   mixed exponential and linear localization characteristics, have not
 been reported in previous SPP studies~---~and, in particular, these phenomenons have not been reported in previous studies of SPP waves involving anisotropic dielectric materials partnered
with isotropic plasmonic materials \c{Sprokel1981,WSC1987,PNL07}.  
The results reported herein have emerged from theoretical and numerical investigations of the 
corresponding canonical boundary-value problem. While the canonical 
boundary-value problem represents an idealization that does not take account of finite thicknesses of the partnering materials or the process(es) of excitation of the surface waves, it does yield useful information on the essential physics of surface-wave propagation. Further study is required to explore the excitation
and propagation
 of multiple SPP waves and SPP--V waves for experimental scenarios.


\vspace{10mm}
\noindent Acknowledgments:
This work was supported by
EPSRC (grant number EP/S00033X/1) and US NSF (grant number DMS-1619901).
AL thanks the Charles Godfrey Binder Endowment at the Pennsylvania State University for ongoing support of his research.

\end{document}